%% file: 0__AFNO_MAIN.tex
\documentclass{article}

\usepackage[utf8]{inputenc}
\usepackage[T1]{fontenc}
\usepackage{anyfontsize}
\AtBeginDocument{\fontsize{10.5}{12.6}\selectfont}
\AtEndDocument{\escapechar=92}

\usepackage{amsmath, amssymb, amsfonts, amsthm}
\usepackage{mathtools}
\usepackage{bbm}
\usepackage{framed}

\usepackage{lmodern}
\usepackage{microtype}
\usepackage{fontawesome}

\usepackage{graphicx}
\usepackage{rotating}
\usepackage{lscape}
\usepackage{adjustbox}
\usepackage{subcaption}
\usepackage{float}
\usepackage{wrapfig}

\usepackage[table,dvipsnames]{xcolor}
\definecolor{bhpurple}{RGB}{250,4,10}   %
\definecolor{bhpurple}{HTML}{9928A0}

\usepackage{booktabs}
\usepackage{multirow}
\usepackage{siunitx}
\usepackage{nicematrix}
\definecolor{oxfordblue}{rgb}{0.0, 0.13, 0.28}
\definecolor{linkblue}{rgb}{0.0, 0.20, 0.40}
\definecolor{limegreen}{rgb}{0.2, 0.8, 0.2}
\definecolor{darkred}{rgb}{0.6, 0.0, 0.0}
\definecolor{bondiblue}{rgb}{0.0, 0.58, 0.71}
\definecolor{britishracinggreen}{rgb}{0.0, 0.26, 0.15}
\definecolor{faintgray}{RGB}{245,245,245}
\definecolor{faintborder}{RGB}{230,230,230}
\definecolor{lightblack}{gray}{0.4}
\definecolor{pinegreen}{rgb}{0.0, 0.47, 0.44}

\usepackage{hyperref}
\hypersetup{
    colorlinks = true,
    linkcolor  = linkblue,
    anchorcolor= linkblue,
    citecolor  = linkblue,
    filecolor  = linkblue,
    urlcolor   = linkblue
}

\usepackage{enumitem}
\usepackage{fancyvrb}
\usepackage{silence}
\usepackage[format=default,font=small,labelfont=it]{caption}
\usepackage[UKenglish]{isodate}
\usepackage{etoolbox}
\usepackage{comment}

\usepackage{appendix}

\usepackage[numbers]{natbib}

\usepackage{tikz}
\usetikzlibrary{arrows.meta,calc,positioning,fit,backgrounds,intersections,decorations.pathreplacing}
\definecolor{nBlue}{RGB}{47,110,173}
\definecolor{nOrange}{RGB}{204,102,34}
\definecolor{nGreen}{RGB}{64,153,115}
\definecolor{nPurple}{RGB}{191,120,170}
\definecolor{nGray}{RGB}{90,90,90}

\usepackage{tikz-cd}
\pgfdeclarelayer{edgelayer}
\pgfdeclarelayer{nodelayer}
\pgfsetlayers{edgelayer,nodelayer,main}
\tikzstyle{new style 0}=[fill={rgb,255: red,255; green,94; blue,247}, draw=black, shape=circle]
\tikzstyle{pointy}=[fill=white, draw=black, shape=circle]
\tikzstyle{pointy}=[->]

\usepackage{algpseudocode}
\usepackage[linesnumbered,lined,boxed,commentsnumbered,ruled,longend]{algorithm2e}

\SetCommentSty{mycommfont}

\usepackage[most]{tcolorbox}
\usepackage[normalem]{ulem}

\usepackage{cleveref}
\newcounter{termcounter}
\renewcommand{\thetermcounter}{\Roman{termcounter}}
\crefname{term}{term}{terms}
\creflabelformat{term}{#2\textup{(#1)}#3}

\makeatletter
\def\term{\@ifnextchar[\term@optarg\term@noarg}%
\def\term@optarg[#1]#2{%
  \textup{#1}%
  \def\@currentlabel{#1}%
  \def\cref@currentlabel{[][2147483647][]#1}%
  \cref@label[term]{#2}}
\def\term@noarg#1{%
  \refstepcounter{termcounter}%
  \textup{(\thetermcounter)}%
  \cref@label[term]{#1}}
\makeatother

\Crefname{lemma}{Lemma}{Lemmata}
\crefname{ass}{Assumption}{assumptions}
\crefname{example}{Example}{Examples}
\crefname{proposition}{Proposition}{Proposition}
\crefname{algorithm}{algorithm}{algorithms}
\Crefname{appendix}{Appendix}{Appendices}

\usepackage[a4paper, top=2.5cm, bottom=2.5cm, left=2.7cm, right=2.7cm]{geometry}

\newcommand\numberthis{\addtocounter{equation}{1}\tag{\theequation}}

\newcommand{\E}{\mathbb{E}}

\renewcommand{\P}{\mathbb{P}}

\newcommand{\R}{\mathbb{R}}

\newcommand{\K}{\mathbb{K}}

\DeclareRobustCommand{\R}{\mathbb{R}}

\renewcommand{\psi}{\varphi}

\newcommand{\eqdef}{\ensuremath{\,\raisebox{-1pt}{$\stackrel{\mbox{\upshape\tiny def.}}{=}$}}\,}

\newcommand{\pushright}[1]{\ifmeasuring@#1\else\omit\hfill$\displaystyle#1$\fi\ignorespaces}
\newcommand{\pushleft}[1]{\ifmeasuring@#1\else\omit$\displaystyle#1$\hfill\fi\ignorespaces}


\theoremstyle{plain}
\newtheorem{theorem}{Theorem}[section]
\newtheorem{proposition}[theorem]{Proposition}
\newtheorem{lemma}[theorem]{Lemma}
\newtheorem{corollary}[theorem]{Corollary}

\theoremstyle{definition}
\newtheorem{ass}[theorem]{Assumption}
\newtheorem{definition}[theorem]{Definition}
\newtheorem{example}[theorem]{Example}

\newtheorem{setting}{Setting}[section]

\theoremstyle{remark}
\newtheorem{remark}[theorem]{Remark}

\newcommand{\addQEDstyle}[2]{%
  \AtBeginEnvironment{#1}{\pushQED{\qed}\renewcommand{\qedsymbol}{#2}}%
  \AtEndEnvironment{#1}{\popQED}}
\addQEDstyle{remark}{$\triangleleft$}
\addQEDstyle{example}{$\triangleleft$}

\newcounter{question}
\newtcolorbox[auto counter, use counter=question]{question}[1][]{
  enhanced,
  colback=faintgray,
  colframe=faintborder,
  boxrule=0.2pt,
  arc=2mm,
  title=\textcolor{lightblack}{\textbf{Question~\thequestion}},
  fonttitle=\bfseries,
  before upper={\centering\itshape},
  after title={\vspace{0.5ex}},
  boxsep=4pt,
  left=6pt,
  right=6pt,
  top=4pt,
  bottom=4pt,
  #1
}

\usepackage{soul}
\usepackage[colorinlistoftodos]{todonotes}

\definecolor{bleudefrance}{rgb}{0.19, 0.55, 0.91}
\definecolor{britishracinggreen}{rgb}{0.0, 0.26, 0.15}
\definecolor{darkcyan}{rgb}{0.0, 0.55, 0.55}
\definecolor{darkgreen}{rgb}{0.0, 0.2, 0.13}
\definecolor{darkspringgreen}{rgb}{0.09, 0.45, 0.27}
\definecolor{darkmidnightblue}{rgb}{0.0, 0.2, 0.4}
\definecolor{MidnightBlue}{RGB}{25,25,112}
\definecolor{MidnightBlueComplementingGreen}{RGB}{25,112,25}
\definecolor{MidnightBlueComplementingPurple}{RGB}{112,25,112}
\definecolor{MidnightBlueComplementingRed}{RGB}{112,25,69}
\definecolor{WowColor}{rgb}{.75,0,.75}
\definecolor{MildlyAlarming}{rgb}{0.85,0.25,0.1}
\definecolor{SubtleColor}{rgb}{0,0,.50}
\definecolor{antiquefuchsia}{rgb}{0.57, 0.36, 0.51}
\definecolor{fashionfuchsia}{rgb}{0.96, 0.0, 0.63}
\definecolor{jade}{rgb}{0.0, 0.66, 0.42}
\definecolor{caribbeangreen}{rgb}{0.0, 0.8, 0.6}
\definecolor{aquamarine}{rgb}{0.5, 0.8, 0.85}
\definecolor{lightseagreen}{rgb}{0.13, 0.7, 0.67}
\definecolor{attentioncolor}{RGB}{152,90,81}
\definecolor{burgred}{RGB}{40,3,22}
\definecolor{AnnieGreen}{RGB}{17,123,92}
\definecolor{Turquoise}{RGB}{64,224,208}
\definecolor{darkjade}{RGB}{0,122,84}
\definecolor{RedAlizarin}{rgb}{0.82, 0.1, 0.26}
\definecolor{Window1}{RGB}{92,150,31}
\definecolor{Window1dark}{RGB}{41,67,13}
\definecolor{Window2}{RGB}{255,168,28}
\definecolor{Window2dark}{RGB}{114,75,12}
\definecolor{Window3}{RGB}{255,96,33}
\definecolor{Window3dark}{RGB}{97,36,12}
\definecolor{InputColor}{RGB}{20,255,177}
\definecolor{InputColorlight}{RGB}{222,237,229}

\title{AFNO: Universal Certifiably-Arbitrage-Free Operator Learning Framework for European Call-Price Surfaces}

\author{%
  \makebox[\textwidth][c]{%
    \textbf{Anastasis~Kratsios}\textsuperscript{3,4}\footnote{Corresponding author.}
    \quad
    \textbf{Yannick~Limmer}\textsuperscript{5}
    \quad
    \textbf{Hans~Buehler}\textsuperscript{1}
  }\\
  \makebox[\textwidth][c]{%
    \textbf{Blanka~Horvath}\textsuperscript{1,2}
    \quad
    \textbf{Raeid~Saqur}\textsuperscript{1,4}
  }\\[0.5em]
  \parbox{0.9\textwidth}{\centering\small
    \textsuperscript{1}Mathematical Institute, University of Oxford \\
    \textsuperscript{2}Oxford-Man Institute for Quantitative Finance, University of Oxford \\
    \textsuperscript{3}Department of Mathematics, McMaster University \\
    \textsuperscript{4}Vector Institute \\
    \textsuperscript{5}DRW
  }
}

\date{ }

\begin{document}

\maketitle

\begin{abstract}
Neural operators (NO) have been introduces as deep learning-powered solutions to the computational bottleneck of resolving families of problems; since, once trained an NO can rapidly output approximate solutions, thus avoiding the need to re-run classical solvers. While standard NO architectures resolve this computation trouble they are inadequate for predicting European call-price surfaces from market or model features cannot guarantee static no-arbitrage conditions and, as we verify experimentally, can yield negative PnL in downstream trading tasks. 

We introduce a certifiably arbitrage-free (AF) and universal encoder-decoder NO architectural framework---\textit{continuously} mapping inputs from a Reproducing Kernel Hilbert Space (RKHS) $\mathcal{H}$ over any fixed subset of $[0,\infty)^2$ to a convex combination of \textit{arbitrage-free} European call-price surfaces; by continuously resolving an approximate nearest neighbour problem.
We show that a nearest arbitrage-free correction of an unconstrained predicted surface need not exist, implying that no-arbitrage cannot, in general, be restored through post-processing. Further, we show that standard NOs such as the FNO and certain encoder-decoder NOs admit parameter configurations whose outputs violate static no-arbitrage. Thus, out-of-the-box NOs, in general, cannot be certifiably restricted to arbitrage-free configurations, but may only be guided towards them through PINN-type regularization penalization during training.
\end{abstract}


\section{Arbitrage-Free Operator Learning}
\label{s:AF_NO}

In computational mathematics, eg.\ computational finance, one often must resolve \textit{large families} of problem \textit{fast}.
\textit{Neural operators} (NO) offer a way out to the computational burden involved in resolving large families of such problems, wherein classical numerical solver can become too costly to re-run several times (even if individual runs are fast, their totality may become unrealistic).  NOs however, can be nearly instantaneously queried, after one costly training round, offering rapid approximate solutions to massive numbers of classical computational problems.  
Briefly, the idea behind NOs is to view the relevant families of problems as input-outputs of some (nonlinear) \textit{operator} between infinite-dimensional topological vector spaces; often Hilbert spaces.

Many problems in mathematical finance are naturally formulated as nonlinear operators between infinite-dimensional Hilbert spaces.  Operator learning extends classical deep learning to this setting, replacing repeated numerical solution by the rapid evaluation of a learned map.  Applications include simultaneously solving families of SDEs and BSDEs~\cite{galimberti2022designing,furuya2024simultaneously,furuya2025onemodel}, modelling term structures and volatility surfaces~\cite{benth2021accuracy,benth2024pricing,gambara2021consistent,wiedemannoperator,choudhary2024funvol,ning2023arbitrage}, learning pricing and optimal-stopping maps~\cite{cui2023learning,bayraktar2025deep}, solving stochastic games and execution problems~\cite{alvarez2025neural,firoozi2025simultaneously,meng2026solving}, and learning risk measures or transition kernels~\cite{diazlozano2026deep,acciaio2024designing} both of which obtain exceptionally-accurate fits discouraging static arbitrage but not exactly prohibiting it.

For financial prediction, however, numerical accuracy alone is not sufficient; as the NO predictions must also satisfy the structural conditions required by economic theory.  In particular, a predicted European call-price surface may be economically meaningless if it violates static no-arbitrage, even when its pointwise prediction error is small, and defective predictions may be erroneously exploited by downstream AI agents which converge to trading strategies generating negative PnL (cf.\ Figure~\ref{fig:badtrades} below).  This is problematic since standard neural-operator architectures do not enforce these conditions uniformly across all parameter configurations; and the state-of-the-art at-best guides it towards no arbitrage by PINN-type penalization; e.g.~\cite{wiedemann2025operator} for neural operator-based and~\cite{choudhary2024funvol,ning2023arbitrage} for neural SDE-based solutions.

This paper addresses this by introducing an \textit{arbitrage-free neural operator} (AFNO): an encoder-decoder neural operator whose decoder maps directly into the nonlinear space of arbitrage-free European call-price surfaces.  The resulting architecture is certifiable, since every parameter configuration produces an arbitrage-free surface, and universal, since it can approximate every Lipschitz operator with the corresponding domain and codomain to arbitrary precision.  Thus, structural certification is achieved without sacrificing expressivity by building on the geometric deep learning results of~\cite{kratsios2022constraints,kratsios2023metricspace} which can handle approximation into non-linear co-domains such as the space of arbitrage-free European call-price surfaces.

\begin{figure}[htp!]
    \centering
    \includegraphics[width=1.0\linewidth]{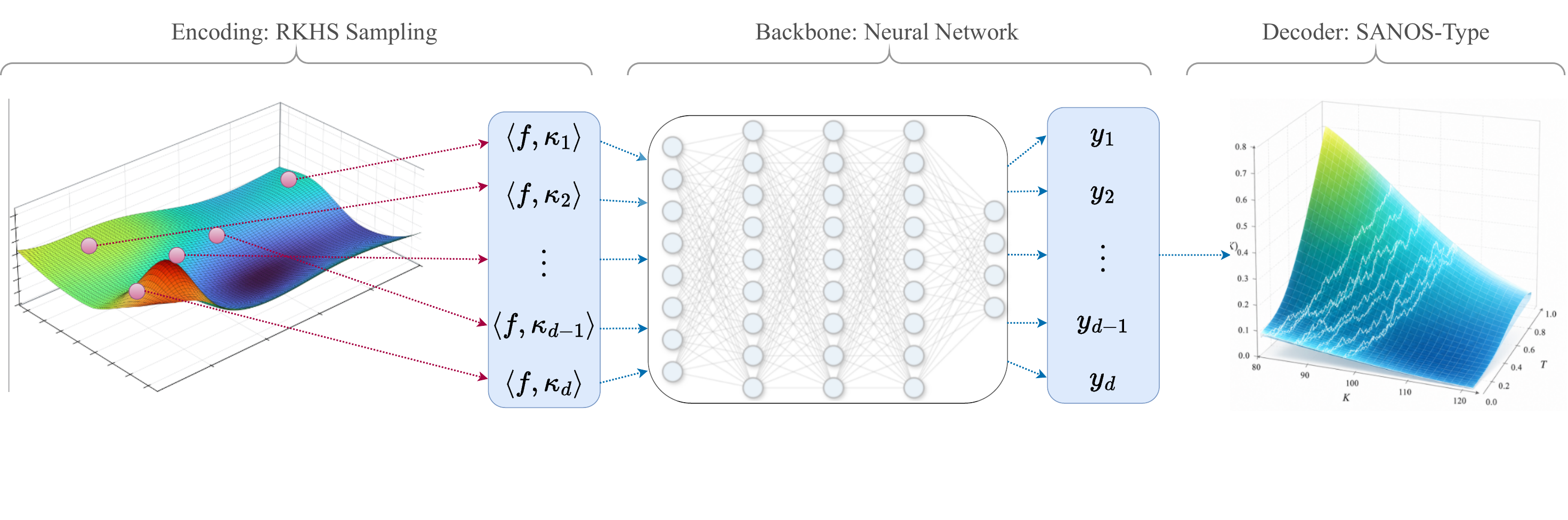}
    \caption{\textbf{AFNO architecture:}
    During the \textit{encoding phase} (cf. \S~\ref{s:EncoderDecoder__ss:Encoder}), input features are encoded from an RKHS over some $\mathcal{D}\subseteq [0,\infty)^2$ with kernel $\kappa$ by projecting them onto a finite collection of, possibly trainable, basis vectors or sampling functionals.  
    The resulting finite-dimensional representation is then passed through a transformer \textit{backbone} (cf. \S~\ref{s:EncoderDecoder__ss:Backbone}); we use a $\operatorname{ReLU}$-MLP, since this class is analytically well understood.  
    Finally, during the \textit{decoding phase} (cf. \S~\ref{s:EncoderDecoder__ss:GoodDecoder}), the backbone output is mapped into the space of arbitrage-free volatility surfaces using any decoder satisfying the structural properties of the $\operatorname{SANOS}$ model; cf.~\cite{buehler2026sanos}.
    Here, $\kappa_n\eqdef \kappa(\cdot,x_n)$ for some \textit{sample points} $x_1,\dots,x_d\in \mathcal{D}$.  Thus, $\langle f,\kappa_n\rangle$ represents ``smoothed sampling''.
    }
    \label{fig:OurNO}
\end{figure}

Our construction, summarized by Figure~\ref{fig:OurNO}, parallels the encoder-decoder NO paradigm cf.~\cite{kovachki2023neural,kovachki2021universal,galimberti2022designing} encompassing PCA-Net~\cite{bhattacharya2021model,lanthaler2023operator}, DeepONet~\cite{lu2021learning}, but not the Fourier neural operator (FNO)~\cite{li2021fourier}.  Rather, we use a new formulation of the geometric attention mechanism of~\cite{kratsios2022constraints,acciaio2024designing}, inspired by the classical attention mechanism of transformers~\cite{vaswani2017attention}, but tailored to the space of arbitrage-free European call-price surfaces.  The main example of our decoder is the $\operatorname{SANOS}$ model, recently introduced in~\cite{buehler2026sanos}.

\subsection{Contributions}
\label{s:Intro__ss:Contributions}

Our contributions are both positive and negative.  

\subsubsection{Primary Contribution}
\label{s:Intro__ss:Contributions___sss:Positive}
On the \textit{positive} side, we prove that every parameter configuration of an AFNO produces an arbitrage-free European call-price surface with controlled right-wing decay (Theorem~\ref{thrm:Cert_AF}).  Moreover, AFNOs are universal: every Lipschitz map from $\mathbb{R}^d$ into the corresponding class of arbitrage-free call surfaces can be approximated to arbitrary precision (Theorem~\ref{prop:UAT_AFNO__technicalversion__FINITEDIMENSIONALVERSION}).  In particular, this applies to parameter-to-surface maps generated by parametric families of SDEs (Corollary~\ref{cor:AF_solmap}).

\paragraph{Why this is challenging.}
Existing universal approximation theorems for neural operators treat operators taking values in linear spaces, typically Banach spaces~\cite{lanthaler2023operator,lanthaler2025nonlocality,schwab2026deep,schmocker2026weighted} or, more generally, Fr\'echet spaces~\cite{galimberti2022designing,benth2023neural}. These results do not enforce membership in nonlinear, nonclosed subsets of such spaces. Consequently, they cannot simultaneously guarantee universal approximation of maps into the space of arbitrage-free call surfaces $\mathcal{C}{\mathrm{AF}}$ and certify that every output of the approximating neural operator remains in $\mathcal{C}{\mathrm{AF}}$. 

A SANOS-type decoder approximates individual arbitrage-free surfaces.  Our contribution is to construct one continuous \textit{rapid to evaluate} input-to-surface map that approximates a target operator uniformly on a compact input set, under the hypotheses of Theorem~\ref{thrm:UAT_AFNO}, while remaining arbitrage-free at every input.  In this way, one does not need to re-fit a surface to any dataset (either real-world or synthetic); e.g.\ avoiding re-running a linear program as with~\cite{buehler2026sanos} or calibrating an SDE-type model, but they simply can simply evaluate a forward pass of our AFNO model.

\subsubsection{Secondary Contributions}
\label{s:Intro__ss:Contributions___sss:Negative}
On the \textit{negative} side, a predicted surface need not admit a nearest arbitrage-free approximation (Theorem~\ref{thrm:No_AF__Proj}); thus, no-arbitrage cannot, in general, be restored by post-processing.  Moreover, standard encoder-decoder neural operators and Fourier neural operators admit parameter configurations whose outputs violate static no-arbitrage (Propositions~\ref{prop:AF_certificatefailure} and~\ref{prop:sigmoid_breaks_convexity_in_K}).  Penalty-based methods may therefore discourage arbitrage during training, but cannot certify its absence.  This motivates enforcing no-arbitrage structurally, at the level of the architecture.

\paragraph{Why this is important.}
These results distinguish empirical constraint enforcement from architectural certification. Penalties may reduce arbitrage violations, and post-processing may repair particular outputs, but neither approach, by itself, guarantees that the resulting operator maps every input into the space of arbitrage-free call surfaces. This motivates enforcing no-arbitrage directly through the architecture. Practically, our experiments demonstrate the consequences of forgoing such a certificate: a downstream learning agent may focus on spurious arbitrage opportunities created by defects in the predicted call-price surfaces rather than learn genuinely profitable trading strategies.

\subsection{Organization of the Paper}
\label{s:Intro__ss:PaperOrganization}
Section~\ref{s:AF_NO__background} introduces the input and output spaces.  
Section~\ref{s:EncodeDecoder} develops the AFNO framework and its arbitrage-free decoders.  
Sections~\ref{s:Theory_Positive} and~\ref{s:Negresults} present our positive and negative theoretical results, respectively.  
Section~\ref{s:Numerics} provides numerical illustrations, while the supplementary diagnostics and proofs are deferred to the appendices.

\section{Preliminaries}
\label{s:AF_NO__background}

We now overview the necessary background and elementary preliminary findings required to formulate our main results.  

\subsection{Input Spaces: Reproducing Kernel Hilbert Spaces with Riesz Bases}
\label{s:RieszBases}
In practice, market data rarely arrives on a fixed and perfectly regular grid.  Available call-price quotes may be irregularly distributed across strikes and maturities, their number may change from one observation time to another, and in many applications one may wish to work not only with individual quotes but also with interpolated slices or entire observed surfaces.  It is therefore useful to work with an input space which is flexible enough to accommodate all of these situations within a single framework.

Reproducing kernel Hilbert spaces provide a natural setting for this purpose.  Point evaluations are continuous, so individual market quotes can be used directly as features, while suitable sampling and interpolation properties allow finite collections of such observations to represent increasingly rich functional inputs.  At the same time, the infinite-dimensional formulation allows us to regard a complete call-price slice or surface as a single input object.  The sampling and interpolation theory introduced below is precisely what allows us to move consistently between these finite observations and their underlying functional representation.

We briefly review the relevant theory of reproducing kernel Hilbert spaces; which have already been extensively used in mathematical finance~\cite{filipovic2001consistency,filipovic2009term,benthKruhner2023filipovic,aronszajn1950theory,berlinetThomasAgnan2004rkhs,wendland2005scattered} and implicitly via Gaussian process regression techniques; e.g.~\cite{ludkovski2018kriging,ludkovski2025gaussian}.
Let $\mathcal{D}\subseteq[0,\infty)^2$ be non-empty, and let $\mathcal{H}$ be a reproducing kernel Hilbert space (RKHS) over $\mathcal{D}$, with reproducing kernel
$
	\kappa:\mathcal{D}\times\mathcal{D}\to\mathbb{R}
$; cf.~\cite{aronszajn1950theory}.  
A countable sequence $\{z_{\lambda}\}_{\lambda\in\Lambda}\subset\mathcal{D}$, indexed by $\Lambda\subseteq\mathbb{N}$, is called a \textit{sampling sequence} for $\mathcal{H}$ if there exists a constant $C\geq 1$ such that, for every $f\in\mathcal{H}$,
\begin{equation}
\label{eq:sampling_def}
	\frac{1}{C}\,\|f\|_{\mathcal{H}}
\le
	\|f\|_{\Lambda}
	\eqdef
	\left(
		\sum_{\lambda\in\Lambda}
		|f(z_{\lambda})|^2
	\right)^{1/2}
\le
	C\,\|f\|_{\mathcal{H}} 
.
\end{equation}
It is called an \textit{interpolating sequence} if, for every square-summable sequence
$v_{\cdot}\eqdef(v_{\lambda})_{\lambda\in\Lambda}\in\ell^2(\Lambda)$, there exists an $f\in\mathcal{H}$ such that: for each $\lambda\in\Lambda$ we have
\begin{equation}
	\label{eq:interpolating_def}
	f(z_{\lambda})
	=
	v_{\lambda}
 .
\end{equation}
If $\{z_{\lambda}\}_{\lambda\in\Lambda}$ is both sampling and interpolating, then we call it a \textit{complete interpolating sampling sequence} (CIS); since that will allow us the feature maps $\{\langle \kappa(z,\cdot),\cdot,\rangle:\,z\in \mathcal{D}\}$ to be trainable (in $z$) set of continuous features which can encode a Riesz basis for $\mathcal{H}$.
Throughout the sequel, we assume that $\mathcal{H}$ admits a CIS.  

Riesz bases of reproducing kernels, sampling, and interpolation in Fock-type RKHSs have been studied in depth; see, for example,~\cite{borichev2010riesz} and~\cite{baranov2015sampling}.  In statistics they are usually referred
to as kernel-regression. Landau-type necessary density conditions in general reproducing kernel Hilbert spaces were developed~\cite{fuhr2017density}, extending the classical density results of~\cite{landau1967necessary}.\\

As a first example we demonstrate sampling and interpolating sequences in the context of quotes on a call price surface. Though samples (quotes) don't arrive in strictly ordered grids, many popular modelling techniques~\cite{DLV,DVL2} rely on grid-based sampling of surfaces, which makes them our first relevant example:
\begin{example}[Finite-Dimensional RKHS: Evaluations on a Fixed Grid on the Call Surface]
\label{ex:FiniteRKHS}
Let $d\in\mathbb{N}_+$ and let
$
        \mathcal{D}
    =
        \{x_n\}_{n=1}^d
    =
        \{(K_n,T_n)\}_{n=1}^d
    \subset
        [0,\infty)^2
$.  
Since $\mathcal{D}$ is finite, a kernel on $\mathcal{D}$ is equivalently a \textit{fixed} $d\times d$ symmetric positive semidefinite matrix
$
        \mathbf{K}
    =
        \big(\kappa(x_i,x_j)\big)_{i,j=1}^d
$.  
If $\mathbf{K}$ is strictly positive definite, then the associated RKHS
$\mathcal{H}$ is $d$-dimensional and consists of \textit{all} functions
$f:\mathcal{D}\to\mathbb{R}$.  Intuitively, this RKHS consists of all call and put features observed on a finite pre-determined set of points, e.g.\ possibly but not necessarily a finite grid.
Moreover, every $f\in\mathcal{H}$ has a unique
representation
\begin{equation}
\label{eq:expansion}
        f
    =
        \sum_{n=1}^d
        \beta_n\,\kappa(\cdot,x_n)
\end{equation}
and the map sending any such $f$ to its vector of coefficients $(\beta_n)_{n=1}^d\in\mathbb{R}^d$ is a vector-space isomorphism  given by the generalized regression formula
$$
    \beta = ( \kappa(x_i,x_j )_{i,j} )^{-1} ( f(x_1,\ldots,x_d) )^T \ .
$$
Thus, one may directly plug in the output of the feature map~\eqref{eq:expansion} into a downstream standard deep learning model; e.g.\ MLP, GNN, or a transformer.
\hfill\\
If $f=\sum_{n=1}^d\beta_n\kappa(\cdot,x_n)$ and
$g=\sum_{n=1}^d\gamma_n\kappa(\cdot,x_n)$, then the RKHS inner product on $\mathcal{H}$ is given by
\[
        \langle f,g\rangle_{\mathcal{H}}
    =
        \beta^\top \mathbf{K}\gamma
.
\]
Thus, the isomorphism mapping any $f$ in~\eqref{eq:expansion} to 
$
\beta = (\beta_n)_{n=1}^d\in\mathbb{R}^d
$
is bi-Lipschitz, and it is an isometry in the special case where $\mathbf{K}=I_d$.
\end{example}

The next example of an RKHS abstracts the idea of having an arbitrarily large, but finite, number of quoted call prices at our disposal. This removes the fixed finite-grid construction of Example~\ref{ex:FiniteRKHS} by extending that example's simple RKHS, which was effectively $\mathbb{R}^d$, to a genuinely infinite-dimensional RKHS. Previously, $d$ represented the fixed number of quotes that could be observed at any given time in the market; in the next example, the infinite-dimensionality instead provides the flexibility to accommodate an arbitrary finite number of quoted prices. Moreover, the infinite-dimensional setting allows one to regard the entire continuous call surface as a feature, rather than merely finitely many quoted prices\footnote{Interestingly, a variant classical RKHS considered for studying forward rate curves, in term structure modelling cf.~\cite{filipovic2001consistency}, can be used to obtain a natural RKHS backdrop for European call surface-price modelling.}.  

We present this RKHS in two steps.  First we introduce a slice-wise RKHS and then we pass to one which can encode both strike and maturity data via tensorization of the following example:

\begin{example}[An RKHS for European Call-price slices]
\label{ex:RKHS_SingleDamirStyle}
Let $\mathcal{H}_0$ denote the anchored Sobolev-type RKHS on $(0,\infty)$ with kernel $\kappa_0:(0,\infty)^2\to \mathbb{R}$ given for every $x,y>0$ by
\begin{equation}
\label{eq:RKHS_DamirType}
        \kappa_0(x,y)
    =
        x\wedge y
,
\end{equation}
and, as in~\citep[Theorem~1.6]{saitohSawano2016rkhs}, this kernel induces the following RKHS on $(0,\infty)$
\[
        \mathcal{H}_0
    =
        \left\{
            h\in AC_{\operatorname{loc}}([0,\infty))
            :
            h(0)=0,\ 
            h'\in L^2(0,\infty)
        \right\}
    \,\,\mbox{ normed by }\,\,
        \|h\|_{\mathcal{H}_0}^2
    =
        \int_0^\infty |h'(u)|^2\,du.
\]
To impose the normalized boundary condition $f(0)=1$, we use the affine translate
\begin{equation}
\label{eq:RKHS_1_BC}
        \mathcal{A}_1
    =
        1+\mathcal{H}_0
    =
        \left\{
            f:[0,\infty)\to\mathbb{R}
            :
            f-1\in\mathcal{H}_0
        \right\}.
\end{equation}
Then every $f\in\mathcal{A}_1$ satisfies $f(0)=1$.  The set $\mathcal{A}_1$ is not a linear RKHS, but it is an affine Hilbert space 
canonically modelled on the RKHS $\mathcal{H}_0$.
\end{example}

To consider European call-price \textit{surfaces}, rather than only one-dimensional slices, as inputs, a convenient construction is to take the Hilbert tensor product of the RKHS from the preceding example with itself.  By the tensor-product theorem for RKHSs~\citep[Theorem~2.20]{saitohSawano2016rkhs}, this again yields an RKHS, now over $[0,\infty)^2$, whose kernel is the product of the kernels in~\eqref{eq:RKHS_DamirType}.
\begin{example}[Tensor-product anchored Sobolev RKHS on the call surface domain]
\label{ex:RKHS_DoubleDamirStyle}
In the setting of Example~\ref{ex:RKHS_SingleDamirStyle}, consider
$
        \mathcal{D}
    =
        [0,\infty)^2
$.  Define the tensor-product kernel
$
        \kappa_{\otimes}
    :
        \mathcal{D}\times\mathcal{D}
        \to
        \mathbb{R}
$
for each $(K_1,T_1),(K_2,T_2)\in[0,\infty)^2$ by
\begin{equation}
\label{eq:RKHS_DoubleDamirType}
        \kappa_{\otimes}\big((K_1,T_1),(K_2,T_2)\big)
    =
        (K_1\wedge K_2)\,(T_1\wedge T_2),.
\end{equation}
Then $\kappa_{\otimes}$ is positive definite and its associated RKHS is the tensor product
$
        \mathcal{H}_{\otimes}
    =
        \mathcal{H}_0
        \widehat{\otimes}
        \mathcal{H}_0
$; where $\mathcal{H}_0$ is as in Example~\ref{ex:RKHS_SingleDamirStyle}.  Thus, $\mathcal{H}_{\otimes}$ is the mixed-smoothness Sobolev-type RKHS
\[
        \mathcal{H}_{\otimes}
    =
        \left\{
            f:\mathcal{D}\to\mathbb{R}
            :
            \text{there exists } g\in L^2(\mathcal{D}) \text{ such that }
            f(K,T)
            =
            \int_0^K\int_0^T g(k,t)\,dt\,dk
        \right\}.
\]
Every
$f\in\mathcal{H}_{\otimes}$ satisfies the anchored boundary conditions: 
$
        f(K,0)=0
$ and $f(0,T)=0$ for each $K,T\ge 0$.
The inner product for any $f,g\in \mathcal{H}$ is 
\[
        \langle f,h\rangle_{\mathcal{H}_{\otimes}}
    =
        \int_0^\infty\int_0^\infty
        \partial_K\partial_T f(K,T)\,
        \partial_K\partial_T h(K,T)
        \,dT\,dK
\qedhere
.
\]
\end{example}

Having established a convenient unified framework for representing both finitely many quotes and the entire call surface as features to be used as inputs to our deep learning models, we are now ready to turn to the structure of the relevant output space; namely, the space of arbitrage-free call surfaces.

\subsection{Output Spaces: Spaces of Arbitrage-Free Call Surfaces}
\label{s:AF_NO__background___AFCallSurfaces}

We now discuss the space of arbitrage-free call surfaces. While input spaces may vary, one point on which there is broad agreement is the value of generating surface-valued outputs that are guaranteed to be arbitrage-free by construction.

While the broad structural properties of arbitrage-free European call-price surfaces are by now well understood in mathematical finance, the space of all such objects is surprisingly less well studied. This is likely because it does not arise naturally in many classical problem formulations in mathematical finance and, moreover, because it is not a particularly tame space. We begin with the following classical result.

\begin{definition}[Arbitrage-Free Call-Price Surfaces]
\label{def:arbfreecall}
We call a function
$
    C
    :
    [0,\infty)^2
    \longrightarrow
    [0,\infty)
$ an \emph{arbitrage-free European call-price surface} if there exist a probability measure $\P$ and a non-negative $\P$-martingale
$
    Z_{\cdot}
\eqdef
    (Z_t)_{t\in[0,\infty)}
$ satisfying
$
    Z_0
    =
    1
$ $\P$-a.s.\ such that: for every $T.K\ge 0$ 
\[
    C(T,K)
    =
    \E_{\P}
    \left[
        (Z_T-K)^+
    \right]
.
\]
In this case, $Z_{\cdot}$ is said to be a
\emph{martingale representation} of $C$.
\end{definition}

It is convenient to give an analytic characterization of the arbitrage-free call-price surfaces introduced in Definition~\ref{def:arbfreecall}.

\begin{proposition}[Analytic Characterization of Arbitrage-Free Call Surfaces]
\label{prop:contarbfree}
A continuous map $C:[0,\infty)^2\rightarrow[0,\infty)$ is an arbitrage-free call surface (cf.\ Definition~\ref{def:arbfreecall}) if and only if:
\begin{enumerate}
    \item[(i)] $
        C(T,0)
        =
        1
    $ for every $T\geq0$.

    \item[(ii)] $
        \partial_K^+C(T,0)
        \eqdef
        \lim_{\delta\downarrow0}
        \frac{
            C(T,\delta)-C(T,0)
        }{
            \delta
        }
        \geq
        -1
    $ for every $T\geq0$.

    \item[(iii)] $
        \lim_{K\uparrow\infty}
        C(T,K)
        =
        0
    $ for every $T\geq0$.

    \item[(iv)] For every $T\geq0$, the map
    $K\mapsto C(T,K)$ is convex.

    \item[(v)] For every $K\geq0$, the map
    $T\mapsto C(T,K)$ is non-decreasing.

    \item[(vi)] For every $K\geq0$,
    $
        C(0,K)
        =
        (1-K)^+
    $.
\end{enumerate}
If $C$ satisfies {\rm (i)}--{\rm (vi)} then, for each $T\geq0$,  every martingale representation $Z_{\cdot}$ of $C$ satisfies
\begin{equation}
\label{eq:call_derivative_zero_atom}
    \partial_K^+C(T,0)
    =
    -1
    +
    \P
    \left(
        Z_T=0
    \right)
.
\end{equation}
\end{proposition}
\begin{proof}
See Supplement~\ref{prop:Proof_of_NA}, for a proof of this standard result.
\end{proof}
\begin{remark}[Default-Free Models]
\label{rem:default_free_models}
Let $C$ be an arbitrage-free call surface and let $Z_{\cdot}$ be any martingale representation of $C$.  Then, for every fixed $T\geq0$,~\eqref{eq:call_derivative_zero_atom} implies that
\[
    \partial_K^+C(T,0)
    =
    -1
\mbox{ if and only if }
    Z_T>0
\quad\P\text{-}a.s.
\qedhere(
\]
\end{remark}

Let $\mathcal{C}_{\mathrm{AF}}\subseteq C([0,\infty)^2,[0,\infty))$ consist of all continuous functions satisfying Definition~\ref{def:arbfreecall}; i.e.\ the space of \textbf{arbitrage-free call surfaces}.  
Since $\mathcal{C}_{\mathrm{AF}}$ does not contain zero it is by no means a vector space (at-least under the familiar pointwise operation on functions), nor is it even a group under addition;  moreover,
the set has a natural attainable lower bound $(1-K)^+$ but no closed upper bound.
Thus, it will be far beyond the limited reach of operator learning~\cite{lu2021deep,li2021fourier,lanthaler2023operator,lanthaler2023curse}%
\footnote{Whose objective is to learn, or at-times approximate, non-linear operators between separable topological vector spaces; e.g.\ Banach spaces.}.  This means, we will have to rely on the more sophisticated tools of infinite-dimensional \textit{geometric deep learning}; e.g.~\cite{kratsios2022constraints,kratsios2023approximation,acciaio2024designing}.

\begin{proposition}[Arbitrage-Free Call Surfaces are Not A Vector Space]
\label{prop:NOVec}
The space $\mathcal{C}_{\mathrm{AF}}$ is not closed under addition, scalar multiplication, nor under pointwise products.
\end{proposition}

The class $\mathcal C_{\mathrm{AF}}$ is convex but is not closed for uniform convergence on compact subsets of $[0,\infty)^2$; see Proposition~\ref{prop:AS_not_closed}\footnote{By contrast, it is closed in the (global) uniform norm. Indeed, if $C_n\in\mathcal C_{\mathrm{AF}}$ and $\|C_n-C\|_\infty\to0$, then continuity, non-negativity, the boundary values, strike convexity, maturity monotonicity, and the strike Lipschitz bound pass to $C$. For each fixed $T\geq0$, $0\leq\limsup_{K\to\infty}C(T,K)\leq\|C-C_n\|_\infty$, so the right-wing limit is zero. Proposition~\ref{prop:contarbfree} yields $C\in\mathcal C_{\mathrm{AF}}$.}.
This is important, as convexity sets up the approximately simplicial property, cf.~\citep[Definition 3.1 and Example 5.1]{acciaio2024designing} and first part of a QAS space~\cite[Definition 3.4]{acciaio2024designing}, required for that paper's main ``static'' approximation theorem in non-linear spaces.  Which, as far as we know is the only available universal approximation theorem in non-linear space holding in that level of generality.  
\begin{proposition}
\label{prop:AS}
The space $\mathcal{C}_{\mathrm{AF}}$ is convex in $C([0,\infty)^2,\mathbb{R})$. 
\end{proposition}
It is already clear that one must be careful here, since in the topology of uniform convergence on compact sets the set $\mathcal{C}_{AF}$ \textbf{fails} to be closed. 
This is problematic, since limit points (and thus approximation) may not stay within $\mathcal{C}_{AF}$; cf.~Proposition~\ref{prop:AS_not_closed} in Appendix~\ref{a:additions}.  However, closedness is not needed for QAS spaces as completeness is ``local'' on the range of the target function's image of a compact set of inputs; which is itself compact, and thus complete, by continuity of the target function.  Both the lack of closeness, and the lack of linearity, means that the classical ``universal'' approximation theorems for neural operators fail.
\subsubsection{Spaces of Call Surfaces with Uniform Right-Wing and Maturity Decay-Rate}
\label{s:AF_NO__background___RightWing}
As a prototype, we ask: what is the \textit{right-wing decay} of the Black-Scholes formula?  For any given maturity $T>0$, it is given by the following. Fix $T>0$ and let
\begin{equation}
\label{eq:BS_Call}
        C_{\mathrm{BS}}(T,K)
    \eqdef
        e^{-rT}\,\mathbb{E}\big[(S_T-K)^+\big]
\end{equation}
be the Black-Scholes call price with strike $K>0$, where $S_{ }\eqdef (S_t)_{t\ge 0}$ follows a geometric Brownian motion with log-mean $\mu\in \mathbb{R}$ and log-volatility $\sigma>0$; i.e.\
\[
        S_T 
    =
        S_0\exp \Big(
            \big(r-\tfrac{1}{2}\sigma^2\big)T + \sigma W_T
        \Big)
.
\]
We thus have the basic right-wing estimate, which shows that for any fixed maturity the right wing $K\mapsto C(T,K)$ decays faster than any power of the reciprocal strike $K^{-1}$. 
{It is straightforward to see that: for every $T,p>0$ there is a constant $C_{T,p}>0$ such that for each strike $K>0$ the Black-Scholes call price decays like
$
        C_{\operatorname{BS}}(T,K)
    \leq
        \frac{C_{T,p}}{(1+K)^p}
$.
}
{
This motivates the following power-law decay rate condition: For a fixed }$p>0$ we write $\mathcal{C}_{\mathrm{AF}}^{p}\subseteq \mathcal{C}_{\mathrm{AF}}$ to consist of all arbitrage-free call surfaces satisfying the following right-wing asymptotic: for every $T\geq 0$, there exists a constant $C_{T,p}>0$ such that
\begin{equation}
\label{eq:right_wing_asymptotic}
        C(T,K)
    \le 
        C_{T,p}\, K^{-p}
.
\end{equation}
{We will consider several examples of families of arbitrage-free call surfaces satisfying this decay condition in \S~\ref{s:AF_NO__background___EXAMPLES}, below.}
Let $\mathcal{C}_{AF}\subseteq C([0,\infty)^2,[0,\infty))$ denote the \textit{metric space} of arbitrage-free call surfaces equipped with the topology of \textit{uniform} convergence (not uniform convergence on compact sets).  
Fix $p,\lambda>0$ and consider the norm\footnote{See~\cite{schmocker2026weighted} for results on weighted spaces.} 
$
\|C\|_{p,\lambda}
\eqdef
\sup_{T,K\ge 0} e^{-\lambda T}(1+K)^p|C(T,K)|
$
on
$$
\mathcal{X}^p_\lambda
\eqdef
\Big\{
C\in C([0,\infty)^2,\mathbb{R})
:\ 
\|C\|_{p,\lambda}<\infty
\Big\}.
$$
Then every $C\in \mathcal{X}^p_\lambda$ automatically satisfies the weighted right-wing bound
$
C(T,K)\le e^{\lambda T}\|C\|_{p,\lambda}(1+K)^{-p}
$
for all $T,K\ge 0$.  In particular, $\mathcal{C}_{AF}\cap \mathcal{X}^p_\lambda\subseteq \mathcal{C}_{AF}^p$.

We will typically consider ``uniformly controlled'' classes of call surfaces, if it has uniform right-wing decay rate, uniform maturity growth rate, and it satisfies ``compactness-type conditions'' akin to those in the Arzela-Ascoli theorems; cf.~\cite{collins2007asymmetric,xia2009geodesic}.  This is formalized as follows:
\begin{definition}[Uniformly Controlled Call-Surface Class]
\label{defn:unif_controlled}
A class of (arbitrage-free) call surfaces $\mathcal{Y}\subseteq \mathcal{C}_{AF}^{p,\lambda}\eqdef \mathcal{C}_{AF}\cap \mathcal{X}_{\lambda}^p$ is called uniformly controlled if there exists a constant $B>0$ and a modulus of continuity $\omega$ such that
\begin{enumerate}
    \item[(a)] \textbf{Uniform weighted bound:}
    $
        \sup_{C\in\mathcal Y}\|C\|_{p,\lambda}
        \le 
        B,
    $
    \item[(b)] \textbf{Equi Uniform Continuity:} $\mathcal{Y}$ is $\omega$-equi-uniformly continuous; i.e.\ for all $T,\tilde{T},K, \tilde{K}\ge 0$
    \[
        |C(T,K)-C(\tilde T,\tilde K)|
        \le
        \omega\big(|T-\tilde T|+|K-\tilde K|\big)
    .
    \]
\end{enumerate}
When needed, we call $(p,\lambda,\omega,B)$ the regularity-data of $\mathcal{Y}$.
\end{definition}
\subsubsection{Examples Families of Surfaces Satisfying the Decay Condition}
\label{s:AF_NO__background___EXAMPLES}
We begin with the Black-Scholes prototype example.
\begin{example}[Black--Scholes Call Surfaces]
\label{ex:BlackScholesPolynomialCallDecay}
Let
$
    Z_t
    \eqdef
    \exp\bigl(\sigma W_t-\tfrac12\sigma^2t\bigr)
$
for some $\sigma>0$, and define the corresponding normalized Black--Scholes call surface by
$
    C_{\operatorname{BS}}(T,K)
    \eqdef
    \mathbb{E}[(Z_T-K)_+].
$
For $T>0$ and $K>0$, the Black--Scholes formula gives
\[
    C_{\operatorname{BS}}(T,K)
    =
    \mathcal{N}(d_+(T,K))
    -
    K\mathcal{N}(d_-(T,K)),
\]
where
$
    d_{\pm}(T,K)
    \eqdef
    -\frac{\log K}{\sigma\sqrt{T}}
    \pm
    \frac{\sigma\sqrt{T}}{2}.
$
In particular,
$
    C_{\operatorname{BS}}(T,K)
    \leq
    \mathcal{N}(d_+(T,K)).
$  Therefore, for any fixed $p>0$.  If $K\geq e^{\sigma^2T/2}$, then $d_+(T,K)\leq0$ and the Gaussian tail bound
$
    \mathcal{N}(-x)\leq e^{-x^2/2}
$
for $x\geq0$ yields
$
    C_{\operatorname{BS}}(T,K)
\leq
    \exp\left(
        -\frac{(\log K-\sigma^2T/2)^2}{2\sigma^2T}
    \right)
$ and noting that $
    -\frac{(\log K-\sigma^2T/2)^2}{2\sigma^2T}
    \leq
    -p\log K
    +
    \frac{p(p+1)}{2}\sigma^2T
$ is enough to deduce that
\[
    C_{\operatorname{BS}}(T,K)
    \leq
    e^{\frac12p(p+1)\sigma^2T}K^{-p}.
\]
If instead $0<K<e^{\sigma^2T/2}$, then
$
    C_{\operatorname{BS}}(T,K)\leq1
$
and
$
    e^{\frac12p(p+1)\sigma^2T}K^{-p}\geq1.
$
The same bound also holds at $T=0$, since
$
    C_{\operatorname{BS}}(0,K)=(1-K)_+\leq K^{-p}.
$
Therefore, for every $p>0$, 
$
    C_{\operatorname{BS}}\in\mathcal{C}_{\mathrm{AF}}^p
$
for every $p>0$.
\end{example}
We now provide examples of popular families of call-price surfaces satisfying condition~\eqref{eq:right_wing_asymptotic}.
\begin{proposition}[Polynomial Decay of Call Payoffs]
\label{prop:EasySufficientCondition_Decay}
Let $S_{\cdot}\eqdef (S_t)_{t\ge 0}$ be a real-valued stochastic process such that there exists a $q>1$ satisfying $\mathbb{E}[|S_t|^q]<\infty$ for every $t\ge 0$. Then, for every $T\ge 0$ and $K>0$,
\begin{equation}
\label{eq:PolynomialCallPayoffDecay}
\mathbb{E}\big[(S_T-K)_+\big]
\le
\frac{(q-1)^{q-1}}{q^q}\,
\mathbb{E}\big[(S_T)_+^q\big]K^{-(q-1)}.
\end{equation}
In particular,~\eqref{eq:right_wing_asymptotic} holds for every $T\ge 0$, with $p=q-1$ and $C_{T,p}=\frac{(q-1)^{q-1}}{q^q}\mathbb{E}[(S_T)_+^q]$.
\end{proposition}
\begin{proof}
See Section~\ref{s:Aux_results}.
\end{proof}
As a simple illustration, the Heston model%
\footnote{The following can be extended to the rough Heston model by~\citep[Theorem 2.4]{gerhold2019moment}.}~%
satisfies the sufficient condition of Proposition~\ref{prop:EasySufficientCondition_Decay}.

\begin{example}[Lognormal Stochastic Volatility]
\label{ex:scope_lognormal}
Let
\[
    dZ_t
    =
    Z_t\sigma_t\,dW_t,
    \qquad
    d\sigma_t
    =
    \eta\sigma_t\,dB_t,
    \qquad
    d\langle W,B\rangle_t
    =
    \rho\,dt,
\]
where $\eta>0$, $\sigma_0>0$, and $\rho\in(-1,0)$.  By~\citep[Theorem~2.3]{LionsMusiela2007}, for every $q>1$,
$
    \sup_{t\in[0,T]}
    \mathbb{E}[Z_t^q]
    <
    \infty
$ for every $T>0$ if and only if $
    \rho
    \leq
    -\sqrt{\frac{q-1}{q}}$; whence, for every
$
    0<p
    \leq
    \frac{\rho^2}{1-\rho^2},
$ if we set $q\eqdef p+1$ then Proposition~\ref{prop:EasySufficientCondition_Decay} implies that: for every $T\geq0$ and $K>0$
\[
    C(T,K)
\eqdef
    \mathbb{E}\big[(Z_T-K)_+\big]
\leq
    \frac{p^p}{(p+1)^{p+1}}
    \,
    \mathbb{E}\big[Z_T^{p+1}\big]
    K^{-p}
.
\]
Thus,~\eqref{eq:right_wing_asymptotic} holds and
$
    C\in\mathcal{C}_{\mathrm{AF}}^p
$.
\end{example}

\begin{example}[Heston Call Surfaces]
\label{ex:HestonPolynomialCallDecay}
Consider the discounted Heston model
\[
dS_t=S_t\sqrt{V_t}\,dW_t,\qquad
dV_t=\kappa(\theta-V_t)\,dt+\sigma\sqrt{V_t}\,dZ_t,
\qquad
d\langle W,Z\rangle_t=\rho\,dt,
\]
where $\kappa,\theta,\sigma>0$ and $\rho\in(-1,1)$. If
$\kappa-3\rho\sigma\geq\sqrt{6}\,\sigma$, then the third-moment explosion time is infinite, and hence $\mathbb{E}[S_T^3]<\infty$ for every $T\geq0$; cf.~\citep[Equation~(2.13) and case~(C)]{gerhold2019moment}.  
Proposition~\ref{prop:EasySufficientCondition_Decay}, applied with $q=3$, therefore gives
\[
    C_{\operatorname{Heston}}(T,K)
\eqdef 
    \mathbb{E}\big[(S_T-K)_+\big]
\leq
    \frac{4}{27}
    \,
    \frac{\mathbb{E}[S_T^3]}{K^2}
,
\]
for each $T\ge 0$ and $K>0$.
Thus,~\eqref{eq:right_wing_asymptotic} holds with $p=2$.
\end{example}
\begin{example}[Rough Bergomi]
\label{ex:RoughBergomi}
Fix a Hurst parameter $H\in(0,\tfrac{1}{2})$ and a correlation $\rho\in(-1,0)$.  We consider the associated fractional rough Bergomi-type model in~\citep[Equation~(1.1)]{bourdon2026moments} with log-normal response $\sigma(t,x)=m(t)e^x$, where $m\in C^1_{\mathrm{loc}}([0,\infty);(0,\infty))$, and fractional kernel
$
K_H(t)
\eqdef
\frac{t^{H-\frac12}}{\Gamma\left(H+\frac12\right)}
$; that is, $S_{\cdot}\eqdef(S_t)_{t\geq0}$ is given by
\begin{equation}
\label{eq:roughBergomi}
S_t
=
S_0+\int_0^t S_s m(s)e^{Y_s}\,dW_s,
\qquad
Y_t
=
y_0+\nu\int_0^t K_H(t-s)\,dB_s,
\qquad
W_t
=
\rho B_t+\sqrt{1-\rho^2}\,B_t^\perp,
\end{equation}
where $S_0>0$, $y_0\in\mathbb{R}$, $\nu>0$, and $B$ and $B^\perp$ are independent Brownian motions.
If $K_H$ and $\sigma(t,x)$ satisfy the technical conditions in\citep[Assumptions~1 and~2]{bourdon2026moments} then~\citep[Theorem~1]{bourdon2026moments} implies that
$
\mathbb{E}[S_T^q]<\infty
$ for every $0<q<\frac{1}{1-\rho^2}$ and for every $T>0$.  Consequently, for every
\begin{equation}
\label{eq:corelation_constraing}
0<p<\frac{\rho^2}{1-\rho^2},
\end{equation}
if we set $q=p+1$ then Proposition~\ref{prop:EasySufficientCondition_Decay} holds, which yields
\[
    C_{\operatorname{rBergomi}}(T,K)
\eqdef 
    \mathbb{E}\big[\max\{S_T-K,0\}\big]
\leq
    \frac{p^p}{(p+1)^{p+1}}
    \mathbb{E}[S_T^{p+1}]K^{-p}
\]
implying that $C_{\operatorname{rBergomi}}\in \mathcal{C}_{\mathrm{AF}}^p$ for every $0<p<\rho^2/(1-\rho^2)$.
\end{example}

\begin{example}[Exponential L\'evy; CGMY]
\label{ex:scope_CGMY}
Let
$
    Z_T
    =
    e^{X_T}
$
for a L\'evy process $X$ satisfying $\mathbb{E}[Z_T]=1$, and consider the associated call surface
$
    C(T,K)
    \eqdef
    \mathbb{E}[(Z_T-K)_+]
$.
For the CGMY model~\cite{carr2002fine}, the L\'evy measure has right tail
$
    \nu(dx)
    =
    C e^{-Mx}x^{-1-Y}\,dx
$
for $x>0$ \textit{for some $M>1$}.
Hence, for every $q<M$,
$
    \mathbb{E}[Z_T^q]
    =
    \mathbb{E}[e^{qX_T}]
    <
    \infty
$
for every $T\geq0$.  Thus for any 
$
    0<p<M-1,
$ we may set $q\eqdef p+1$ and apply Proposition~\ref{prop:EasySufficientCondition_Decay} to conclude that 
$
    C\in\mathcal{C}_{\mathrm{AF}}^p
$; again for every $0<p<M-1$.
\end{example}

\paragraph{Examples in the SSVI Framework}
We note that condition~\eqref{eq:right_wing_asymptotic} need \textit{not} be directly interpreted and explicit moment condition.  Rather, it is simply an analytic decay condition on the call surface.  We illustrate it by showing that it holds for the SSVI model of~\cite{gatheral13}, for which a stochastic representation is not known to the authors.  
\begin{proposition}[SSVI Call Surfaces Satisfy~\eqref{eq:right_wing_asymptotic} for every $p>0$]
\label{prop:SSVIVerification}
Fix smooth functions $\varphi ,\theta \in C((0,\infty),(0,\infty))$ with $\theta_{\cdot}$ increasing, and let $C_{\mathrm{SSVI}}\in\mathcal{C}_{\mathrm{AF}}$ be an arbitrage-free call surface induced by the SSVI total implied variance
\[
w(k,\theta_T)
=
\frac{\theta_T}{2}
\left\{
1+\rho\varphi(\theta_T)k
+
\sqrt{\bigl(\varphi(\theta_T)k+\rho\bigr)^2+1-\rho^2}
\right\},
\]
where $\rho\in(-1,1)$.  For any fixed $p>0$, $C_{\mathrm{SSVI}}\in\mathcal{C}_{\mathrm{AF}}^{p}$ holds if
\begin{equation}
\label{eq:SSVI_DecayCondition}
    \sup_{T>0}
    \,
    \frac{1+\rho}{2}\theta_T\varphi(\theta_T)
\leq
    2\bigl(\sqrt{p+1}-\sqrt{p}\bigr)^2
.
\end{equation}
\end{proposition}
\begin{proof}
See Section~\ref{s:Aux_results}.
\end{proof}
\begin{example}[Heston-like SSVI]
\label{ex:HestonLikeSSVI}
In the context of Proposition~\ref{prop:SSVIVerification}, consider the Heston-like SSVI model in~\cite[Example~4.1]{gatheral2014arbitrage}; i.e.\ let $\lambda>0$ and for each $\theta>0$
\[
\varphi(\theta)
\eqdef
\frac{1}{\lambda\theta}
\left(
1-\frac{1-e^{-\lambda\theta}}{\lambda\theta}
\right)
.
\]
Then $\theta\mapsto\theta\varphi(\theta)$ is increasing and $
\sup_{\theta>0}\theta\varphi(\theta)
=
\lim_{\theta\to\infty}\theta\varphi(\theta)
=
\frac{1}{\lambda}.
$
Consequently, condition~\eqref{eq:SSVI_DecayCondition} holds if $\tfrac{1+\rho}{2}\,\tfrac{1}{\lambda}\le 2(\sqrt{p+1}-\sqrt{p})^2$; or simply 
\begin{equation}
\label{eq:Stronger_SSVI_SANA}
    \lambda
\geq
    \frac{1+\rho}
    {4\bigl(\sqrt{p+1}-\sqrt p\bigr)^2}
.
\end{equation}
Note that, if $\rho\ge 0$ then~\eqref{eq:Stronger_SSVI_SANA} implies the no butterfly arbitrage condition in~\citep[Theorem 4.2]{gatheral2014arbitrage}, namely $(1+|\rho|)\theta_T\varphi(\Theta_T) < 4$, since if $p>0$ then $4\bigl(\sqrt{p+1}-\sqrt p\bigr)^2<4$.  
In either case, if~\eqref{eq:Stronger_SSVI_SANA} holds then $C_{\mathrm{SSVI}}\in\mathcal{C}_{\mathrm{AF}}^p$.
\end{example}
We wrap up this set of examples with a modified power-law decay SSVI model since, as noted in~\cite[Remark 4.4]{gatheral2014arbitrage}, power law decay SSVI model of~\citep[Example 4.3]{gatheral2014arbitrage} may not be arbitrage-free for unbounded strikes.  Instead, we consider their ``modified power-law'' decaying SSVI.
\begin{example}[Modified power-law SSVI]
\label{ex:ModifiedPowerLawSSVI}
In the context of Proposition~\ref{prop:SSVIVerification}, consider the modified power-law SSVI specification in~\cite[Remark 4.4, Equation (4.5)]{gatheral2014arbitrage}; i.e.\ for $\eta>0$ and $\gamma\in(0,\tfrac12]$, for every $\theta>0$ let 
\[
\varphi(\theta)
\eqdef
\frac{\eta}
{\theta^\gamma(1+\theta)^{1-\gamma}}.
\]
Then
$
\theta\varphi(\theta)
=
\eta
\left(
\frac{\theta}{1+\theta}
\right)^{1-\gamma}
$
and therefore $
\sup_{\theta>0}\theta\varphi(\theta)
=
\eta$.  Condition~\eqref{eq:SSVI_DecayCondition} thus holds whenever
\begin{equation}
\label{eq:ModifiedPowerLawSSVI}
    \eta
\leq
    \frac{4}{1+\rho}
    \bigl(\sqrt{p+1}-\sqrt p\bigr)^2
.
\end{equation}
Moreover,~\cite[Remark~4.4]{gatheral2014arbitrage} shows that the corresponding SSVI surface is free of static arbitrage whenever $\eta(1+|\rho|)\leq2$.  Combining this with~\eqref{eq:ModifiedPowerLawSSVI} we deduce that $C_{\mathrm{SSVI}}\in\mathcal{C}_{\mathrm{AF}}^p$ if
\begin{equation}
\label{eq:SSVI_modpowerlaw_parameterization}
    \eta
\leq
    \min\left\{
    \frac{2}{1+|\rho|},
    \frac{4}{1+\rho}
    \bigl(\sqrt{p+1}-\sqrt p\bigr)^2
    \right\}
.
\end{equation}
\end{example}

We close our examples section by noting that satisfying condition~\eqref{eq:right_wing_asymptotic} does not imply being free of static arbitrage.
\begin{example}[Condition~\eqref{eq:right_wing_asymptotic} does not imply no arbitrage]
\label{ex:NonAF_ReLU_Decay}
Fix $\beta>0$ and set
$
    C_{\beta}:[0,\infty)^2\to[0,\infty)
$
by
\[
    C_{\beta}(T,K)
    \eqdef
    (1-\beta K)_+.
\]
Then $C_{\beta}$ satisfies~\eqref{eq:right_wing_asymptotic} for every $p>0$.  Indeed,
$
    C_{\beta}(T,K)=0
$
whenever $K\geq\beta^{-1}$, while for $0<K<\beta^{-1}$,
$
    C_{\beta}(T,K)\leq1\leq\beta^{-p}K^{-p}.
$
We note that: if $\beta>1$, then
$
    \partial_K^+C_{\beta}(T,0)
    =
    -\beta
    <
    -1,
$
so $C_{\beta}$ violates condition~{\rm(ii)} of Proposition~\ref{prop:contarbfree}.
\end{example}

We now present our AFNO framework, and provide several examples of neural operator architecture variants which are covered by our main guarantees.

\section{The Arbitrage-Free Neural Operator Framework}
\label{s:EncodeDecoder}

We will consider encoder-processor-decoder, sometimes called encoder-decoder, neural network models which are informed with the $\operatorname{SANOS}:\mathbb{R}^{F_{\mathrm{out}}}\to C_{\operatorname{AF}}$, cf.~\cite{buehler2026sanos}, as decoding head.  We first start by abstracting this into a class of ``good decoders''.

\subsection{Phase 1: Sample-Based Encoder on RKHS}
\label{s:EncoderDecoder__ss:Encoder}

Unlike neural operators built from fixed encoders, our model allows the sampling locations and the subsequent linear mixing of the sampled values to be trained, thereby permitting the encoder to adapt to the geometry relevant to the RKHS $\mathcal{H}$.  At an architectural level, this is closest in spirit to DeepONets~\cite{lu2021learning}, whose trunk network learns a basis for the output functions, and to PCA-Net~\cite{bhattacharya2021model}, whose linear input and output encoders are constructed using functional PCA.  A related trainable sampling encoder was recently considered in the sample neural operator (SNO) for deep functional clustering~\cite{li2026neural}, while related trainable decoder constructions were also noted in~\citep[Example~5.2 and Remark 5.4]{acciaio2024designing}.

For every $K\in\mathbb{N}_+$ satisfying $K\le\max\{1,\dim(\mathcal{H})\}$, we define the trainable encoder
\begin{equation}
\label{eq:trainable_encoder}
\begin{aligned}
    \mathcal{E}_K:
        \mathcal{H}
        \times
        \mathcal{D}^K
        \times
        \mathbb{R}^{K\times K}
    &\longrightarrow
        \mathbb{R}^K,
\\
    \big(
        C,
        (x_k)_{k=1}^K,
        A
    \big)
    &\longmapsto
        A
        \Big(
            \big\langle
                C,
                \kappa(\cdot,x_k)
            \big\rangle_{\mathcal{H}}
        \Big)_{k=1}^K
\end{aligned}
\end{equation}
Here $C\in\mathcal{H}$ denotes the input feature function, and
$\kappa$ is the reproducing kernel of the input RKHS. Thus,
\[
    \mathcal{E}_K(C,(x_k)_{k=1}^K,A)
    =
    A\big(C(x_k)\big)_{k=1}^K.
\]
These input coordinates are unrestricted real numbers, not convex
weights. The kernel sections used for input reconstruction are
distinct from the arbitrage-free surfaces used in the output decoder.

The trainable matrix $A$ allows the encoder to linearly mix, and in particular to precondition, the sampled values.  The following result shows that suitable choices of the sampling locations and of $A$ yield reconstructions converging uniformly to the identity on compact subsets of $\mathcal{H}$.

\begin{proposition}[Consistent Reconstruction]
\label{prop:encoding}
Suppose that $\mathcal{H}$ admits a complete interpolating sequence $\{x_i\}_{i\in I}$ (for some non-empty countable indexing set $I$).
Then, for every non-empty compact set $\mathcal{K}\subseteq\mathcal{H}$ and every $\varepsilon>0$, there exist $K_{\varepsilon}\in\mathbb{N}_+$,
$
    (x_k)_{k=1}^{K_{\varepsilon}}
    \in
    \mathcal{D}^{K_{\varepsilon}}
$,
and
$
    A_{\varepsilon}
    \in
    \mathbb{R}^{K_{\varepsilon}\times K_{\varepsilon}}
$
such that
\begin{equation}
\label{eq:projec}
    \sup_{C\in\mathcal{K}}
    \biggl\|
        C
        -
        \sum_{k=1}^{K_{\varepsilon}}
        \mathcal{E}_{K_{\varepsilon}}
        \big(
            C,
            (x_j)_{j=1}^{K_{\varepsilon}},
            A_{\varepsilon}
        \big)_k
        \kappa(\cdot,x_k)
    \biggr\|_{\mathcal{H}}
    \le
        \varepsilon.
\end{equation}
\end{proposition}
\begin{proof}
See Section~\ref{s:Proofs__ss:ToInfinitedimensions}.
\end{proof}

 While the above statement is straightforward from the \textit{bounded approximation property} (BAP) for any Riesz basis, it has only shown that each $C$ in $H$ can be written
as linear combination of basis functions with weights in $\tilde \beta\in\R^{K_\epsilon}$. it is \emph{not} evident that the same statement remains true for convex weights $\beta\in[0,1]^{K_\epsilon}$ such that $|\beta|=1$; which we use below.

\subsection{Phase 2: Transformer Backbone}
\label{s:EncoderDecoder__ss:Backbone}
 
The backbone of our model will be the standard transformer architecture.

\begin{definition}[Transformers with multi-head attention]
Let $\lambda>0$, and let $d_{\rm in},d_{{\rm key}},d_{\rm out},N\in \mathbb N$. 
Let
${\bf Q},{\bf K}\in \mathbb{R}^{d_{{\rm key}}\times d_{\rm in}}$
and
${\bf V}\in \mathbb{R}^{d_{\rm out}\times d_{\rm in}}$. 
We define the associated attention mechanism
$\operatorname{Attn}(\cdot|{\bf Q},{\bf K},{\bf V},\lambda):\mathbb{R}^{N\times d_{\rm in}}\to \mathbb{R}^{N\times d_{\rm out}}
$
by
\begin{equation} \label{eq:def_attention}
    [\operatorname{Attn}(\mathbf{X}|{\bf Q},{\bf K},{\bf V},\lambda)]_n
    \eqdef 
    \sum_{m=1}^N
    \frac{e^{\lambda \langle {\bf Q}[\mathbf{X}]_n, {\bf K}[\mathbf{X}]_m\rangle/\sqrt{d_{{\rm key}}}}}{
    \sum_{l=1}^N e^{\lambda \langle {\bf Q}[\mathbf{X}]_n, {\bf K}[\mathbf{X}]_l \rangle/\sqrt{d_{{\rm key}}}}}
    {\bf V}[\mathbf{X}]_m, \qquad n=1,\dots,N.
\end{equation}
Here $[\mathbf{X}]_j \in \mathbb{R}^{d_{\rm in}}$ denotes the $j$th row of $\mathbf{X}$, interpreted as a column vector when multiplied by ${\bf Q},{\bf K},{\bf V}$.
A transformer network operates by iterating layers, each consisting of a row-wise activation–bias map followed by an attention mechanism.
Concretely, let $d,D,L\in\mathbb{N}$, and let $d_0,d_1,\dots,d_L, H_0,H_1,\dots,H_L\in\mathbb{N}$ such that $d_0=d$ and $d_L=D$.
For $j=0,1,\dots,L-2$ and $\mathtt{h}=1,\dots,H_j$, let $d^{(j)}_{{\rm key}, \mathtt{h}}, d^{(j)}_{{\rm value}, \mathtt{h}}\in\mathbb{N}$ such that $d_{j+1} = \sum_{{\mathtt{h}}=1}^{H_j} d^{(j)}_{{\rm value}, \mathtt{h}}$.
A $j$th \emph{transformer layer with multi-head attention}, for $j=0,1,\dots,L-2$, is a map
$
\mathcal{T}_j:\mathbb{R}^{N\times d_j}\to \mathbb{R}^{N\times d_{j+1}}
$
that sends an input matrix $\mathbf{X}^{(j)}\in \mathbb{R}^{N\times d_j}$ to
$
\mathcal{T}_j(\mathbf{X}^{(j)})\eqdef  \mathbf{X}^{(j+1)}\in \mathbb{R}^{N\times d_{j+1}},
$
where 
\begin{alignat}{2}
    \nonumber \mathbf{Z}^{(j)}
&\eqdef 
    \underbrace{
        \bigoplus_{\mathtt{h}=1}^{H_j}
            \operatorname{Attn}\big(
                \mathbf{X}^{(j)}
                |{\bf Q}_{\mathtt{h}}^{(j)},{\bf K}_{\mathtt{h}}^{(j)},{\bf V}_{\mathtt{h}}^{(j)},\lambda
            \big)
    }_{\text{multi-head attention}} && \\
    \label{eq:representation_transformer} [\mathbf{X}^{(j+1)}]_n
&\eqdef  
        \sigma\bullet
        \big(
            [\mathbf{Z}^{(j)}]_n + b^{(j)}
        \big),
    &&\qquad n=1,\dots,N.
\end{alignat}
Here $\bigoplus$ denotes concatenation along the second (column) dimension, $
{\bf Q}_{\mathtt{h}}^{(j)},{\bf K}_{\mathtt{h}}^{(j)}\in \mathbb{R}^{d_{{\rm key}, \mathtt{h}}^{(j)}\times d_j}$ and 
$
{\bf V}_{\mathtt{h}}^{(j)}\in \mathbb{R}^{d_{{\rm value}, \mathtt{h}}^{(j)}\times d_j}$, and $b^{(j)}\in\mathbb{R}^{d_{j+1}}$.
Further, the addition in \eqref{eq:representation_transformer} is applied row-wise.
A \emph{transformer with multi-head attention} with depth $L$, obtained by composing the layers $\mathcal{T}_0,\mathcal{T}_1,\dots,\mathcal{T}_{L-2}$, is a map $\mathcal{T}:\mathbb{R}^{N\times d}\to\mathbb{R}^{N\times D}$ sending an input matrix ${\bf X}^{(0)}\eqdef {\bf X}\in\mathbb{R}^{N\times d}$ to $\mathcal{T}({\bf X})\in\mathbb{R}^{N\times D}$, defined row-wise by $[\mathcal{T}({\bf X})]_n\eqdef {\bf A}^{(L-1)}[{\bf X}^{(L-1)}]_n+b^{(L-1)}$ for $n=1,\ldots,N$, for some ${\bf A}^{(L-1)}\in\mathbb{R}^{D\times d_{L-1}}$ and $b^{(L-1)}\in\mathbb{R}^D$.
\end{definition}

We will derive all our results for $\operatorname{ReLU}$-multilayer perceptrons (MLPs) backbones and then convert them into transformer backbones using the transfer principle of~\citep[Proposition A.1]{kratsios2026adaptivity}; which follows in similar suit to the conversion results for CNNs in~\cite{petersen2020equivalence} or for transformers with probabilistic formulations of context~\cite[Proposition 11]{kratsios2025context}.   For completeness, we record the definition of the standard MLP model in Appendix~\ref{s:ExtrasAI}.

\subsection{Phase 3: Arbitrage-Free Decoders}
\label{s:EncoderDecoder__ss:GoodDecoder}

We will consider pointwise decoders, mapping latent parameters to call surfaces of similar type to the $\operatorname{SANOS}$ model of~\cite{buehler2026sanos}.  We refer to these as arbitrage-free decoders, and defined them as follows.
We will consider pointwise decoders which map finite-dimensional latent parameters to arbitrage-free call surfaces.  These decoders are modelled on the SANOS architecture of~\cite{buehler2026sanos}, but the following definition isolates only the structural properties needed in the sequel.  

We first concisely present the abstract definitions which are enough for our framework.  Shortly after we provide piecewise linear and smooth examples of arbitrage-free decoders in $\S$~\ref{s:EncoderDecoder__ss:Examples_AFDecoders} towards the end of this section before passing to our main results.

\begin{definition}[Arbitrage-Free (AF) Surface-Decoder]
\label{defn:Good_Decoder}
Let $\mathcal{Y}\subseteq\mathcal{C}_{AF}^{p,\lambda}$ be a uniformly controlled call-surface class with regularity-data $(p,\lambda,B,\omega)$.  We say that $\mathcal{Y}$ is admitted by an arbitrage-free decoder if, for every $R,S>0$ and every $N,M\in\mathbb{N}_+$, there exist $D_{N,M,R,S}\in\mathbb{N}_+$ and a map
\[
    \phi^{(N,M,R,S)}
    :
    \mathbb{R}^{D_{N,M,R,S}}\times [0,\infty)^2
    \to
    [0,\infty)
\]
such that, for every $C\in\mathcal{C}_{AF}^{p,\lambda}$
and $\varepsilon>0$ 
there is a $\theta_C\in\mathbb{R}^{D_{N,M,R,S}}$ such that 
$\phi_{\theta_C}\eqdef \phi(\theta_C,\cdot)$  satisfies:
\begin{enumerate}
    \item[(i)] \textbf{Approximate-interpolation on the rectangular grid:}
    on the uniform grid $\{(t_m,K_n)\}_{m,n=0}^{M,N}$, 
    where $t_m \eqdef \frac{mS}{M}$ and $K_n \eqdef \frac{nR}{N}$, one has: for each $m\in [M]$ and $n\in [N]$
    \[
        \big|
            \phi^{(N,M,R,S)}_{\theta_C}(t_m,K_n)
            -
            C(t_m,K_n)
        \big|
        <
        \varepsilon
    \]

    \item[(ii)] \textbf{No-arbitrage:}
    for every parameter $\theta\in\mathbb{R}^{D_{N,M,R,S}}$,
    $
        \phi^{(N,M,R,S)}_{\theta}\in\mathcal{C}_{AF}
    $,
    \item[(iii)] \textbf{Locally uniform right-wing control:}
    For every $A>0$ and every compact maturity interval
    $
        [T_{\min},T_{\max}]
    \subseteq
        [0,\infty),
    $
    there exist $\rho>p$ and
    $
        B_{\rho,A,R,S,N,M,T_{\min},T_{\max}}>0
    $
    such that, for every
    $
        \theta
    \in
        [-A,A]^{D_{N,M,R,S}},
    $
    every $T\in[T_{\min},T_{\max}]$, and every $K\ge R+1$,
    \begin{equation}
    \label{eq:rightwing_decaycondition}
        \phi^{(N,M,R,S)}_{\theta}(T,K)
        \le
        B_{\rho,A,R,S,N,M,T_{\min},T_{\max}}
        (1+K)^{-\rho}.
    \end{equation}
\end{enumerate}
When clear from the context, we abbreviate $\Phi_{\theta_C}\eqdef \phi^{(N,M,R,S)}_{\theta_C}$ for notational convenience.
\end{definition}

The arbitrage-free decoder should be understood as finitely parameterized individual arbitrage-free call surfaces which are extremely flexible; i.e.\ non-parametric.  These are then convexly combined together to construct our arbitrage-free decoder\footnote{Each individual call price surfaces can, but need not, be trained.}. 
\begin{definition}[Arbitrage-Free Decoder]
\label{defn:AFDECODER}
An \textit{arbitrage-free decoder} combines arbitrage-free
surface-decoder outputs using nonnegative weights summing to one.
For each input, these weights are independent of strike and maturity.
\end{definition}
In~\eqref{eq:Dysanos}, the surface parameters may depend on the input.
In the fixed-dictionary specialization~\eqref{eq:AFNO_definition},
the parameters $\theta_i$ are shared across inputs; they may be
pre-fitted or trained jointly with the backbone.
We propose here to learn the convex combinations with the attention mechanism introduced earlier, i.e.:
\begin{equation}
\label{eq:Dysanos}
\begin{aligned}
\Phi:
\mathbb{R}^{F_{\mathrm{mkt}}} & \rightarrow C([0,\infty)^2,[0,\infty))
\\
    \Phi
    (\theta)
& \mapsto 
        \operatorname{Softmax}\big(
            \tfrac{1}{\lambda}\alpha(\vartheta)
        \big)^{\top}
        \,
        [\phi(v(\vartheta)_k)]_{k=1}^K
,
\end{aligned}
\end{equation}
where $F_{\mathrm{mkt}},K\in \mathbb{N}_+$, $\alpha:\mathbb{R}^{F_{\mathrm{mkt}}}\to \mathbb{R}^K$ is a transformer with multi-head attention an $\operatorname{ReLU}$-activation, in~\eqref{eq:representation_transformer}, $\lambda>0$, $v:\mathbb{R}^{F_{\mathrm{mkt}}}\to \mathbb{R}^{K\times F}$ is another $\varsigma$-MLP, the product in~\eqref{eq:Dysanos} is applied point-wise in $x$, and $\phi$ is a AF-Decoder model%
\footnote{Cf.~Definition~\ref{defn:Good_Decoder}.}.  

In Definition~\ref{defn:AFDECODER}, $\alpha$ represents adaptively computed alignment scores, testing how similar the input market features are to each of the arbitrage-free call surfaces generated by feeding the rows of $v(x)$ into the AF-Decoder model $\Phi$.

\subsection{The End-to-End Framework: Arbitrage-Free Neural Operator}
\label{s:Model__ss:AFNO}

We may now close this section by formally defining the class of architectures illustrated in Figure~\ref{fig:OurNO}.  We will show that every such architecture is compatible with risk-neutral pricing theory, uniformly over all parameter specifications, and remains expressive: suitable parameter choices can approximate any sufficiently regular map from an RKHS into the space of arbitrage-free European call-price surfaces.

We emphasize that the Arbitrage-Free Neural Operator (AFNO) architecture is best viewed as an architectural framework.  
The encoder extracts finitely many real-valued input features. The backbone then computes simplex weights by softmax, and the decoder combines a separate dictionary of arbitrage-free output surfaces using those weights. Convexity of $\mathcal{C}_{AF}$ provides the no-arbitrage certificate. The output dictionary need not be a basis and does not determine the input RKHS.

We can \textit{model} the RKHS used to model the input data, as well as the arbitrage-free decoder most appropriate for the particular use case or downstream application, provided that it satisfies the basic requirements of Definition~\ref{defn:Good_Decoder}.  Examples include the $\operatorname{SANOS}$ model of~\cite{buehler2026sanos} and the simpler piecewise-linear decoder introduced in Definition~\ref{def:Parallel_ReLU_CallSurface_Irregular}.

\begin{definition}[Arbitrage-Free Neural Operator]
\label{defn:AFNO}
Let $\mathcal{H}$ be an RKHS over a non-empty set
$
    \mathcal{D}
\subseteq
    [0,\infty)^2
$.
A map
$
    \Psi:
    \mathcal{H}
    \to
    C\big([0,\infty)^2,[0,\infty)\big)
$
is called an \emph{arbitrage-free neural operator} (AFNO) if it admits a representation of the form
\begin{equation}
\label{eq:AFNO_definition}
\begin{aligned}
    \Psi:
    \mathcal{H}
    &\longrightarrow
    C\big([0,\infty)^2,[0,\infty)\big),
\\
    C
    &\longmapsto
    \sum_{i=1}^I
        \operatorname{Softmax}
        \left(
            \frac{1}{\tau}
            \alpha
            \left(
                \mathcal{E}_K
                \big(
                    C,
                    \mathbf{x},
                    A
                \big)
            \right)
        \right)_i
        \phi_{\theta_i},
\end{aligned}
\end{equation}
where $K,I\in\mathbb{N}_+$,
$
    \mathbf{x}
    \eqdef
    (x_k)_{k=1}^K
\in
    \mathcal{D}^K
$,
$
    A
\in
    \mathbb{R}^{K\times K}
$,
and
$
    \alpha:
    \mathbb{R}^K
    \to
    \mathbb{R}^I
$
is either a transformer with multi-head attention or a
$\operatorname{ReLU}$-MLP.  Moreover, $\tau>0$, and
$
    \phi_{\theta_1},
    \dots,
    \phi_{\theta_I}
\in
    \mathcal{C}_{AF}
$
are outputs of an arbitrage-free decoder in the sense of
Definition~\ref{defn:Good_Decoder}.
\end{definition}

In what follows, we refer to both \textit{AF surface-decoders} and \textit{AF decoders} simply as \textit{AF decoders} to keep terminology light, as it will be conceptually clear.

\subsection{Examples of AF-Decoders}
\label{s:EncoderDecoder__ss:Examples_AFDecoders}
We now provide two examples of AF-decoders showing the reader how they can build their own AFNO.  The first is an intuitive model which is piecewise linear in each strike and along maturities for a fixed strike.  The second is the SANOS model of~\cite{buehler2026sanos} which is effectively a shallow neural network with Black-Scholes European call-price as is activation function.

\subsubsection{A Simple Example of an Arbitrage-Free Decoder}
\label{s:EncoderDecoder__ss:Examples_AFDecoders___sss:PWL}

\begin{figure}[H]
    \centering
    \begin{minipage}[t]{0.65\textwidth}
        \vspace{0pt}
        While one may consider more sophisticated arbitrage-free decoders, such as the $C^{\infty}$-regularity $\operatorname{SANOS}$ model in~\cite{buehler2026sanos}, we find it most illustrative to focus here on piecewise-linear examples, since our aim is not to advocate for a particular decoder, but rather to demonstrate how such decoders can be used to construct a certifiably arbitrage-free neural operator.  We also note that our chosen decoder admits a tensorized $\operatorname{ReLU}$-MLP representation, thereby further connecting our construction to a ``structured version'' of DeepONets; cf.~\cite{lu2021deep}.  
One could also build an arbitrage-free decoder using the $C^{1,1}$-regularity reconstruction formula of~\cite{carr2017local}.
    \end{minipage}
    \hfill
    \begin{minipage}[t]{0.275\textwidth}
        \vspace{0pt}
        \centering
        \includegraphics[width=\linewidth]{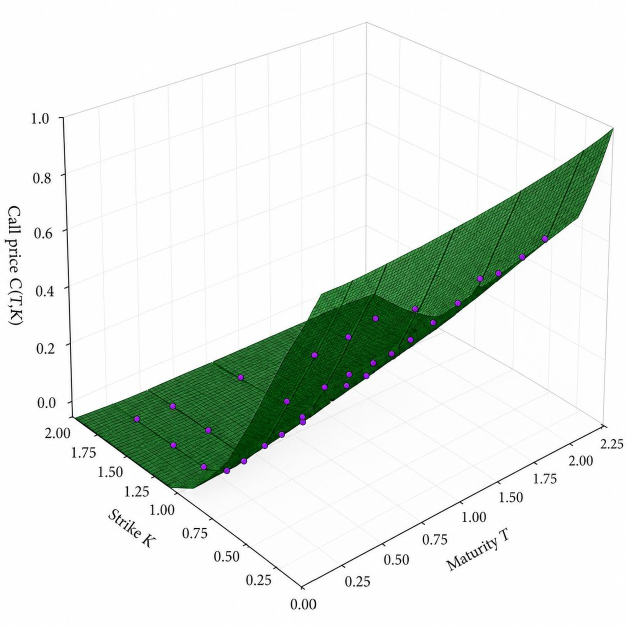}
    \end{minipage}
   
        \caption{Illustration of the structured $\operatorname{ReLU}$-MLP arbitrage-free decoder with $M=5$ maturities and $N=8$ strikes; the dots indicate the interpolated grid quotes.}
        \label{fig:Simple_AF_Decoder}
\end{figure}

The following is the ``Discrete Local Vol'' interpolation scheme for
homogeneous strikes described in~\cite{DLV}. This amounts to linear interpolation between strikes:

\begin{proposition}[Arbitrage-Free Discrete Polytope]
\label{prop:admissible_polytope}
Let $M,N\in\mathbb{N}_+$.  Fix
$
    0<T_1<\dots<T_M
$
and
$
    0<K_1<\dots<K_N
$.
Fix, moreover, two virtual strikes $K_{\mathrm{L}},K_{\mathrm{R}}>0$ satisfying
$
    0
<
    K_{\mathrm{L}}
<
    \min\{K_1,1\}
$ and $
    K_{\mathrm{R}}
>
    \max\{K_N,1\}
$.
Write
$
    \mathbf{T}\eqdef(T_m)_{m=1}^M
$, 
$
    \mathbf{K}\eqdef(K_n)_{n=1}^N
$, and set
\[
    \widehat K_0
    \eqdef
    0,
    \qquad
    \widehat K_1
    \eqdef
    K_{\mathrm{L}},
    \qquad
    \widehat K_{n+1}
    \eqdef
    K_n
    \quad(n\in[N]_+),
    \qquad
    \widehat K_{N+2}
    \eqdef
    K_{\mathrm{R}}.
\]
Given any matrix of quotes
$
    y=(y_{m,n})_{m,n=1}^{M,N}
    \in\mathbb{R}^{M\times N},
$ we defined the extended matrix of quotes by
\[
    \widehat y_{m,0}
    \eqdef
    1,
    \qquad
    \widehat y_{m,1}
    \eqdef
    1
    +
    \frac{K_{\mathrm{L}}}{K_1}
    \big(
        y_{m,1}-1
    \big)
    \qquad
    \widehat y_{m,n+1}
    \eqdef
    y_{m,n}
    \quad(n\in[N]_+),
    \qquad
    \widehat y_{m,N+2}
    \eqdef
    0
\]
and for each $j\in [N+1]$, define $
    s_{m,j}(y)
\eqdef
    \tfrac{
        \widehat y_{m,j+1}
        -
        \widehat y_{m,j}
    }{
        \widehat K_{j+1}
        -
        \widehat K_j
    }$.  
\hfill\\
\noindent
Then, for each
$
    y\in\mathbb{R}^{M\times N},
$
the following are equivalent:
\begin{enumerate}
    \item[(i)] \textbf{Existence of an Arbitrage-Free Interpolator:}
    There exists a continuous map
    $
        \mathfrak{I}:[0,\infty)^2\to[0,\infty)
    $
    such that: $\mathfrak{I}(T,\cdot)$ and $\mathfrak{I}(\cdot,K)$ are piecewise affine for each $T,K\ge 0$ such that:
    \[
        \mathfrak{I}
        \big(
            T_m,\widehat K_j
        \big)
        =
        \widehat y_{m,j}
        \qquad
        \mbox{for every }
        m\in[M]_+
        \mbox{ and }
        j\in[N+2]
    .
    \]
    and such that $\mathfrak{I}$ is an arbitrage-free call-price surface (cf.~Proposition~\ref{prop:contarbfree}),
    \item[(ii)] \textbf{Polytope Membership:}
    $y$ belongs to the compact and convex polytope
    \begin{equation}
    \label{eq:AF_polytope}
    \begin{aligned}
        \mathfrak{C}_{\mathbf{T},\mathbf{K}}
    \eqdef
        \Big\{
            y\in\mathbb{R}^{M\times N}
            :\;&
            -1
            \le
            s_{m,0}(y)
            =
            s_{m,1}(y)
            \le
            \dots
            \le
            { s_{m,N+1}(y)\le 0,\qquad m\in[M]_+}
\\
            &
            y_{m,n}
            \le
            y_{m+1,n},
            \quad
            m\in[M-1]_+,
            \quad
            n\in[N]_+
        \Big\}.
    \end{aligned}
    \end{equation}
\end{enumerate}
Moreover,
$
    \mathfrak{C}_{\mathbf{T},\mathbf{K}}
$
is non-empty and there is
a unique orthogonal projection
$
    P_{\mathbf{T},\mathbf{K}}
    :
    \mathbb{R}^{M\times N}
    \rightarrow
    \mathfrak{C}_{\mathbf{T},\mathbf{K}}
$.
\end{proposition}
\begin{proof}
See Section~\ref{s:AF_Decoder_Proofs}.
\end{proof}

The polytope in~\eqref{eq:AF_polytope} characterizes no-arbitrage for `low-regularity'' European call-price interpolators, by which we mean interpolators that are piecewise linear in each argument separately.  This characterization allows us to define the following arbitrage-free decoder on a single parameter space.

\begin{definition}[Structured $\operatorname{ReLU}$-MLP Arbitrage-Free Decoder]
\label{def:Parallel_ReLU_CallSurface_Irregular}
Fix $\delta>0$ and $
M,N\in\mathbb{N}_2$.  Consider the decoder 
\begin{equation}
\label{eq:def:Parallel_ReLU_CallSurface_Irregular}
\begin{aligned}
    \phi^{(M,N)}_{\delta}:
    \mathbb{R}^{M}
    \times
    \mathbb{R}^{N}
    \times
    \mathbb{R}^{M\times N}
    &\longrightarrow
    C\big([0,\infty)^2,[0,\infty)\big),
\\
    (\mathbf{t},\mathbf{k},\mathbf{z})
    &\longmapsto
    \Psi_{\mathbf{T},\mathbf{K},\mathbf{y}},
\end{aligned}
\end{equation}
where $\Psi_{\mathbf{T},\mathbf{K},\mathbf{y}}$ is defined as follows:
Write
$
    \mathbf{t}\eqdef(t_m)_{m=1}^M
$
and
$
    \mathbf{k}\eqdef(k_n)_{n=1}^N,
$
and set
$
    t_0\eqdef k_0\eqdef0.
$
Define the ordered grids
\[
\begin{aligned}
        (T_1,\dots,T_M)
    \eqdef
        \left(
            \sum_{j=1}^{m}
                \max\big\{
                    |t_j-t_{j-1}|,
                    \delta
                \big\}
        \right)_{m=1}^M
\,\mbox{ and }\,
        (K_1,\dots,K_N)
    \eqdef
        \left(
            \sum_{j=1}^{n}
                \max\big\{
                    |k_j-k_{j-1}|,
                    \delta
                \big\}
        \right)_{n=1}^N
\end{aligned}
\]
Define
$
    K_{\mathrm{L}}
\eqdef
    \frac{1}{2}\min\{\delta,1\}
$, $K_{\mathrm{R}}
    \eqdef
    K_N+1+\delta^{-1}
$,  
and define the corrected quote data by
$
    \mathbf{y}
=
    (y_{m,n})_{m,n=1}^{M,N}
    \eqdef
    P_{\mathbf{T},\mathbf{K},\delta}(\mathbf{z})
$ 
where
$
    P_{\mathbf{T},\mathbf{K},\delta}
    :
    \mathbb{R}^{M\times N}
    \mapsto
    \mathfrak{C}_{\mathbf{T},\mathbf{K},\delta}
$ 
is the projection onto the polytope
$
    \mathfrak{C}_{\mathbf{T},\mathbf{K},\delta},
$
defined as in Proposition~\ref{prop:admissible_polytope}%
\footnote{Using the
fixed virtual strikes $K_{\mathrm{L}}$ and $K_{\mathrm{R}}$.}%
.
Next, for each $m\in[M]_+$, define the augmented strike and quote data, where $n\in[N]_+$, by
\begin{equation*}
\begin{aligned}
    (\widehat K_0,\widehat y_{m,0})
& \eqdef
    (0,1)
\\
(\widehat K_1,\widehat y_{m,1})
&
\eqdef\left(K_{\mathrm L},1+\frac{K_{\mathrm L}}{K_1}(y_{m,1}-1)\right)
\\
    (\widehat K_{n+1},\widehat y_{m,n+1})
& \eqdef
    (K_n,y_{m,n})
\\
    (\widehat K_{N+2},\widehat y_{m,N+2})
& \eqdef
    (K_{\mathrm{R}},0)
.
\end{aligned}
\end{equation*}
For each $j\in[N+2]_+$, define
$
    \Delta\widehat K_j
    \eqdef
    \widehat K_j-\widehat K_{j-1}
$,
and the decreasing secant slopes
$
    S_{m,j}
    \eqdef
    \frac{
        \widehat y_{m,j-1}-\widehat y_{m,j}
    }{
        \Delta\widehat K_j
    }
$.
Set
$
    S_{m,N+3}
    \eqdef
    0
$.
Set $T_0\eqdef 0 $, and let $e_0,\dots,e_M$ denote the canonical basis of $\mathbb{R}^{M+1}$, indexed from $0$ to $M$.  Consider the $\operatorname{ReLU}$-MLPs $\boldsymbol{\Psi},\Lambda:\mathbb{R}\to\mathbb{R}^{M+1} $ defined, for each $K,T\in\mathbb{R}$, by
\begin{equation}
\label{eq:ReLUInterpolator}
\begin{aligned}
    \boldsymbol{\Psi}(K)_0
    &\eqdef
    \operatorname{ReLU}(1-K),
\\
    \boldsymbol{\Psi}(K)_m
    &\eqdef
    \operatorname{ReLU}
    \Biggl(
        \sum_{j=1}^{N+2}
            \big(
                S_{m,j}-S_{m,j+1}
            \big)
            \operatorname{ReLU}
            \big(
                \widehat K_j-K
            \big)
    \Biggr)
\qquad
    m\in[M]_+,
\\
    \Lambda(T)
    &\eqdef
    \operatorname{ReLU}
    \Biggl(
        e_0
        +
        \sum_{m=0}^{M-1}
            \frac{
                \operatorname{ReLU}(T-T_m)
                -
                \operatorname{ReLU}(T-T_{m+1})
            }{
                T_{m+1}-T_m
            }
            \big(
                e_{m+1}-e_m
            \big)
    \Biggr).
\end{aligned}
\end{equation}
Finally, define
$
    \Psi_{\mathbf{T},\mathbf{K},\mathbf{y}}
    :
    [0,\infty)^2
    \to
    [0,\infty)
$
by
$
    \Psi_{\mathbf{T},\mathbf{K},\mathbf{y}}(T,K)
    \eqdef
    \Lambda(T)^\top
    \boldsymbol{\Psi}(K)
$.
\end{definition}

The next proposition shows that, for sufficiently small ordering tolerance, the decoder exactly interpolates any arbitrage-free call surface on the prescribed grid while remaining arbitrage-free globally.

\begin{proposition}[Universal Arbitrage-Free Interpolation]
\label{prop:DecoderWorks}
Let $C:[0,\infty)^2\to[0,\infty)$ be an arbitrage-free call surface,
let $M,N\in\mathbb{N}_2$, and fix grids
$
    0<T_1<\dots<T_M
$
and
$
    0<K_1<\dots<K_N.
$
Set
$
    z_{m,n}
    \eqdef
    C(T_m,K_n)
$
for each $m\in[M]_+$ and $n\in[N]_+$.  Then there exists
$\delta_0>0$ such that, for every $0<\delta\le\delta_0$,
\[
    \phi_{\delta}^{(M,N)}
    \big(
        \mathbf{T},
        \mathbf{K},
        \mathbf{z}
    \big)
    (T_m,K_n)
    =
    C(T_m,K_n)
\]
for each $m\in[M]_+$ and $n\in[N]_+$.  Moreover,
$
    \phi_{\delta}^{(M,N)}
    (
        \mathbf{T},
        \mathbf{K},
        \mathbf{z}
    )
$
is an arbitrage-free call surface.
\end{proposition}
\begin{proof}
See Section~\ref{s:AF_Decoder_Proofs}.
\end{proof}

The following proposition records that the resulting ReLU construction is an arbitrage-free decoder in the sense of Definition~\ref{defn:Good_Decoder}.

{ 
\begin{proposition}[The Structured $\operatorname{ReLU}$-MLP Family Induce Arbitrage-Free Decoders]
\label{prop:ReLU_Is_AF_Decoder}
For $R,S>0$ and $N,M\in\mathbb N_+$, set $\widehat N\eqdef\max\{N,2\}$ and $\widehat M\eqdef\max\{M,2\}$. The family
\[
\phi^{(N,M,R,S)}_\theta\eqdef\phi^{(\widehat M,\widehat N)}_{e^{-\xi}}(\mathbf t,\mathbf k,\mathbf z),\qquad \theta=(\xi,\mathbf t,\mathbf k,\mathbf z)\in\mathbb R^{1+\widehat M+\widehat N+\widehat M\widehat N},
\]
satisfies Definition~\ref{defn:Good_Decoder} for every $p,\lambda>0$. In particular, the ordering tolerance is a selectable parameter.
\end{proposition}
}
\begin{proof}
See Section~\ref{s:AF_Decoder_Proofs}.
\end{proof}

Next, we show that the $\operatorname{SANOS}$ model is a smooth decoder and the approximate interpolation property gives us just enough slack to obtain \textit{closed-form} parameter specifications certifying no static arbitrage.

\subsubsection{SANOS is an Arbitrage-Free Decoder}
\label{s:EncoderDecoder__ss:Examples_AFDecoders___sss:SANOS}
We now construct a smooth counterpart of the structured
$\operatorname{ReLU}$ decoder in
Definition~\ref{def:Parallel_ReLU_CallSurface_Irregular}.  The construction
is inspired by the Black-Scholes smoothing mechanism underlying the
$\operatorname{SANOS}$ parameterization of~\cite{buehler2026sanos}.  A
direct replacement of each hinge
$
    (\widehat K_j-K)^+
$
by a smooth approximation of $\operatorname{ReLU}(\widehat K_j-K)$ does not,
in general, preserve the two left-wing conditions
$
    C(T,0)=1
$ and $
    \partial_K^+C(T,0)=-1
$.  
We therefore first introduce a smooth Black-Scholes approximation of
$\operatorname{ReLU}$ function, defined 
for every $\nu>0$ and each $x\in \mathbb{R}$  by
$
    \operatorname{BS}_{\nu}:\mathbb{R}\to[0,\infty)
$
by
\begin{equation}
\label{eq:BS_activation__primitive}
    \operatorname{BS}_{\nu}(x)
\eqdef
    \begin{cases}
        (1+x)
        \Phi\left(
            \dfrac{\log(1+x)+\frac{1}{2}\nu^2}{\nu}
        \right)
        -
        \Phi\left(
            \dfrac{\log(1+x)-\frac{1}{2}\nu^2}{\nu}
        \right)
        &
        \mbox{ if: }x>-1
        \\[8pt]
        0
        &
        \mbox{ if: } x\le-1.
    \end{cases}
\end{equation}
\begin{figure}[H]
    \centering
    \begin{minipage}[t]{0.55\textwidth}
        \vspace{0pt}
    We extend~\eqref{eq:BS_activation__primitive} it to values of $\nu\le 0$ by setting
    $
        \operatorname{BS}_{\nu}
    \eqdef
        \operatorname{ReLU}
    $ for every $\nu\le 0$.  It can be shown (Proposition~\ref{prop:BS_activation}) that doing so makes $\operatorname{BS}$ continuous in both $\nu$ and $x$ on all of $\mathbb{R}^2$ since $\operatorname{BS}_{\nu}$ converges to $\operatorname{ReLU}$ uniformly, over all of $\mathbb{R}$, as $\nu$ tends to $0$ from the right.
    We highlight that, for $\nu>0$ and $x>-1$,
    $
        \operatorname{BS}_{\nu}(x)
    $
    is the Black-Scholes put price with spot and maturity equal to $1$,
    zero interest rate, strike $1+x$, and volatility $\nu$ and that $\operatorname{BS}_{\nu}$ is smooth on $(0,\infty)$ whenever $\nu>0$ and it is in $C^2(\mathbb{R})$. 
    \end{minipage}
    \hfill
    \begin{minipage}[t]{0.4\textwidth}
        \vspace{0pt}
        \centering
        \includegraphics[width=\linewidth]{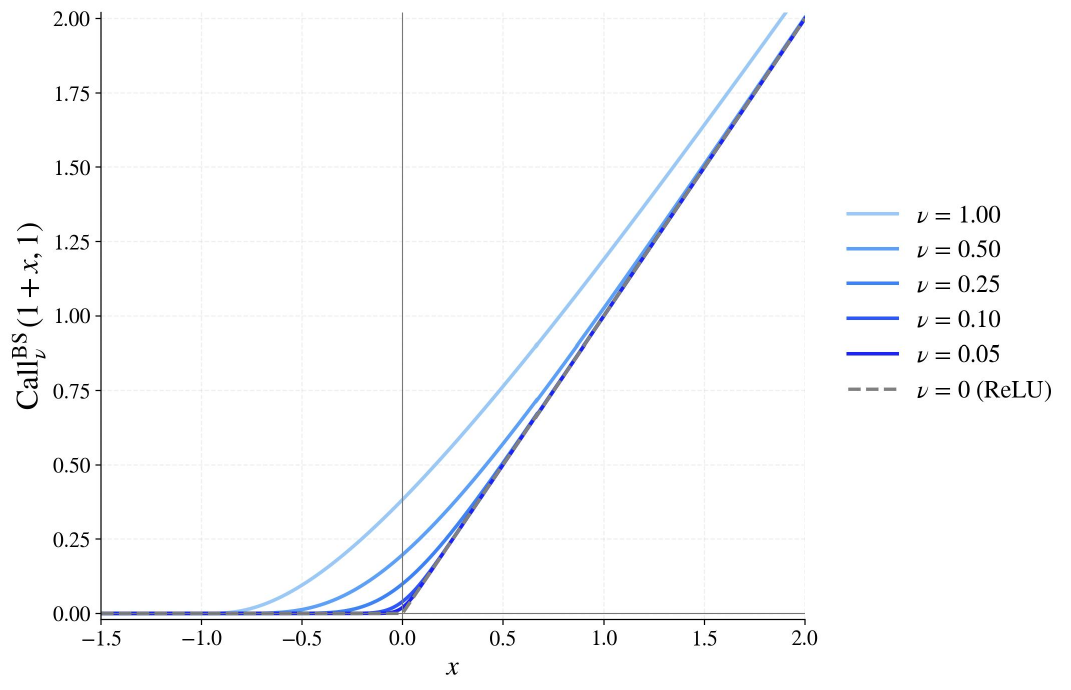}
    \end{minipage}
    \caption{Illustration of the normalized Black--Scholes activation
    $
        \operatorname{BS}_{\nu}(x)
        =
        \operatorname{Call}^{\operatorname{BS}}_{\nu}(1+x,1)
    $,
    corresponding to the homogeneous Black--Scholes call hinge
    \eqref{eq:BS_call_hinge}.  As $\nu$ decreases towards $0$ $\operatorname{Call}_{\nu}^{\operatorname{BS}}$ converges uniformly to $\operatorname{ReLU}$.}
    \label{fig:ConvergenceActivations}
\end{figure}
The strike-side smoothing must respect the homogeneity of call prices.  For
$\nu\ge0$, $s>0$, and $K\ge0$, define the Black-Scholes call hinge by
\begin{equation}
\label{eq:BS_call_hinge}
    \operatorname{Call}^{\operatorname{BS}}_{\nu}(s,K)
\eqdef
    s-K
    +
    s\,
    \operatorname{BS}_{\nu}
    \left(
        \frac{K}{s}-1
    \right)
\end{equation}
Moreover, the zero-volatility limit recovers the intrinsic-value; i.e.\ 
$
    \operatorname{Call}^{\operatorname{BS}}_{0}(s,K)
=
    (s-K)^+
$.

\begin{definition}[SANOS Decoder]
\label{def:Smooth_AF_Decoder}
Fix $\delta,\overline\nu>0$ and $M,N\in\mathbb{N}_2$.  Retain the grid
correction, polytope projection, augmented strikes, augmented quotes, and
secant slopes from
Definition~\ref{def:Parallel_ReLU_CallSurface_Irregular}.  Define
\begin{equation}
\label{eq:def_Smooth_AF_Decoder}
\begin{aligned}
    \phi_{\delta,\overline\nu}^{(M,N)}
    :
    \mathbb{R}^{M}
    \times
    \mathbb{R}^{N}
    \times
    \mathbb{R}^{M\times N}
    \times
    \mathbb{R}
    &\longrightarrow
    C\big([0,\infty)^2,[0,\infty)\big)
    \cap
    C^{\infty}\big((0,\infty)^2\big),
\\
    (\mathbf t,\mathbf k,\mathbf z,\eta)
    &\longmapsto
    \Psi_{\mathbf T,\mathbf K,\mathbf y,\nu},
\end{aligned}
\end{equation}
where
$
    \nu\eqdef\nu_{\overline\nu}(\eta)
$
{ 
and the surface on the right-hand side is defined as follows. Here $e_1,\ldots,e_M$ denote the canonical basis of $\mathbb R^M$. For each $K,T\geq0$, set
}
\begin{align}
\label{eq:smooth_BS_strike_slices}
    \boldsymbol{\Psi}_{\nu}(K)_m
& \eqdef
    \sum_{j=1}^{N+2}
        q_{m,j}
        \operatorname{Call}^{\operatorname{BS}}_{\nu}
        \big(
            \widehat K_j,K
        \big)
\qquad\mbox{ for } 
    m\in [M]_+
\\
\label{eq:smooth_maturity_coefficients}
  a_{m,\nu}(T)
&\eqdef
    \frac{
        \operatorname{BS}_{\nu}(T-T_m)
        -
        \operatorname{BS}_{\nu}(T-T_{m+1})
    }{
        T_{m+1}-T_m
    }
\qquad\mbox{ for } 
    m\in[M-1]_+
\\
    \Lambda_{\nu}(T)
& \eqdef
    e_1
    +
    \sum_{m=1}^{M-1}
        a_{m,\nu}(T)
        \big(
            e_{m+1}-e_m
        \big)
\end{align}
where, for every
$
    m\in[M]_+
$
and
$
    j\in[N+2]_+,
$
set
$
    q_{m,j}
\eqdef
    S_{m,j}-S_{m,j+1}
$.
For every $\nu,T\geq0$ set 
$
    b_{\nu}(T)
\eqdef
    \frac{
        \operatorname{BS}_{\nu}
        \left(
            \frac{T}{T_1}
        \right)
        -
        \operatorname{BS}_{\nu}
        \left(
            \frac{T}{T_1}-1
        \right)
        -
        \operatorname{BS}_{\nu}(0)
    }{
        1-\operatorname{BS}_{\nu}(0)
    }
$\footnote{Since $\operatorname{BS}_{\nu}(-1)=0,$ we have $b_{\nu}(0)=0$ and  at $\nu=0$, $
    b_0(T)
    =
    \min
    \{
        \frac{T}{T_1},
        1
    \}$.},%
$
    B_{\nu}(T,K)
\eqdef
    \operatorname{Call}^{\operatorname{BS}}_{\nu b_{\nu}(T)}(1,K)
$, $
    \Gamma_{\mathbf T,\mathbf K,\mathbf y,\nu}(T,K)
\eqdef
    \Lambda_{\nu}(T)^\top
    \boldsymbol{\Psi}_{\nu}(K)
$ and define 
$
    \Psi_{\mathbf T,\mathbf K,\mathbf y,\nu}
    :
    [0,\infty)^2
    \to
    [0,\infty)
$
by
\begin{equation}
\label{eq:smooth_AF_surface}
\begin{aligned}
    \Psi_{\mathbf T,\mathbf K,\mathbf y,\nu}(T,K)
\eqdef
    \big(
        1-b_\nu(T)
    \big)
    B_\nu(T,K)
    +
    b_\nu(T)
    \Gamma_{\mathbf T,\mathbf K,\mathbf y,\nu}(T,K).
\end{aligned}
\end{equation}
\end{definition}
\begin{proposition}[Smooth arbitrage-free approximation]
\label{prop:Smooth_AF_Decoder}
In the setting of Definition~\ref{def:Smooth_AF_Decoder}, let
$
    \Psi_{\mathbf T,\mathbf K,\mathbf y,0}
$
denote the structured $\operatorname{ReLU}$ surface obtained at $\nu=0$, and
set
$
    \Delta_T^{\min}
\eqdef
    \min_{m\in[M-1]_+}
    \big(
        T_{m+1}-T_m
    \big)
>0
$.
Then, for every $\nu\in(0,\overline\nu)$, the following assertions hold:
\begin{enumerate}
    \item[(i)] \textbf{Regularity:}
    $
        \Psi_{\mathbf T,\mathbf K,\mathbf y,\nu}
        \in
            C^{\infty}\big((0,\infty)^2\big)
        \cap 
            C\big([0,\infty)^2,[0,\infty)\big)
        .
    $

    \item[(ii)] \textbf{No static arbitrage:}
    $
        \Psi_{\mathbf T,\mathbf K,\mathbf y,\nu}
        \in
        \mathcal C_{AF}.
    $

    \item[(iii)] \textbf{Uniform perturbation of PWL Surface:}
    For any $\varepsilon>0$, if $
    0<\nu \leq
    \min
    \left\{
        \frac{\sqrt{2\pi}}{2},
        \frac{
            \sqrt{2\pi}\,\Delta_T^{\min}
        }{
            10\Delta_T^{\min}+4(M-1)
        }
        \,\varepsilon
    \right\}$
    \begin{equation}
    \label{eq:smooth_decoder_global_approximation}
    \begin{aligned}
        \sup_{T,K\ge 0}
        \,
        \big|
            \Psi_{\mathbf T,\mathbf K,\mathbf y,\nu}(T,K)
            -
            \Psi_{\mathbf T,\mathbf K,\mathbf y,0}(T,K)
        \big|
    \le
        \varepsilon
        .
    \end{aligned}
    \end{equation}
\end{enumerate}
{ 
Moreover, fix $\overline\nu>0$. For $R,S>0$ and $N,M\in\mathbb N_+$, set $\widehat N\eqdef\max\{N,2\}$ and $\widehat M\eqdef\max\{M,2\}$. The family
\[
\phi^{(N,M,R,S)}_\theta\eqdef\phi^{(\widehat M,\widehat N)}_{e^{-\xi},\overline\nu}(\mathbf t,\mathbf k,\mathbf z,\zeta),\qquad \theta=(\xi,\mathbf t,\mathbf k,\mathbf z,\zeta)\in\mathbb R^{2+\widehat M+\widehat N+\widehat M\widehat N},
\]
is an arbitrage-free surface-decoder in the sense of Definition~\ref{defn:Good_Decoder}.
}
\end{proposition}
\begin{proof}
See Section~\ref{s:Proofs__ss:AFDecodersSANOS}.
\end{proof}

\begin{remark}
The main difference between the formula in~\eqref{def:Smooth_AF_Decoder} and the formulation of the $\operatorname{SANOS}$ model in~\cite{buehler2026sanos} is that here we choose the parameters for approximate interpolation, cf.~\eqref{eq:smooth_decoder_global_approximation}, which allow us enough flexibility access to the closed-form expression in~\eqref{eq:smooth_BS_strike_slices}-\eqref{eq:smooth_maturity_coefficients}; while in~\cite{buehler2026sanos} exact interpolation is achieved which is solved implicitly via a linear program.
\end{remark}

We are now ready to showcase our positive results first.

\section{Positive Results: AF-NO}
\label{s:Theory_Positive}

We present our first positive result for our AFNO model.  Namely, we show that it necessarily always produces arbitrage-free European call-price surfaces.

{ 
\begin{theorem}[AF-NO is Certifiably Arbitrage-Free]
\label{thrm:Cert_AF}
Every AF-NO maps into $\mathcal{C}_{AF}$; i.e.\ for every input $x$ in its domain, $(T,K)\mapsto\Phi(x)(T,K)$ satisfies the conditions in Proposition~\ref{prop:contarbfree}.
\end{theorem}
}
\begin{proof}
See Appendix~\ref{s:Proof_certificate}.
\end{proof}

The approximation proof chooses finitely many anchor surfaces for the target operator and tolerance, then combines their decoder approximations using continuous softmax weights. Unlike hard nearest-neighbour selection, this yields a continuous map while preserving no-arbitrage. The theorem does not assert a smaller dictionary or a better worst-case approximation rate than nearest-neighbour approximation.

\begin{setting}
\label{set:thesetting}
Fix an RKHS $\mathcal{H}$ over a non-empty set
$
    \mathcal{D}
\subseteq
    [0,\infty)^2
$
and suppose that $\mathcal{H}$ admits a complete interpolating sequence.
Fix $0<q<p$ and let $l,\lambda>0$ satisfy
$
    \frac{q}{p}\lambda<l<\lambda
$.
Let $\mathcal{K}\subseteq\mathcal{H}$ be non-empty and compact (set if inputs on which we will approximate), and let
$
    \mathcal{Y}
\subseteq
    \mathcal{C}_{AF}^{p,\lambda}
$
be a uniformly controlled call-surface class admitted by the arbitrage-free decoder of Definition~\ref{defn:Good_Decoder}, and assume that this class and decoder satisfy Assumption~\ref{ass:weighted_decoder_selection}. 
Let $\mathcal{Q} \subset \mathcal{C}_{AF}^{p,\lambda}$ be compact and non-empty.
Let
$
    \mathcal{F}:\mathcal{H}\to\mathcal{Y}
$
be $L$-Lipschitz as a map from
$
    (\mathcal{H},\|\,\|_{\mathcal{H}})
$
into
$
    (\mathcal{C}_{AF}^{q,l},\|\,\|_{q,l})
$.
\end{setting}

\begin{theorem}[Universal Arbitrage-Free Approximation]
\label{thrm:UAT_AFNO}
Under Setting~\ref{set:thesetting}, for every $0<\varepsilon<1$, there exist
$I,K\in\mathbb{N}_+$, points
$
    x_1,\dots,x_K\in\mathcal{D}
$,
a matrix
$
    A\in\mathbb{R}^{K\times K}
$,
numbers $R,S>0$, integers $N_0,M_0\in\mathbb{N}_+$,
parameters
$
    \theta_1,\dots,\theta_I
\in
    \mathbb{R}^{D_{N_0,M_0,R,S}}
$,
a temperature $\tau>0$, and a $\operatorname{ReLU}$-MLP
$
    \alpha_{\mathbf{X}}:\mathbb{R}^K\to\mathbb{R}^I
$
such that the map $\Phi:\mathcal{H}\to\mathcal{C}_{AF}$ defined by
\begin{equation}
\label{eq:approximation_gaurantee_construction0}
    \Phi(C)
\eqdef
    \sum_{i=1}^I
        \operatorname{Softmax}
        \left(
            \frac{1}{\tau}
            \alpha_{\mathbf{X}}
            \left(
                \mathcal{E}_K
                \big(
                    C,
                    (x_k)_{k=1}^K,
                    A
                \big)
            \right)
        \right)_i
        \widetilde{C}_i
\quad
    \mbox{ where }
\quad
    \widetilde{C}_i
\eqdef
    \phi^{(N_0,M_0,R,S)}_{\theta_i}
\end{equation}
satisfies:
\begin{enumerate}
    \item[(i)] \textbf{Uniform Approximation:} $
        \|\Phi(C)-\mathcal{F}(C)\|_{q,l}
        \le
        \varepsilon$
        for all $C\in\mathcal{K}$,
    \item[(ii)] \textbf{Interpretable Range:} $
        \Phi(\mathcal{H})
    \subseteq
        \mathcal{C}_{AF}
        \cap
        \operatorname{co}
        \{
            \widetilde{C}_1,\dots,\widetilde{C}_I
        \}$.
\end{enumerate}
\end{theorem}
\begin{proof}[\textit{Proof Sketch}]
For full details, see Appendix~\ref{s:Proofs__ss:ToInfinitedimensions}; at a high-level, the idea of the proof is the following:
\begin{enumerate}
    \item[Step 1]: For $\delta>0$, compactness of $\mathcal K$ gives a finite $\delta$-net $\{x_n\}_{n=1}^N\subseteq\mathcal K$. Setting $y_n\eqdef\mathcal F(x_n)$, Lipschitz continuity gives
    \[
    \sup_{x\in\mathcal K}\min_{1\leq n\leq N}\|\mathcal F(x)-y_n\|_{q,l}\leq L\sup_{x\in\mathcal K}\min_{1\leq n\leq N}\|x-x_n\|_{\mathcal H}\leq L\delta.
    \]
    Proposition~\ref{prop:interpol_2_approx} then provides decoder approximations of these finitely many output surfaces with common decoder hyperparameters.
    \item[Step 2]: The next step is to interpolate the a \textit{continuous} approximate relaxation of the \textit{nearest neighbourhoods problem}\footnote{Which itself is typically discontinuous, and thus not desirable as it would yield a network not trainable by standard gradient-based optimization algorithms in practice.}; i.e. the map sending any $x\in \mathcal{K}$ to a choice of closest $x_n$ and then, assigning $x$ to $\mathcal{F}(x_n)$.
\end{enumerate}
\end{proof}

\subsection{Illustration: Simultaneous Call Surface Prediction}
\label{s:Positive__ss:Implicaitons_math_fin}

We now apply our theory to show how one can learn the \textit{parameter-to-European Call-price surface} map for a broad class of parametric families of SDEs; possibly with non-Brownian drivers.  
We recall that any finite-dimensional RKHS is isometrically isomorphic to $\mathbb{R}^d$, and we treat it as such henceforth.

The basic idea here is that of a pre-trained model for call prices: we use sufficiently parameterized
general diffusion processes to generate a large number of $\tilde C$ basis call price surfaces. We then show, first,
that a surface generated by any new set of diffusion parameters is approximated to precision.

\begin{ass}[Latent Process]
\label{ass:latent_gen}
{ 
Let $d_{\nu},F_{\mathrm{mkt}}\in\mathbb{N}_+$, $r\geq0$, and let $(\Omega,\mathcal{F},\mathbb{F},\mathbb{Q})$ be a filtered probability space. Let $M=(M_t)_{t\geq0}$ be a continuous square-integrable $(\mathbb{Q},\mathbb{F})$-martingale generating $\mathbb{F}$, and assume that $d\langle M\rangle_t=\chi_t\,dt$ for a predictable process $\chi$ satisfying $0\leq\chi_t\leq\overline c<\infty$. Let $\nu$ be a fixed $\mathbb{R}^{d_{\nu}}$-valued predictable process, let $\Theta\subseteq\mathbb{R}^{F_{\mathrm{mkt}}}$ be non-empty and compact, and let $\sigma:[0,\infty)\times\mathbb{R}^{d_{\nu}}\times\Theta\times\mathbb{R}\to\mathbb{R}$ be Borel measurable. Assume that there exist finite constants $\overline\sigma,L_x,L_\theta\geq0$ such that, for all $t\geq0$, $v\in\mathbb{R}^{d_{\nu}}$, $\theta,\theta'\in\Theta$, and $x,y\in\mathbb{R}$,
}
\begin{equation}
\label{eq:latent_coefficient_bounds}
\begin{aligned}
|\sigma(t,v,\theta,x)|&\leq\overline\sigma ,
\\
|x\sigma(t,v,\theta,x)-y\sigma(t,v,\theta,y)|
& \leq L_x|x-y|,
\\
|\sigma(t,v,\theta,x)-\sigma(t,v,\theta',x)| & \leq L_\theta\|\theta-\theta'\|_2.
\end{aligned}
\end{equation}
and 
for each $\theta\in \Theta$, let $X_{\,}^{\theta}$ be the unique adapted solution of
\begin{equation}
\label{eq:sigma_martingale}
    X_t^\theta
=
    1+\int_0^t rX_s^\theta\,ds+\int_0^t X_s^\theta\sigma(s,\nu_s,\theta,X_s^\theta)\,dM_s,\qquad t\geq0.
\end{equation}
Assume, moreover, that, for every $\theta\in\Theta$,
$
        \inf_{0\le t\le T} X_t^{\theta}>0
$
$\mathbb{Q}$-a.s. 
\end{ass}
Our objective is to approximate the solution operator corresponding to the family of latent price generating process $X_{\,}$ above; namely, the map
{ 
\begin{equation}
\label{eq:sol_op}
\operatorname{Sol}:\Theta\longrightarrow C([0,\infty)^2,[0,\infty)),\qquad \operatorname{Sol}(\theta)(T,K)\eqdef\mathbb{E}_{\mathbb{Q}}\big[(e^{-rT}X_T^\theta-K)_+\big].
\end{equation}
}
Let us first note that our problem is well-posed before approximately solving it.
{ 
\begin{proposition}[Well-posedness]
\label{prop:Solution_MapRegularity}
Under Assumption~\ref{ass:latent_gen}, put $a\eqdef\overline c\,\overline\sigma^2$, $b\eqdef\max\{2\overline cL_x^2,a\}$, $\kappa_u\eqdef\frac a2u(u-1)^+$ for $u>0$, and $\gamma_q\eqdef\frac12(b+\kappa_{2q})$ for $q>0$. The solution operator in~\eqref{eq:sol_op} is well-defined and takes values in $\mathcal C_{AF}$. For every $p,\lambda>0$ satisfying $\lambda\geq\kappa_{p+1}$, its image $\mathcal Y\eqdef\operatorname{Sol}(\Theta)$ is uniformly controlled with $B=2^p$ and common modulus $\omega(h)=h+\max\{1,\sqrt{e^a-1}\}\sqrt h$. For every $q>0$ and $l>\gamma_q$, for each $\theta,\theta'\in\Theta$ we have
\[
\|\operatorname{Sol}(\theta)-\operatorname{Sol}(\theta')\|_{q,l}\leq\frac{2^{q+1}\sqrt{\overline c}\,L_\theta}{\sqrt{2e(l-\gamma_q)}}\|\theta-\theta'\|_2.
\]
\end{proposition}
}

Upon specializing to Proposition~\ref{ass:latent_gen}, we obtain the following \textit{universal approximation} result with no-arbitrage constraints.  Briefly, the result 
{says that under mild conditions, for a parametric family of price processes, the parameter-to-call surface map can be approximated to arbitrary precision without ever violating the static no-arbitrage conditions.}

\begin{corollary}[Universal Arbitrage-Free Pricing for Families of Pricing Models of the Form~\eqref{eq:sigma_martingale}]
\label{cor:AF_solmap}
Under Assumption~\ref{ass:latent_gen}, choose $0<q<p$, $\lambda>\max\{\kappa_{p+1},\gamma_q\}$, and $\max\{q\lambda/p,\gamma_q\}<l<\lambda$, with the constants from Proposition~\ref{prop:Solution_MapRegularity}. Let $\mathcal Y\eqdef\operatorname{Sol}(\Theta)$ and use an AF surface-decoder\footnote{Satisfying Definition~\ref{defn:Good_Decoder} and Assumption~\ref{ass:weighted_decoder_selection}; e.g.\ the PWL and SANOS families satisfy these requirements by Proposition~\ref{prop:PWL_weighted_selection} and Corollary~\ref{cor:SANOS_weighted_selection}.} for $\mathcal Y$
then, for every $0<\varepsilon<1$ and $H>0$, there exists an AF-NO $\Phi:\mathbb R^{F_{\mathrm{mkt}}}\to\mathcal C_{AF}$ satisfying
\[
\sup_{\theta\in\Theta}\max_{0\leq T,K\leq H}\left|\mathbb E_{\mathbb Q}\big[(e^{-rT}X_T^\theta-K)_+\big]-\Phi(\theta)(T,K)\right|\leq\varepsilon.
\]
\end{corollary}

We now showcase our negative results.

\section{Negative Results: \textit{Un}-certifiability for Projections and Standard NOs do not Have No-Arbitrage}
\label{s:Negresults}

In this section, we show that the natural alternative tools available either fail all-together to be arbitrage-free in general, or at-the very least, they cannot be \textbf{certified} to always yield \textbf{exactly} arbitrage-free call surfaces.

\subsection{No Arbitrage-Free Projections}
\label{s:Negresults__ss:NoProjections}
We first show that one simply cannot learn a surface which fits market data well and then ``project'' it onto the space of arbitrage-free surfaces; where by projection we mean minimizing \textit{any} metric which is similar to an $\|\cdot\|_{p,\lambda}$ metric on the natural space $\mathcal{X}_p^{\lambda}\subseteq C([0,\infty)^2,[0,\infty))$.  I.e.\ we show that there is some surface which is ``un-projectable'' onto $\mathcal{C}_{AF}\cap \mathcal{X}_p^{\lambda}$.
\begin{theorem}[No \textit{Nearest} Arbitrage-Free Projection]
\label{thrm:No_AF__Proj}
Fix $p,\lambda>0$, and let $\rho$ be any metric on $C([0,\infty)^2,[0,\infty))$ which induces a topology which is no stronger than the 
uniform convergence on compact sets 
topology on $C([0,\infty)^2,[0,\infty))$.  Then, there exists surface $C_{\operatorname{bad}}\in 
C([0,\infty)^2,[0,\infty))
$ such that there is no $\Pi_C\in \mathcal{C}_{AF}
$ satisfying
\[
        \rho(\Pi_C,C_{\operatorname{bad}})
    =
        \inf_{\tilde{C}\in \mathcal{C}_{AF}
        }\,
        \rho(\tilde{C},C_{\operatorname{bad}})
.
\]
\end{theorem}
\begin{proof}
See Section~\ref{s:NoProj}.
\end{proof}

\begin{figure}[H]
    \centering
    \begin{minipage}[t]{0.45\textwidth}
        \vspace{0pt}
        \caption{\textbf{No Nearest Arbitrage-Free Projection:} Theorem~\ref{thrm:No_AF__Proj} states that there exists a candidate European call-price surface $C_{\operatorname{bad}}:[0,\infty)^2\to[0,\infty)$ ({\color{red}{red}} dot) which admits no best approximation in $\mathcal{C}_{AF}$.  Indeed, $C_{\operatorname{bad}}$ lies in the topological closure of $\mathcal{C}_{AF}$, represented by the dashed green boundary of the green region, but does not itself belong to $\mathcal{C}_{AF}$.  Consequently, its distance from $\mathcal{C}_{AF}$ is zero, while this infimum is not attained by any arbitrage-free call surface.  This is possible because $\mathcal{C}_{AF}$ is not closed in the topology of uniform convergence on compact subsets, and therefore is not closed in any weaker metrizable topology considered in Theorem~\ref{thrm:No_AF__Proj}.}
        \label{fig:NoAFProjections}
    \end{minipage}
    \hfill
    \begin{minipage}[t]{0.45\textwidth}
        \vspace{0pt}
        \centering
        \includegraphics[width=.95\linewidth]{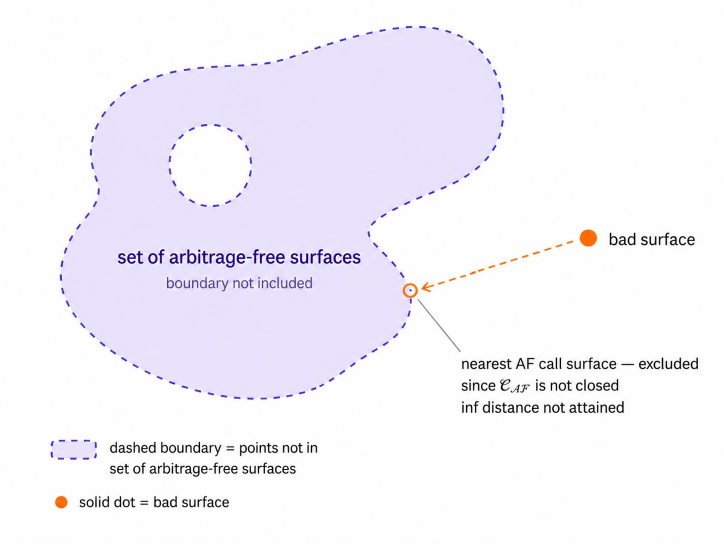}
    \end{minipage}
\end{figure}

\subsection{Neural Operators Admit Arbitrage}
\label{s:Negresults__ss:NONoCertificate}

Our last set of results show that standard NO builds cannot generally be certified to be free of arbitrage.  We consider encoder-decoder type builds where decoding is done by projecting onto basis functions; and then we consider the FNO build.

\paragraph{Encoder-Decoder Architectures}
We consider encoder-decoder NO builds; cf.\cite{galimberti2026designing,lanthaler2023operator}.

\begin{proposition}[Failure of strike-convexity certificate for decoder-type NOs]
\label{prop:AF_certificatefailure}
Fix $F,M\in \mathbb{N}_+$, fix pointwise-defined representatives $(e_n)_{n=1}^{\infty}$ of an orthonormal basis of $L^2([0,1])$, and define
$$
{ 
E_{n,m}(T,K)\eqdef e_n(K)e_m(T),
}
$$
{ 
for any $(T,K)\in[0,1]^2$.
}
Fix any activation function $\sigma\in C(\mathbb{R})$, fix pairwise distinct indices $n_1,n_2,n_3\in \mathbb{N}$ and some $m_\star\in \mathbb{N}$, and consider the decoder
\begin{equation}
\label{eq:decoder_NO}
f_{D\text{-}NO}(x)
\eqdef
\sum_{k=1}^3
[f_{NN}(x)]_k\,E_{n_k,m_\star}.
\end{equation}
Then there exists a $\sigma$-MLP $f_{NN}:\mathbb{R}^F\to \mathbb{R}^3$ such that: for every $x\in \mathbb{R}^F$
$$
{ 
f_{D\text{-}NO}(x)\notin\{C|_{[0,1]^2}:C\in\mathcal C_{AF}\}
}
.
$$
\end{proposition}

\paragraph{Fourier Neural Operators (FNOs)}

We also consider the mainstream Fourier Neural Operator (FNO) architecture~\cite{li2021fourier,li2023fourier}, e.g.\ considered in~\cite{wiedemann2025operator} for call surface smoothing, are \textit{not certifiably arbitrage-free}: their parameter spaces contain configurations whose outputs violate strike convexity.  Thus, without an additional architectural restriction or certified decoder, the FNO architecture itself cannot guarantee no-arbitrage.
Before stating our \textit{un-certifiability} result, we recall the definition of the FNO architecture.
\begin{definition}[Fourier Neural Operator]
\label{def:FNO}
Let $D\eqdef [0,1]^2\subseteq \mathbb{R}^2$ be a rectangular domain, let $d_a,d_u,m\in \mathbb{N}_+$, and let $L\in \mathbb{N}_+$.  
Fix a pointwise \textit{lifting} map
$
Q:\mathbb{R}^{d_a}\to \mathbb{R}^m
$,
a pointwise \textit{projection} map
$
P:\mathbb{R}^m\to \mathbb{R}^{d_u}
$,
linear maps
$
W_{\ell}:\mathbb{R}^m\to \mathbb{R}^m
$
for $\ell\in\{0,\dots,L-1\}$,
Fourier multipliers $R_{\ell}$ acting on finitely many Fourier modes,
and an activation function
$
\sigma:\mathbb{R}\to \mathbb{R}
$
acting componentwise on $\mathbb{R}^m$.
Given an input field
$
a:D\to \mathbb{R}^{d_a}
$,
define the latent fields $(v_{\ell})_{\ell=0}^L$ recursively by
\begin{equation}
\label{eq:linear_lift}
v_0(x)\eqdef Q(a(x))
\end{equation}
and, for $\ell\in\{0,\dots,L-1\}$, for each $x\in D$ by
\begin{equation}
\label{eq:FNO_recusions}
v_{\ell+1}(x)
\eqdef
\sigma\Big(
    W_{\ell}v_{\ell}(x)
    +
    \mathcal{F}^{-1}\big(
        R_{\ell}\cdot \mathcal{F}(v_{\ell})
    \big)(x)
    +
    b_{\ell}(x)
\Big),
\end{equation}
where $\mathcal{F}$ denotes the Fourier transform on $D$ and $b_{\ell}:D\to \mathbb{R}^m$ is a bias field.  
The associated Fourier Neural Operator is the map
$
\Phi:[L^2(D)]^{d_a}\to [L^2(D)]^{d_u}$ sending any input field $a\in [L^2(D)]^{d_a}$ to 
$\Phi(a)(x)
\eqdef
P(v_L(x))
$; where $x\in D$.
\end{definition}

\begin{figure}[t]
\centering
\includegraphics[width=.9\linewidth]{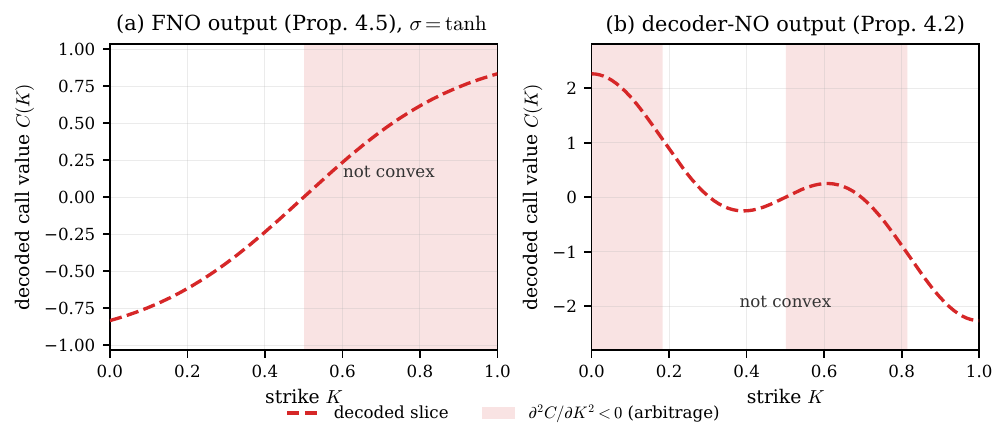}
\caption{\label{fig:counterexamples}Standard neural operators are not certifiable.  Left: a scalar Fourier neural operator with an activation whose second derivative changes sign, as in Proposition~\ref{prop:sigmoid_breaks_convexity_in_K}, produces a non-convex strike slice.  Right: a decoder-type neural operator on a tensorized basis, as in Proposition~\ref{prop:AF_certificatefailure}, does the same.  The shaded regions mark violations of strike convexity.}
\end{figure}

The limitations of the FNO for predictions of European call-price surfaces which are strictly free of arbitrage is illustrated by Figure~\ref{fig:counterexamples} plots the explicit parameter configurations used in Propositions~\ref{prop:AF_certificatefailure} and~\ref{prop:sigmoid_breaks_convexity_in_K}.  These are not training failures.  They are fixed admissible parameter choices of the corresponding architectures whose decoded strike slices are non-convex, and therefore contain butterfly arbitrage for every input.  

\begin{proposition}[FNO architecture is not certifiably arbitrage-free]
\label{prop:sigmoid_breaks_convexity_in_K}
Let $\sigma\in C^2(\mathbb{R})$ and suppose that there exist $\xi\in\mathbb{R}$ and $\eta>0$ such that
$\sigma''$ takes both positive and negative values on $(\xi-\eta,\xi+\eta)$.
Consider the scalar FNO architecture of Definition~\ref{def:FNO} on $D=[0,1]^2$ with
{ 
$d_a=d_u=m=1$ and pointwise lifting and projection $Q=P=\operatorname{Id}$.
}
Let $Q^{\uparrow}:\mathbb{R}^F\to L^2([0,1]^2)$ be any injective linear map.

If $L\ge 1$, then there is a choice of parameters
$
    (W_{\ell},R_{\ell},b_{\ell})_{\ell=0}^{L-1}
$
such that, for every $a\in\mathbb{R}^F$, the continuous representative of
$
    \Phi(Q^{\uparrow}(a))
$
is \textbf{not} the restriction to $[0,1]^2$ of an element of $\mathcal{C}_{AF}$.
\end{proposition}

\section{Numerical Illustration}
\label{s:Numerics}

We illustrate the AFNO construction on the parameter-to-surface maps generated by two stochastic-volatility families, Heston and rough Bergomi with $H<1/2$, both viewed as instances of the solution operator~\eqref{eq:sol_op}.  For each family, parameters are sampled from a compact set, a reference European-call surface is computed, and the learned operator maps model parameters to the corresponding call-price surface.  We compare two instances of Definition~\ref{defn:AFDECODER}: the single-decoder SANOS-DLV instance, for which $K=1$ and the network predicts the decoder parameters, and a fixed-anchor AFNO mixture, for which the network learns only the softmax weights over a finite dictionary of arbitrage-free decoder outputs.  The baselines are standard decoder-type neural operators with, respectively, a direct implied-volatility head, a direct price head, and a direct price head trained with a soft no-arbitrage penalty.  Appendix~\ref{s:NumericsSupplement} gives the complete data-generation, pricing, architecture, optimization, diagnostic, and reproducibility protocol, together with the Heston results.

Accuracy is reported as implied-volatility error in basis points on the liquid $10\Delta$-$90\Delta$ band.  Static-arbitrage diagnostics are computed on call prices.  On a non-uniform strike grid $K_1<\cdots<K_L$, we use the grid-aware butterfly value
\[
    b_\ell(T)
    \eqdef
    \frac{K_{\ell+1}-K_\ell}{K_{\ell+1}-K_{\ell-1}}\,C(T,K_{\ell-1})
    -
    C(T,K_\ell)
    +
    \frac{K_\ell-K_{\ell-1}}{K_{\ell+1}-K_{\ell-1}}\,C(T,K_{\ell+1}),
\]
for $2\leq \ell\leq L-1$; non-negativity of all $b_\ell(T)$ is precisely the discrete convexity condition on this grid.  This avoids the equal-spacing second difference, which is not a valid arbitrage diagnostic on delta-based strike grids.

\begin{table}[t]
\centering
\scriptsize
\setlength{\tabcolsep}{3.5pt}
\caption{\label{tab:numerics-headline}Accuracy and static-arbitrage compliance on the rough-Bergomi test set.  The first two rows are AFNO instances and are arbitrage-free by construction.  Some direct baselines improve pointwise implied-volatility accuracy, but their price surfaces are not certified and become arbitrageable under the off-grid and LP diagnostics of Figures~\ref{fig:irregular-out}-\ref{fig:exploit-profit}.  LP profit is the mean over the 256-surface dense-grid diagnostic; definitions and fuller summaries are in Appendix~\ref{s:NumericsSupplement}.}
\begin{tabular}{@{}lccccc@{}}
\toprule
Model & IV MAE & IV p95 & Arb.\ surf. & Max.\ arb. & LP profit \\
& (bps) & (bps) & (\%) & & \\
\midrule
SANOS-DLV (ours) & $14.2$ & $59.6$ & $0.00$ & $0$ & $4.80{\times}10^{-3}$ \\
AFNO mixture & $35.9$ & $162$ & $0.00$ & $0$ & $4.89{\times}10^{-3}$ \\
Direct-IV & $5.58$ & $17.7$ & $0.0^{\dagger}$ & $8.91{\times}10^{-9}$ & $1.26{\times}10^{-2}$ \\
Direct-Price & $161$ & $2.05{\times}10^{3}$ & $100$ & $4.51$ & $2.64{\times}10^{-1}$ \\
Direct-Price+arb & $259$ & $2.51{\times}10^{3}$ & $99.8$ & $3.43$ & $2.86{\times}10^{-1}$ \\
\midrule
Per-surface SANOS fit & $40.1$ & $143$ & $0.00$ & $0$ & - \\
rBergomi reference & - & - & - & - & $5.23{\times}10^{-3}$ \\
\bottomrule
\end{tabular}
\par\smallskip
\raggedright
$^\dagger$Canonical-grid diagnostic only; the dense-grid violations of the Direct-IV head are shown in Figure~\ref{fig:irregular-out}.  LP profit denotes the per-surface extractable profit of Figure~\ref{fig:exploit-profit}, in units of forward notional.
\end{table}

\paragraph{Learning from irregular observation grids.}

\begin{figure}[H]
\centering
\includegraphics[width=\linewidth]{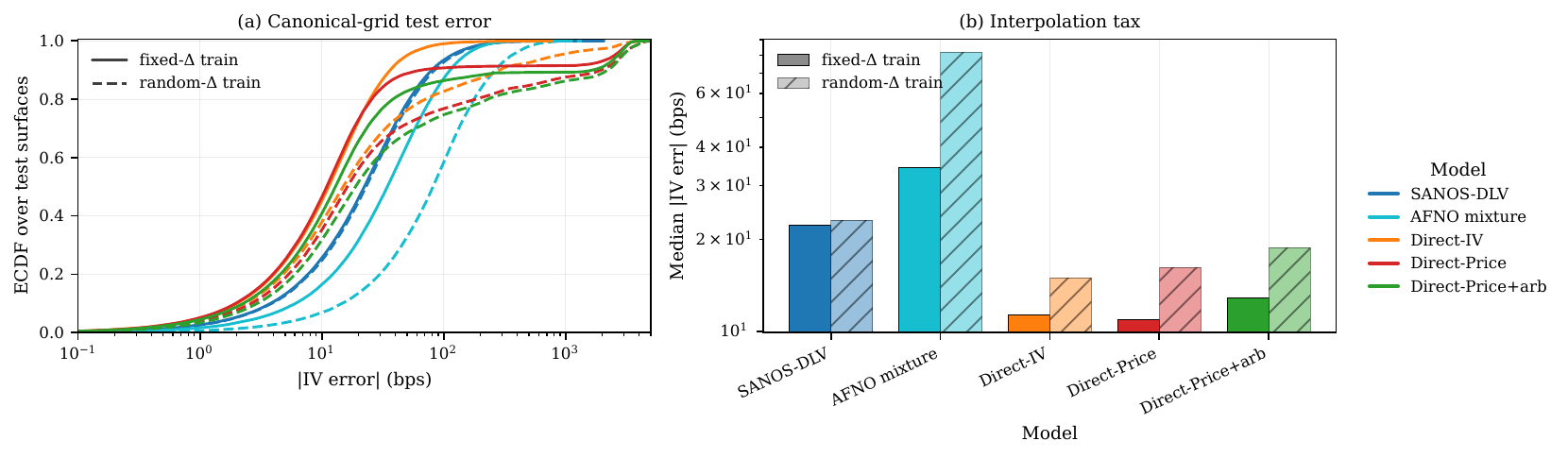}
\caption{\label{fig:irregular-in}Grid-agnostic training.  Each model is trained once on a fixed canonical strike grid and once on per-sample random strike grids, then evaluated on the same canonical test set.  Panel (a) reports the empirical distribution of absolute implied-volatility errors; panel (b) reports the median error.  The single-decoder AFNO, SANOS-DLV, ingests random grids without first imposing a lattice, while fixed-output models incur the interpolation cost.}
\end{figure}

Figure~\ref{fig:irregular-in} tests the universality claim of Theorem~\ref{thrm:UAT_AFNO} and Corollary~\ref{cor:AF_solmap} in the data regime most relevant for market surfaces: each training sample is observed on its own random strike grid.  The SANOS-DLV instance accepts these grids directly and its canonical-grid median error is essentially unchanged, increasing by $1.6\%$.  Models with a fixed output lattice, including the fixed-anchor AFNO mixture and the direct heads, must first interpolate the targets onto a canonical grid and incur a $1.7$--$6.3$ times increase in median error.

\begin{wrapfigure}{r}{0.65\textwidth}
  \begin{center}
    \includegraphics[width=1.0\linewidth]{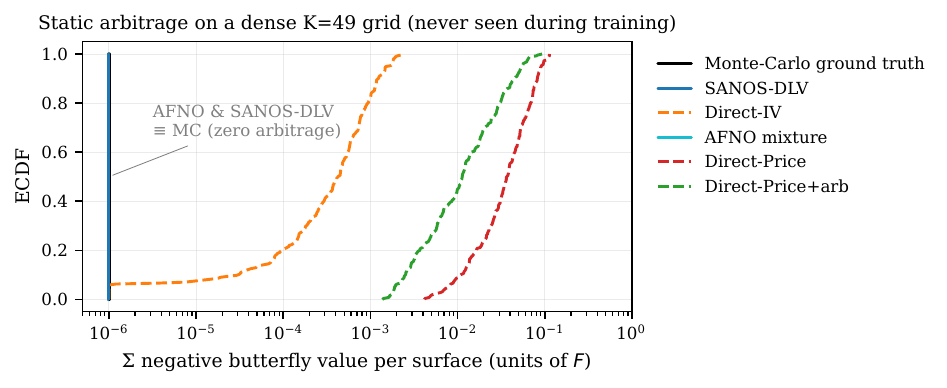}
  \end{center}
\end{wrapfigure}

\begin{figure}[H]
\centering
\includegraphics[width=\linewidth]{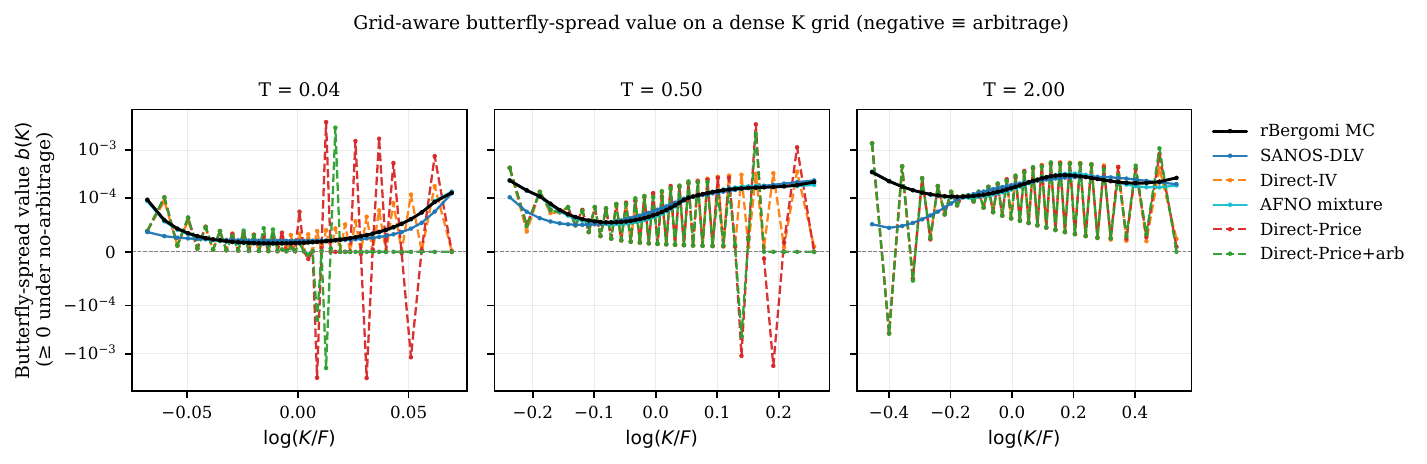}
\caption{\label{fig:irregular-out}Emitting irregular data while preserving no-arbitrage.  All operators are queried on a dense $K=49$ grid unseen during training and scored by the grid-aware butterfly value $b_\ell(T)$.  Top: at three maturities, SANOS-DLV and the AFNO mixture track the reference pricer through non-negative butterfly curves, whereas the direct price heads produce negative values.  Bottom: empirical distribution of the total negative butterfly value per surface; the AFNO instances and the reference pricer concentrate at zero.}
\end{figure}

\paragraph{Off-grid emission and the arbitrage certificate.}
Figure~\ref{fig:irregular-out} queries the trained operators on a dense strike grid not used during training.  The two AFNO instances retain non-negative butterfly values, as required by Theorem~\ref{thrm:Cert_AF}, because the decoder produces an element of $\mathcal{C}_{AF}$ at every query point.  By contrast, the direct price heads develop negative butterfly values between training strikes.  The Direct-IV head has no canonical-grid violations in Table~\ref{tab:numerics-headline}, but the dense-grid diagnostic exposes small and persistent off-grid convexity violations.

\begin{figure}[H]
    \centering
    \begin{minipage}[t]{0.38\textwidth}
        \vspace{0pt}
        \caption{\label{fig:exploit-profit}Economic magnitude of static-arbitrage violations.  Each rough-Bergomi test surface is passed to a bounded scenario LP; positive values indicate negative-cost portfolios with nonnegative payoff on every tested path.  The AFNO instances are comparable with the finite-scenario Monte Carlo reference level, whereas the direct heads are shifted to substantially larger profits.  Figure~\ref{fig:exploit-profit-grid} gives the corresponding Heston and rough-Bergomi grid of diagnostics.}
        \label{fig:badtrades}
    \end{minipage}
    \hfill
    \begin{minipage}[t]{0.6\textwidth}
        \vspace{0pt}
        \centering
        \includegraphics[width=1\linewidth]{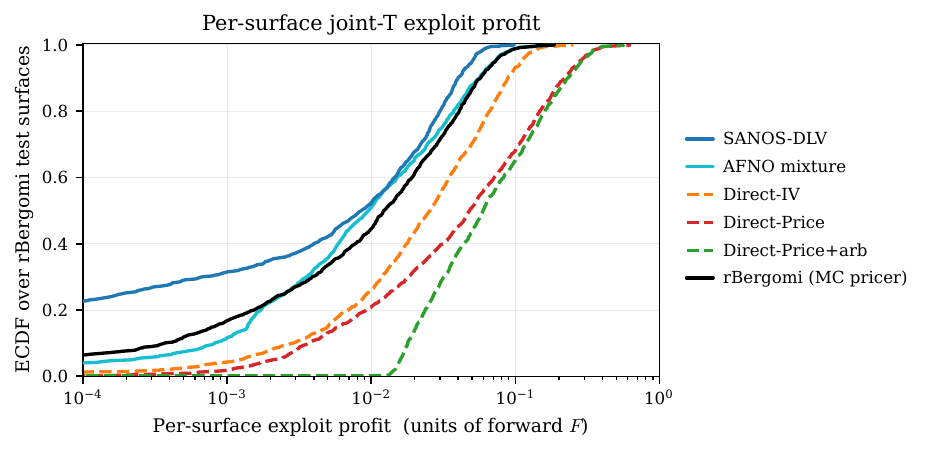}
    \end{minipage}
\end{figure}

\paragraph{Economic size of residual violations.}
Figure~\ref{fig:exploit-profit} measures the economic value of the remaining violations.  For each quoted surface, we solve a bounded scenario-based static-arbitrage linear program; the objective value is the riskless profit extractable from that surface on the tested scenarios, in units of forward notional.  The AFNO outputs remain close to the reference-pricer range, while the direct price heads have mean profits of $26\%$--$29\%$ and maxima near $70\%$ of forward notional.  The particular soft lattice penalty used here does not remove this defect.  This is consistent with, but is not by itself a proof of, the structural distinction in Theorem~\ref{thrm:No_AF__Proj}; the exact LP and penalty are specified in Appendix~\ref{s:NumericsSupplement}.

\paragraph{Comparison scope.}
The controlled baselines share the same input normalization and MLP backbone, isolating the effect of the output parameterization and the soft penalty.  We do not include SVI or related per-surface smile calibrators: those methods solve a different, non-amortized calibration/interpolation problem, and their no-arbitrage guarantees depend on additional cross-slice constraints.  The per-surface SANOS fit in Table~\ref{tab:numerics-headline} provides the closest like-for-like calibration baseline.  We therefore make no claim here that AFNO dominates all classical surface calibrators; the experiment tests whether an amortized parameter-to-surface operator retains accuracy while carrying a global architectural certificate.

\section{Conclusion}
\label{s:conclusion}

We introduced AFNO, an encoder-processor-decoder framework for learning maps into the space of arbitrage-free European call-price surfaces.  No-arbitrage is imposed structurally by the decoder rather than encouraged through training penalties or repaired after prediction.  Consequently, every parameter configuration produces a globally arbitrage-free call surface (Theorem~\ref{thrm:Cert_AF}).  This guarantee is available with both an explicit piecewise-linear decoder (Proposition~\ref{prop:ReLU_Is_AF_Decoder}) and a smooth Black--Scholes-based decoder (Proposition~\ref{prop:Smooth_AF_Decoder}).

From a financial perspective, the resulting model may be queried at maturities and strikes that were not present during training without losing the static no-arbitrage conditions required for consistent pricing, interpolation, and downstream risk calculations.  At the same time, architectural certification does not sacrifice expressivity: AFNOs approximate Lipschitz maps into uniformly controlled classes of arbitrage-free call surfaces (Theorem~\ref{thrm:UAT_AFNO}).  In particular, they approximate parameter-to-call-surface maps generated by broad parametric families of risk-neutral martingale models (Corollary~\ref{cor:AF_solmap}).

The negative results explain why this structural approach is necessary.  A closest arbitrage-free correction need not exist for a surface produced by an unconstrained model (Theorem~\ref{thrm:No_AF__Proj}), while standard decoder-type neural operators (Proposition~\ref{prop:AF_certificatefailure}) and Fourier neural operators (Proposition~\ref{prop:sigmoid_breaks_convexity_in_K}) admit parameter configurations producing butterfly arbitrage.  The numerical results in Section~\ref{s:Numerics} reflect this distinction: AFNO retains its certificate on unseen grids, whereas unconstrained price and volatility heads may exhibit economically exploitable off-grid violations.

Taken together, the results support a simple modelling principle for financial operator learning: when every predicted surface must remain economically admissible, no-arbitrage should be encoded in the range of the architecture itself.  AFNO provides such a certificate while retaining the flexibility needed for accurate and computationally efficient surface prediction.

\section*{Acknowledgements and funding}
\label{s:Acknow}
A.\ Kratsios acknowledges financial support from the Natural Sciences and Engineering Research Council of Canada (NSERC) through Discovery Grant Nos.\ RGPIN-2023-04482 and DGECR-2023-00230.
We further acknowledge that resources used in the preparation of this research were provided, in part, by the Province of Ontario, the Government of Canada through CIFAR, and the industry sponsors of the Vector Institute\footnote{\href{https://vectorinstitute.ai/partnerships/current-partners/}{https://vectorinstitute.ai/partnerships/current-partners/}}.

\appendix

\section{Notation}
\label{s:Notation}

We use the following notation throughout our manuscipt.
\begin{itemize}[leftmargin=*,itemsep=0.25em]

    \item
    $\mathbb{N}\eqdef\{0,1,2,\dots\}$,
    $\mathbb{N}_+\eqdef\{1,2,\dots\}$, and
    $\mathbb{N}_2\eqdef\{n\in\mathbb{N}:n\ge2\}$.

    \item
    For each $n\in\mathbb{N}_+$,
    $[n]\eqdef\{0,\dots,n\}$ and
    $[n]_+\eqdef\{1,\dots,n\}$.

    \item
    For each $M\in\mathbb{N}_2$,
    $
        \mathbb{R}_{\le}^{M}
        \eqdef
        \big\{
            (x_1,\dots,x_M)\in\mathbb{R}^M
            :
            x_1<\dots<x_M
        \big\}.
    $
    \item
    For each $n\in\mathbb{N}_+$,
    $\{e_i\}_{i=1}^n$ denotes the standard orthonormal basis of
    $\mathbb{R}^n$, $\mathbf{1}_n\eqdef(1,\dots,1)\in\mathbb{R}^n$,
    and $I_n\in\mathbb{R}^{n\times n}$ denotes the identity matrix.

    \item
    For $x\in\mathbb{R}$, $x^+\eqdef\max\{x,0\}$, and for
    $x,y\in\mathbb{R}$, $x\wedge y\eqdef\min\{x,y\}$.

    \item
    If $\mathcal{H}$ is a Hilbert space, then
    $\langle\cdot,\cdot\rangle_{\mathcal{H}}$ and
    $\|\cdot\|_{\mathcal{H}}$ denote its inner product and norm,
    respectively.

    \item
    Throughout, $\mathcal{D}\subseteq[0,\infty)^2$ is the input
    domain, $\mathcal{H}$ is an RKHS over $\mathcal{D}$, and
    $\kappa:\mathcal{D}\times\mathcal{D}\to\mathbb{R}$ is its
    reproducing kernel.

    \item
    The variables $T\ge0$ and $K\ge0$ denote maturity and strike,
    respectively, and $C(T,K)$ denotes a European call price.

    \item
    $\mathcal{C}_{\mathrm{AF}}$ denotes the space of continuous
    arbitrage-free European call-price surfaces.

    \item
    For $p,\lambda>0$, $\mathcal{C}_{\mathrm{AF}}^{p,\lambda}
    \eqdef
    \mathcal{C}_{\mathrm{AF}}\cap\mathcal{X}_{\lambda}^p$ is normed by
    \[
        \|C\|_{p,\lambda}
        \eqdef
        \sup_{T,K\ge0}
        e^{-\lambda T}(1+K)^p|C(T,K)|
    \]
    where $\mathcal{X}_{\lambda}^p
        \eqdef
        \big\{
            C\in C([0,\infty)^2,\mathbb{R})
            :
            \|C\|_{p,\lambda}<\infty
        \big\}$.

    \item
    For $N\in\mathbb{N}_+$,
    $
        \Delta_N
    \eqdef
        \big\{
            w\in[0,1]^N:
            \sum_{n=1}^N w_n=1
        \big\}
    $
    denotes the probability simplex, and
    $\operatorname{co}(A)$ denotes the convex hull of a set $A$.

    \item
    For a Lipschitz map $f$, $\operatorname{Lip}(f)$ denotes its
    optimal Lipschitz constant, and $\operatorname{diam}(A)$ denotes
    the diameter of a non-empty set $A$.

    \item
    $\operatorname{ReLU}(x)\eqdef\max\{x,0\}$,
    $\sigma\bullet x$ denotes component-wise application of
    $\sigma$, and $\operatorname{Softmax}$ denotes the usual
    softmax map.

    \item
    For an MLP, $\operatorname{depth}(\cdot)$,
    $\operatorname{width}(\cdot)$, and $\operatorname{size}(\cdot)$
    denote its number of layers, maximal hidden-layer width, and
    total number of affine parameters, respectively.

    \item
    We write $a\lesssim b$ if $a\le cb$ for a constant $c>0$
    independent of the quantities under consideration.

\end{itemize}

\section{Numerical Protocol and Supplementary Results}
\label{s:NumericsSupplement}

This appendix specifies the full-profile experiment implemented in the companion repository\footnote{\texttt{arbitrage-free-neural-operator}}.  All reported quantities below are computed from the saved prediction arrays rather than transcribed into the experiment code.  Prices are undiscounted forward prices: the forward and money-market account are both normalized to one, rates are zero, and strikes are reported as $K/F$.  Thus, a call price lies between $(1-K)^+$ and one.  Total variance, rather than instantaneous variance or volatility, is the variance argument of every Black--Scholes map.

\subsection{Data-generating distributions and grids}
\label{s:NumericsSupplement:data}

For each model family we draw parameters independently and uniformly over the boxes in Table~\ref{tab:numerical-parameter-boxes}.  The rough-Bergomi parameter vector is $(\xi_0,H,\eta,\rho)$, where $\xi_0$ is the flat forward-variance level, $H$ is the Hurst parameter, $\eta$ is vol-of-vol, and $\rho$ is spot--volatility correlation.  The Heston vector is $(v_0,\kappa,\vartheta,\sigma,\rho)$, with long-run variance denoted by $\vartheta$ to distinguish it from the full parameter vector.  The Heston box is not restricted by the Feller condition.  Each family has $60{,}000$ training, $4{,}000$ validation, and $8{,}000$ test surfaces; no sampled surface was removed by the implemented hygiene filter.

\begin{table}[t]
\centering
\small
\caption{\label{tab:numerical-parameter-boxes}Parameter distributions for the numerical study.  Every coordinate is sampled independently and uniformly on the displayed interval.}
\begin{tabular}{@{}lll@{}}
\toprule
Family & Parameter & Sampling interval \\
\midrule
rough Bergomi & $\xi_0$ & $[0.015,0.12]$ \\
 & $H$ & $[0.05,0.20]$ \\
 & $\eta$ & $[0.8,2.5]$ \\
 & $\rho$ & $[-0.95,-0.10]$ \\
\addlinespace
Heston & $v_0$ & $[0.005,0.16]$ \\
 & $\kappa$ & $[0.3,4.0]$ \\
 & $\vartheta$ & $[0.005,0.16]$ \\
 & $\sigma$ & $[0.10,0.80]$ \\
 & $\rho$ & $[-0.90,-0.10]$ \\
\bottomrule
\end{tabular}
\end{table}

Every surface contains the seven maturities
\[
    \mathcal{T}=\{2/52,1/12,1/4,1/2,1,3/2,2\}\ \text{years}
\]
and 25 strikes per maturity.  Given a reference volatility $\sigma_{\rm ref}$ and call delta $\Delta$, the strike is
\[
 K(T,\Delta;\sigma_{\rm ref})
 =\exp\!\left(-\Phi^{-1}(\Delta)\sigma_{\rm ref}\sqrt{T}
              +\tfrac12\sigma_{\rm ref}^{2}T\right).
\]
The canonical grid uses 25 linearly spaced deltas from $0.98$ down to $0.02$ and $\sigma_{\rm ref}=0.20$; across all maturities its strikes range from $0.5822$ to $1.8606$.  The irregular-grid experiment draws one $\sigma_{\rm ref}$ per surface log-uniformly on $[0.08,0.45]$ and uses it at every maturity of that surface.  This gives sample-dependent strikes ranging from $0.3314$ to $4.5242$ in the saved training set.  SANOS-DLV is trained directly on these grids.  For every fixed-output model, irregular targets are first linearly interpolated in log-strike onto the canonical grid.  The fixed- and random-grid rough-Bergomi training sets use the same parameter draws, isolating the effect of the observation grid.

The parameter/data seeds are $0,1,2$ for rough-Bergomi train/validation/test splits and $10,11,12$ for Heston.  The irregular rough-Bergomi train/validation sets again use seeds $0,1$; the canonical test set with seed $2$ is shared by both training regimes.  The 256 frozen mixture anchors are selected without replacement from the relevant training set using seed $17$.

\subsection{Reference prices and numerical inversion}
\label{s:NumericsSupplement:pricing}

\paragraph{Heston.}
Reference calls are computed by a Fang--Oosterlee COS expansion for put prices followed by put--call parity.  The expansion uses 256 cosine terms and truncation interval
$[c_1-12\sqrt{c_2},c_1+12\sqrt{c_2}]$, where the first two log-forward cumulants are recovered from five-point characteristic-function differences with step $10^{-3}$.  Computations use double precision, and final prices are clipped to the model-independent interval $[(1-K)^+,1]$.

\paragraph{Rough Bergomi.}
The Volterra driver is simulated with the $\kappa=1$ hybrid discretization on 256 uniform time steps over $[0,2]$.  Each surface uses 32,768 paths, processed in four chunks of 8,192, with antithetic Gaussian pairs.  Calls are evaluated by conditional Monte Carlo after conditioning on the Brownian driver correlated with spot.  Each surface has its own saved Monte Carlo seed.  For numerical stability, the implementation clips the log-variance ratio to $[-8,8]$ and the conditional log-forward to $[-4,4]$, then clips call prices to $[(1-K)^+,1]$.  The same specification is used when repricing the off-grid targets and mixture anchors.

Implied total variance is recovered from price by 60 bisection steps on log-variance over $[10^{-10},30]$, followed by eight safeguarded Newton steps; implied volatility is $\sqrt{v/T}$.  Pricing and inversion are performed in double precision, after which the dataset's prices, total variances, and parameters are stored in single precision.  The rough-Bergomi prices are Monte Carlo labels, not exact ground truth.  We did not attach Monte Carlo confidence intervals to individual quotes; their residual noise is visible in the nonzero LP profit of the reference row in Tables~\ref{tab:numerics-headline} and~\ref{tab:lp-complete}.  Likewise, the saved study contains no separate COS- or time-step-convergence sweep, so the stated discretizations are part of the definition of the numerical targets.

\subsection{Models, objectives, and optimization}
\label{s:NumericsSupplement:training}

All five operators first map every input coordinate affinely to $[-1,1]$ and then use a five-hidden-layer, width-256 SiLU MLP.  Their output heads differ as follows.

\begin{table}[t]
\centering
\scriptsize
\setlength{\tabcolsep}{4pt}
\caption{\label{tab:model-sizes}Output parameterization and trainable parameter count.  The small difference between families is due to the four-dimensional rough-Bergomi versus five-dimensional Heston input.}
\begin{tabular}{@{}lp{7.2cm}rr@{}}
\toprule
Model & Output head & rBergomi & Heston \\
\midrule
SANOS-DLV & 217 positive interior DLV values and seven positive total variances; 33 decoder anchors & $322{,}016$ & $322{,}272$ \\
AFNO mixture & softmax weights over 256 frozen training-surface anchors; temperature one & $330{,}240$ & $330{,}496$ \\
Direct-IV & 175 positive implied volatilities, converted pointwise to Black--Scholes prices & $309{,}423$ & $309{,}679$ \\
Direct-Price & 175 prices, clipped to $[0,1]$ & $309{,}423$ & $309{,}679$ \\
Direct-Price+arb & same as Direct-Price & $309{,}423$ & $309{,}679$ \\
\bottomrule
\end{tabular}
\end{table}

SANOS-DLV uses the strike anchors
\[
 \{0\}\cup\{\exp(x):x\ \text{is one of 31 equally spaced points in }[-0.975,0.975]\}\cup\{8\},
\]
which include the forward strike exactly.  Softplus maps impose a floor of $10^{-5}$ on interior DLV values and $10^{-6}$ on total variances.  The 256-anchor AFNO mixture learns only simplex weights.  Its frozen dictionary consists of training surfaces selected with seed $17$; at a new query grid, the anchors are repriced on that grid before applying the learned weights.  Direct-IV applies softplus plus $10^{-4}$ to each output volatility.  The direct price heads are clipped pointwise to $[0,1]$ but have no structural strike or maturity constraint.

The basic loss is the equally weighted price mean-square error over all $7\times25$ quotes.  The penalized baseline minimizes
\[
 \mathcal{L}_{\rm price}+0.003\,\mathcal{P}_{\rm lattice},
\]
where, writing $\langle\cdot\rangle$ for the mean over every valid batch and lattice index,
\begin{align*}
 \mathcal{P}_{\rm lattice}
 ={}&\left\langle[C_{j,\ell+1}-C_{j,\ell}]_+\right\rangle
 +\left\langle[-(C_{j,\ell+2}-2C_{j,\ell+1}+C_{j,\ell})]_+\right\rangle\\
 &+\left\langle[C_{j,\ell}-C_{j+1,\ell}]_+\right\rangle
 +\left\langle[(1-K_{j,\ell})^+-C_{j,\ell}]_+\right\rangle
 +\left\langle[C_{j,\ell}-1]_+\right\rangle.
\end{align*}
This is deliberately a soft, lattice-local baseline.  In particular, its convexity term is an unweighted second difference rather than the non-uniform-grid butterfly used for evaluation, and equal strike indices at adjacent maturities need not represent equal strikes on a delta grid.  The weight $0.003$ is the fixed experiment configuration; the code base does not record a penalty-weight sweep or select it on the test set.

Every model is trained for exactly 80 epochs with AdamW, learning rate $2\times10^{-3}$, zero weight decay, batch size 512, and gradient-norm clipping at one.  A cosine schedule decreases the learning rate to $10^{-4}$ at epoch 80.  Validation metrics are recorded after every epoch, but there is no early stopping or best-epoch selection: evaluation uses the final checkpoint.  Network initialization and batch shuffling use seed $2026$.  Thus the reported tables are one deterministic training run, not a mean over repeated initializations; no training-seed error bars should be inferred from them.  The per-surface SANOS fit in Table~\ref{tab:numerics-headline} independently optimizes a 49-anchor decoder on the first 256 rough-Bergomi test surfaces for 1,500 Adam iterations at learning rate $0.05$ and seed $77$.  It is an optimization baseline, not a certified global optimum or a statistical oracle.

\subsection{Metrics and arbitrage diagnostics}
\label{s:NumericsSupplement:metrics}

Price RMSE is computed over all test quotes.  Implied-volatility MAE, median, p95, and maximum are computed only where the \emph{target-price-implied} Black--Scholes delta lies in $[0.10,0.90]$; one volatility basis point is $10^{-4}$.  For each surface, the canonical-grid severity is the maximum of: a positive adjacent price increment, a decrease in adjacent strike slopes, an intrinsic-value shortfall, and an upper-bound excess.  A surface is counted as arbitrageable when severity exceeds $10^{-4}$.  Because the delta grid has maturity-dependent strikes, the canonical table does not compare calendar prices at unequal strikes; cross-maturity effects are instead probed by the joint LP below.  Off-grid butterfly counts use the grid-aware value in Section~\ref{s:Numerics} and tolerance $10^{-6}$.

The off-grid sample comprises 256 test parameters drawn without replacement with seed $202604$.  In addition to the canonical 25-strike grid, the saved evaluation contains: 49 deltas from $0.98$ to $0.02$ (the dense grid used in Figures~\ref{fig:irregular-out} and~\ref{fig:exploit-profit}); 25 deltas from $0.995$ to $0.005$; and a 25-strike grid constructed with $\sigma_{\rm ref}=0.35$.  The dense grid contains $256\times7\times47=84{,}224$ interior butterflies.  Neither AFNO instance nor the reference pricer has a violation below $-10^{-6}$ among them; the minimum butterfly values are respectively $4.86\times10^{-7}$, $5.18\times10^{-6}$, and $5.00\times10^{-7}$.  Direct-IV has 701 violating butterflies across $73.8\%$ of surfaces, while both direct-price variants violate at least one butterfly on every dense-grid surface.  SANOS-DLV and the mixture are evaluated natively on the new grid; direct heads are linearly interpolated in log-strike from their canonical price outputs.

\paragraph{Bounded scenario LP.}
For a quoted surface, let $z=(z_0,z_S,(z_{j\ell}))$ hold cash, the terminal underlying, and the calls.  With quote vector $q=(1,1,(C(T_j,K_{j\ell})))$, the diagnostic is
\[
 \Pi(C)=\max_{-1\le z_i\le1}\{-q^\top z:A z\ge0\},
 \qquad
 (Az)_r=z_0+z_S S^{(r)}_{T_M}
       +\sum_{j,\ell}z_{j\ell}(S^{(r)}_{T_j}-K_{j\ell})^+.
\]
Hence $\Pi$ is the largest negative initial cost among portfolios whose aggregate payoff is nonnegative on every tested path.  The box $[-1,1]$ applies separately to every instrument, so the reported number is in units of forward notional and scales with this portfolio bound.

The joint test uses 800 as its nominal scenario count.  Its deterministic additions give 954 actual paths for seven maturities: 154 one-maturity spot scans, 160 constant paths, 160 one-jump paths, and 480 drift-corrected GBM paths with volatility uniform on $[0.08,0.65]$.  Spot scans cover $\exp([-1.5,1.5])$; the single-maturity version additionally includes zero, 100, and every strike kink.  Scenario randomness uses seed $202604$.  We solve with SciPy/HiGHS, the solver's default feasibility tolerances, position bound one, and a four-second limit per LP.  The implementation records zero if HiGHS does not return success.  Rough-Bergomi LP summaries use all 256 dense off-grid surfaces; Heston summaries use the first 1,000 canonical-grid test surfaces.  This finite-scenario bounded LP measures economic severity but is not itself a global no-arbitrage certificate.

\subsection{Complete tabulations}
\label{s:NumericsSupplement:results}

Table~\ref{tab:heston-complete} supplies the Heston counterpart of Table~\ref{tab:numerics-headline}.  The ordering is consistent across families: Direct-IV is most accurate pointwise, the structured models have no detected grid arbitrage, and the unconstrained price heads have large tail errors and economically material LP values.

\begin{table}[t]
\centering
\scriptsize
\setlength{\tabcolsep}{2.5pt}
\caption{\label{tab:heston-complete}Complete Heston test accuracy and compliance results on 8,000 surfaces.  IV errors use the liquid band; LP mean is the joint-maturity bounded profit over 1,000 surfaces.}
\resizebox{\linewidth}{!}{%
\begin{tabular}{@{}lrrrrrrrr@{}}
\toprule
Model & Price RMSE & IV MAE & IV median & IV p95 & IV max & Arb. surf. & Max. arb. & LP mean \\
 & & \multicolumn{4}{c}{(bps)} & (\%) & & \\
\midrule
SANOS-DLV (ours) & $2.10{\times}10^{-4}$ & $8.39$ & $2.97$ & $35.6$ & $392$ & $0$ & $4.57{\times}10^{-7}$ & $8.81{\times}10^{-4}$ \\
AFNO mixture & $9.78{\times}10^{-4}$ & $30.5$ & $14.3$ & $112$ & $856$ & $0$ & $0$ & $6.79{\times}10^{-4}$ \\
Direct-IV & $9.43{\times}10^{-5}$ & $3.05$ & $1.67$ & $10.4$ & $169$ & $0$ & $3.50{\times}10^{-5}$ & $9.36{\times}10^{-4}$ \\
Direct-Price & $3.63{\times}10^{-3}$ & $191$ & $2.06$ & $2462$ & $3987$ & $99.1$ & $5.31$ & $1.14{\times}10^{-1}$ \\
Direct-Price+arb & $3.61{\times}10^{-3}$ & $234$ & $3.35$ & $2667$ & $3988$ & $97.5$ & $4.19$ & $1.03{\times}10^{-1}$ \\
Heston reference & -- & -- & -- & -- & -- & -- & -- & $1.02{\times}10^{-3}$ \\
\bottomrule
\end{tabular}}
\end{table}

The irregular-input comparison is shown numerically in Table~\ref{tab:irregular-complete}.  These are median absolute implied-volatility errors on the same canonical rough-Bergomi test set, so the ratio isolates the training-grid change.  Only SANOS-DLV consumes the random grids natively.

\begin{table}[t]
\centering
\small
\caption{\label{tab:irregular-complete}Effect of per-sample irregular training grids on canonical-grid test accuracy.}
\begin{tabular}{@{}lrrr@{}}
\toprule
Model & Fixed-grid train & Random-grid train & Ratio \\
 & \multicolumn{2}{c}{median absolute IV error (bps)} & \\
\midrule
SANOS-DLV (ours) & $5.79$ & $5.88$ & $1.02$ \\
AFNO mixture & $14.9$ & $92.9$ & $6.25$ \\
Direct-IV & $3.55$ & $6.53$ & $1.84$ \\
Direct-Price & $3.93$ & $8.36$ & $2.13$ \\
Direct-Price+arb & $6.76$ & $11.5$ & $1.70$ \\
\bottomrule
\end{tabular}
\end{table}

Table~\ref{tab:lp-complete} reports the full joint-LP distribution underlying the mean column in the main text.  The nonzero reference levels quantify the diagnostic's numerical/scenario floor, especially for Monte Carlo rough-Bergomi prices.

\begin{table}[t]
\centering
\scriptsize
\setlength{\tabcolsep}{2.8pt}
\caption{\label{tab:lp-complete}Joint-maturity bounded LP profit in units of forward notional.  Rough-Bergomi uses 256 dense-grid surfaces; Heston uses 1,000 canonical-grid surfaces.}
\resizebox{.6\linewidth}{!}{%
\begin{tabular}{@{}lrrrrrrrr@{}}
\toprule
& \multicolumn{4}{c}{rough Bergomi} & \multicolumn{4}{c}{Heston} \\
\cmidrule(lr){2-5}\cmidrule(l){6-9}
Model & Mean & Median & p95 & Max & Mean & Median & p95 & Max \\
\midrule
SANOS-DLV (ours) & $.00480$ & $.00299$ & $.0122$ & $.0211$ & $.000881$ & $0$ & $.00524$ & $.0240$ \\
AFNO mixture & $.00489$ & $.00287$ & $.0135$ & $.0219$ & $.000679$ & $0$ & $.00370$ & $.0245$ \\
Direct-IV & $.0126$ & $.0114$ & $.0204$ & $.0308$ & $.000936$ & $0$ & $.00536$ & $.0283$ \\
Direct-Price & $.264$ & $.236$ & $.539$ & $.698$ & $.114$ & $.109$ & $.240$ & $.314$ \\
Direct-Price+arb & $.286$ & $.266$ & $.575$ & $.709$ & $.103$ & $.0916$ & $.213$ & $.278$ \\
Numerical reference & $.00523$ & $.00295$ & $.0136$ & $.0278$ & $.00102$ & $0$ & $.00576$ & $.0397$ \\
\bottomrule
\end{tabular}}
\end{table}

\begin{figure}[t]
\centering
\includegraphics[width=\linewidth]{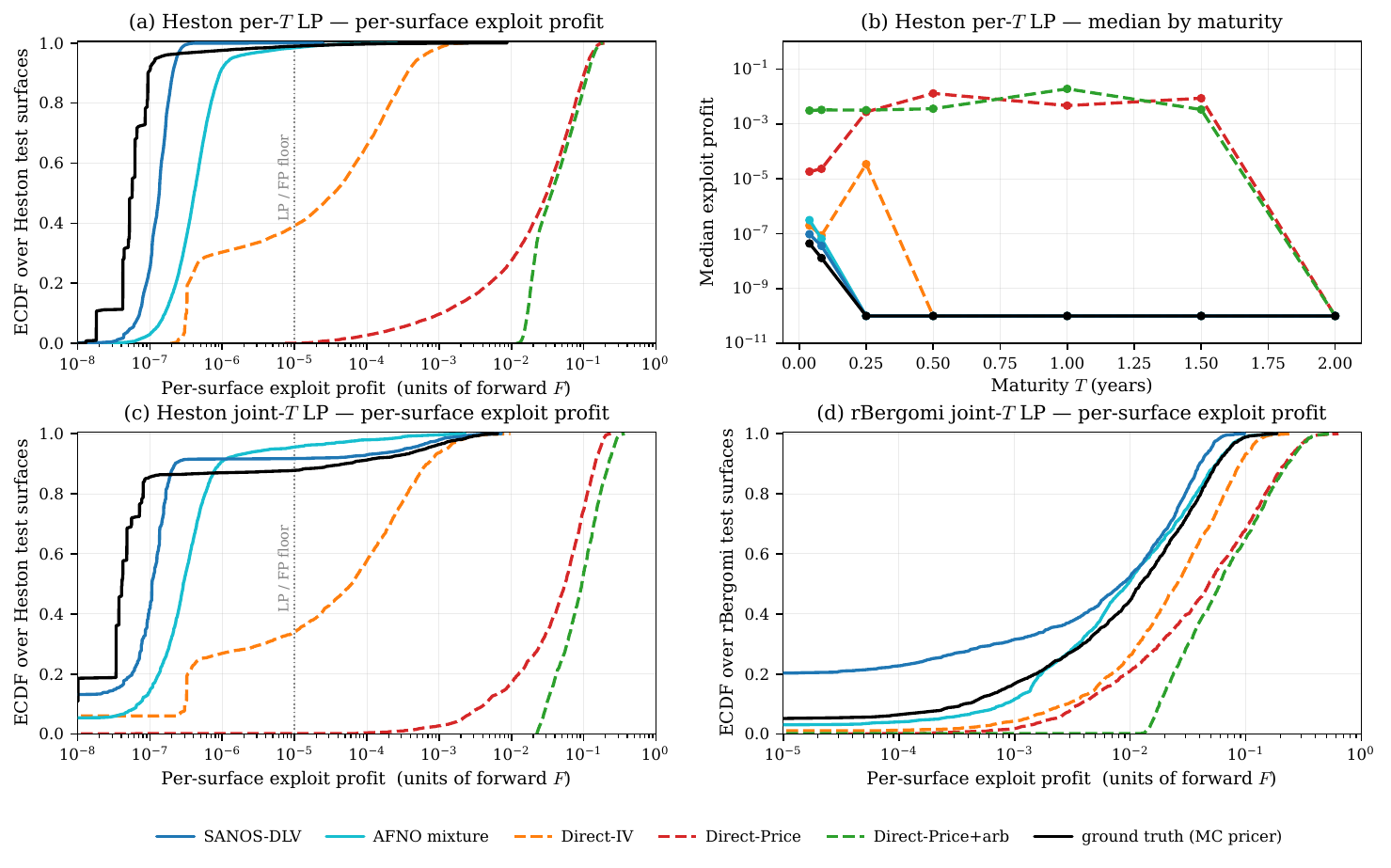}
\caption{\label{fig:exploit-profit-grid}Full static-arbitrage LP diagnostic.  The panels report per-surface exploit-profit distributions and maturity-wise medians for Heston and rough-Bergomi test sets.  AFNO instances stay close to the numerical reference level; direct decoder heads remain materially arbitrageable.}
\end{figure}

Figure~\ref{fig:exploit-profit-grid} complements Figure~\ref{fig:exploit-profit} with the maturity-by-maturity and joint distributions.  In the Heston per-maturity LP, SANOS-DLV, the mixture, and the Heston reference are zero to reported precision; Direct-IV has mean $5.31\times10^{-11}$ and maximum $4.17\times10^{-8}$.  The corresponding means for Direct-Price and Direct-Price+arb are $0.0875$ and $0.0632$.

\subsection{Runtime and reproducibility}
\label{s:NumericsSupplement:runtime}

The saved manifest records Python 3.12.0 on macOS 15.2 arm64, with NumPy 2.5.1, SciPy 1.18.0, PyTorch 2.13.0, Matplotlib 3.11.1, pandas 3.0.3, and joblib 1.5.3.  It does not record the processor model, so the timings below are indicative rather than hardware-normalized.  Table~\ref{tab:numerical-runtime} reports wall time from each final training history.  The entire 15-model training stage (fixed-grid rough Bergomi, random-grid rough Bergomi, and Heston) took $11{,}795$ seconds.  For inference, we loaded the saved fixed-grid rough-Bergomi checkpoints, used one PyTorch CPU thread and batches of 1,024, warmed up once, timed five passes over all 8,000 test surfaces, and report the median.

\begin{table}[t]
\centering
\small
\caption{\label{tab:numerical-runtime}Recorded 80-epoch training time and batched CPU inference time.  Inference produces all $7\times25$ quotes for one surface.}
\begin{tabular}{@{}lrrr@{}}
\toprule
Model & rBergomi train (min) & Heston train (min) & Inference ($\mu$s/surface) \\
\midrule
SANOS-DLV (ours) & $34.9$ & $35.9$ & $89.2$ \\
AFNO mixture & $6.91$ & $6.95$ & $2.24$ \\
Direct-IV & $8.06$ & $7.91$ & $3.39$ \\
Direct-Price & $6.82$ & $6.93$ & $1.78$ \\
Direct-Price+arb & $7.93$ & $8.06$ & $1.84$ \\
\bottomrule
\end{tabular}
\end{table}

From a clean \texttt{afno-code} checkout, the full pipeline is invoked by
\begin{center}
\verb|uv run afno-reproduce --profile full|.
\end{center}
It generates data and anchors, trains all models, evaluates canonical and off-grid predictions, fits the per-surface decoder, solves the LPs, renders figures, and finally writes \texttt{artifacts/manifest.json}.  The manifest stores the full profile, global seed, package versions, stage times, SHA-256 hashes of every artifact, and the minimum dense-grid butterfly of each certified model.  Pipeline stages are cached and resumable; rough-Bergomi generation checkpoints the completed-surface mask and rejects a checkpoint whose simulation configuration differs.  The included test suite is run with \verb|uv run pytest|.  A reduced \verb|--profile quick| is provided only as a smoke test and is not the source of any number in this paper.

\section{Proofs}
\label{s:Proofs}

We now prove our main results.

\subsection{Auxiliary Results}
\label{s:Aux_results}

\begin{proof}[{Proof of Proposition~\ref{prop:EasySufficientCondition_Decay}}]
Fix $T\ge 0$ and $K>0$. Since $\mathbb{E}[|S_T|^q]<\infty$, we also have $\mathbb{E}[(S_T)_+^q]<\infty$. We first establish the required estimate pointwise.  For $x\le K$, one has $(x-K)_+=0$. Suppose therefore that $x>K$. In particular, $x>0$, and, upon setting $y\eqdef x/K>1$, we obtain
$
(x-K)_+=K(y-1)=K^{-(q-1)}x^q\frac{y-1}{y^q}
$.  Consider $g:(1,\infty)\to\mathbb{R}$ given by $g(y)\eqdef (y-1)y^{-q}$. Its derivative is
$
g'(y)=y^{-q-1}\bigl(q-(q-1)y\bigr)$.
Since $q>1$, it follows that $g$ is increasing on $(1,q/(q-1))$ and decreasing on $(q/(q-1),\infty)$. Hence $g$ attains its maximum at $y=q/(q-1)$, where
$
g\left(\frac{q}{q-1}\right)=\frac{(q-1)^{q-1}}{q^q}
$.
Consequently, for every $x\in\mathbb{R}$,
$
(x-K)_+\le \frac{(q-1)^{q-1}}{q^q}K^{-(q-1)}(x_+)^q
$.
Applying this deterministic inequality with $x=S_T(\omega)$ gives
$
(S_T-K)_+\le \frac{(q-1)^{q-1}}{q^q}K^{-(q-1)}(S_T)_+^q
$ $\mathbb{P}$-a.s.  
The random variable on the right-hand side is integrable because $(S_T)_+^q\le |S_T|^q$ and $\mathbb{E}[|S_T|^q]<\infty$. Taking expectations therefore yields~\eqref{eq:PolynomialCallPayoffDecay}.  
Finally, set $p\eqdef q-1>0$. Then~\eqref{eq:PolynomialCallPayoffDecay} becomes
$
\mathbb{E}\big[(S_T-K)_+\big]\le C_{T,p}K^{-p}
$ and $C_{T,p}\eqdef \frac{(q-1)^{q-1}}{q^q}\mathbb{E}\big[(S_T)_+^q\big]<\infty$; which is precisely~\eqref{eq:right_wing_asymptotic}.
\end{proof}

\begin{proof}[{Proof of Proposition~\ref{prop:NOVec}}]
Consider the ``bad'' call surface $C^{\star}\in C([0,\infty)^2,[0,\infty))$ defined for every $T,K\ge 0$ by 
\[
        C^{\star}(T,K)
    \eqdef
       (1-K)_+
.
\]
First, observe that call surface generated by the deterministic strictly-positive martingale $Z_T\equiv 1$.
Hence $C^{\star}\in \mathcal{C}_{\mathrm{AF}}$.  
However, $
    \widetilde{C}
    \eqdef
    C^{\star} + C^{\star}
    =
    2(1-K)_+.
$, and so, $
    \widetilde{C}(T,0)=2 \neq 1
$ for all $
T\geq 0.
$; violating Theorem~\ref{prop:contarbfree}~(i).  
Let $\lambda=0$ and define
$
    \widehat{C}
    \eqdef
    0C^{\star}
    =
    0
$; which violates Theorem~\ref{prop:contarbfree}~(i).
Finally, consider $
    \overline{C}
    \eqdef
    C^{\star}   C^{\star}
    =
    (1-K)_+^2
$ and observe that, for $K\in [0,1)$, $
    \overline{C}(T,K)=(1-K)^2,
$.  Whence $
    \partial_K \overline{C}(T,0)
    =
    -2.
$ and Theorem~\ref{prop:contarbfree}~(ii) is violated.  
\end{proof}

\begin{proof}[{Proof of Proposition~\ref{prop:AS}}]
Let $C_1,C_2\in\mathcal{C}_{\mathrm{AF}}$ and let $\alpha\in[0,1]$.  We prove that $\alpha C_1+(1-\alpha)C_2\in\mathcal{C}_{\mathrm{AF}}$ directly from Definition~\ref{def:arbfreecall}.
For each $i\in\{1,2\}$, since $C_i\in\mathcal{C}_{\mathrm{AF}}$, there exists a filtered probability space $(\Omega_i,\mathcal{F}^i,(\mathcal{F}^i_T)_{T\ge 0},\mathbb{P}_i)$ and a non-negative $(\mathbb{P}_i,\mathcal{F}^i_T)$-martingale $Z^i=(Z^i_T)_{T\ge 0}$ with $Z^i_0=1$ a.s. such that
$
    C_i(T,K)
=
    \mathbb{E}_{\mathbb{P}_i}
    \big[
        (Z^i_T-K)^+
    \big]
$ for every $T,K\ge 0$.
We now construct a filtered probability space which randomly selects one of these two martingale models.  Consider the disjoint union
$
    \Omega
\eqdef
    \big(\{1\}\times\Omega_1\big)
    \sqcup
    \big(\{2\}\times\Omega_2\big)
$, define the probability measure $\mathbb{P}$ on $\Omega$ by
\[
    \mathbb{P}\big(\{1\}\times A_1\big)
    \eqdef
    \alpha\,\mathbb{P}_1(A_1)
\mbox{ and }
    \mathbb{P}\big(\{2\}\times A_2\big)
    \eqdef
    (1-\alpha)\,\mathbb{P}_2(A_2),
\]
for $A_i\in\mathcal{F}^i$, and for every $T\ge 0$, define the terminal time $\sigma$-algebra by
\[
    \mathcal{F}_T
    \eqdef
    \Big\{
        \big(\{1\}\times A_1\big)
        \cup
        \big(\{2\}\times A_2\big)
        :
        A_i\in\mathcal{F}^i_T,\ i=1,2
    \Big\}.
\]
Then $(\mathcal{F}_T)_{T\ge 0}$ is a filtration on $\Omega$.  Define $Z=(Z_T)_{T\ge 0}$ by
$
    Z_T(i,\omega_i)
    \eqdef
    Z^i_T(\omega_i)
$ for each $i\in\{1,2\}$.  
The process $Z$ is nonnegative and satisfies $Z_0=1$ a.s.  We verify that $Z$ is a $(\mathbb{P},\mathcal{F}_T)$-martingale.  Let $0\le S\le T$ and let
$
    A
=
    \big(\{1\}\times A_1\big)
    \cup
    \big(\{2\}\times A_2\big)
    \in\mathcal{F}_S
$, 
where $A_i\in\mathcal{F}^i_S$.  Since each $Z^i$ is a martingale, we find that
\[
\begin{aligned}
    \int_A Z_T\,d\mathbb{P}
    &=
    \alpha
    \int_{A_1} Z^1_T\,d\mathbb{P}_1
    +
    (1-\alpha)
    \int_{A_2} Z^2_T\,d\mathbb{P}_2
\\
    &=
    \alpha
    \int_{A_1} Z^1_S\,d\mathbb{P}_1
    +
    (1-\alpha)
    \int_{A_2} Z^2_S\,d\mathbb{P}_2
\\
    &=
    \int_A Z_S\,d\mathbb{P}.
\end{aligned}
\]
Hence $\mathbb{E}_{\mathbb{P}}[Z_T\mid\mathcal{F}_S]=Z_S$; whence, $Z$ is a nonnegative martingale.
\hfill\\
It only remains to compute the expectation of $(Z_t-K)^+$ under the mixture laws; indeed, for every $T,K\ge 0$, we compute
\[
\begin{aligned}
    \mathbb{E}_{\mathbb{P}}\big[(Z_T-K)^+\big]
    &=
    \alpha
    \mathbb{E}_{\mathbb{P}_1}
    \big[
        (Z^1_T-K)^+
    \big]
    +
    (1-\alpha)
    \mathbb{E}_{\mathbb{P}_2}
    \big[
        (Z^2_T-K)^+
    \big]
\\
    &=
    \alpha C_1(T,K)+(1-\alpha)C_2(T,K).
\end{aligned}
\]
Thus $\alpha C_1+(1-\alpha)C_2$ admits a martingale representation in the sense of Definition~\ref{def:arbfreecall}.  Hence $\alpha C_1+(1-\alpha)C_2\in\mathcal{C}_{\mathrm{AF}}$, proving that $\mathcal{C}_{\mathrm{AF}}$ is convex.
\end{proof}

\begin{proof}[{Proof of Proposition~\ref{prop:SSVIVerification}}]
Use normalized spot $S_0=1$, fix $T>0$, abbreviate $\phi_T\eqdef\varphi(\theta_T)$, and set
$
\beta_T
\eqdef
\frac{1+\rho}{2}\theta_T\phi_T.
$
Since $\theta_T>0$, $\phi_T>0$, and $\rho\in(-1,1)$, we have $\beta_T>0$. Moreover, since $p>0$,
$
2(\sqrt{p+1}-\sqrt p)^2<2,
$
and hence the assumed inequality
$
\beta_T\leq2(\sqrt{p+1}-\sqrt p)^2
$
implies $\beta_T<2$.
\hfill\\
\noindent
We first identify the right-wing asymptotics of the SSVI total variance. Recall that
\[
w(k,\theta_T)
=
\frac{\theta_T}{2}
\left(
1+\rho\phi_Tk+
\sqrt{(\phi_Tk+\rho)^2+1-\rho^2}
\right).
\]
This is consistent with the asymptotic statement in~\citep[Remark~4.3]{gatheral2014arbitrage}, but we record the required remainder explicitly. Since $\phi_T>0$, there exists $k_T^{(0)}>0$ such that $\phi_Tk+\rho>0$ whenever $k\geq k_T^{(0)}$. For such $k$, rationalization gives
\[
\sqrt{(\phi_Tk+\rho)^2+1-\rho^2}-(\phi_Tk+\rho)
=
\frac{1-\rho^2}
{\sqrt{(\phi_Tk+\rho)^2+1-\rho^2}+\phi_Tk+\rho}.
\]
The denominator on the right-hand side is of order $k$ and is bounded from below by a positive multiple of $k$ for all sufficiently large $k$. Consequently,
$
\sqrt{(\phi_Tk+\rho)^2+1-\rho^2}
=
\phi_Tk+\rho+O(k^{-1})
$
as $k\to\infty$. Substituting this identity into the SSVI formula yields
\[
w(k,\theta_T)
=
\beta_Tk+\frac{1+\rho}{2}\theta_T+O(k^{-1}),
\qquad k\to\infty.
\]
In particular,
$
w(k,\theta_T)=\beta_Tk+O(1)
$
and
$
w(k,\theta_T)/k\to\beta_T
$
as $k\to\infty$.
Write, as usual,
$
d_{\pm}(k)
=
-k/\sqrt{w(k,\theta_T)}
\pm
\sqrt{w(k,\theta_T)}/2.
$
For every $k$ for which $w(k,\theta_T)>0$, direct algebra gives
\[
\frac{d_+(k)^2}{2}
=
\frac{(2k-w(k,\theta_T))^2}{8w(k,\theta_T)}
=
\frac{k^2}{2w(k,\theta_T)}-\frac{k}{2}+\frac{w(k,\theta_T)}{8}.
\]
Since $w(k,\theta_T)=\beta_Tk+O(1)$ and $\beta_T>0$, we also have
$
w(k,\theta_T)^{-1}=(\beta_Tk)^{-1}+O(k^{-2})
$
as $k\to\infty$. Therefore,
\[
\frac{d_+(k)^2}{2}
=
\left(
\frac{1}{2\beta_T}
-\frac{1}{2}
+\frac{\beta_T}{8}
\right)k+O(1)
=
\frac{(2-\beta_T)^2}{8\beta_T}\,k+O(1).
\]
Define
$
q_T\eqdef(2-\beta_T)^2/(8\beta_T).
$
Thus there exist constants $M_T>0$ and $k_T^{(1)}>0$ such that
\[
\left|
\frac{d_+(k)^2}{2}-q_Tk
\right|
\leq M_T,
\qquad k\geq k_T^{(1)}.
\]
Notice also that the corresponding expression involving $d_-$ is not merely asymptotically equal to the preceding one: it is identical to it. Indeed,
\[
\frac{d_-(k)^2}{2}-k
=
\frac{(2k+w(k,\theta_T))^2}{8w(k,\theta_T)}-k
=
\frac{(2k-w(k,\theta_T))^2}{8w(k,\theta_T)}
=
\frac{d_+(k)^2}{2}.
\]
Furthermore,
\[
d_+(k)
=
\frac{w(k,\theta_T)-2k}{2\sqrt{w(k,\theta_T)}}
=
-\frac{2-\beta_T}{2\sqrt{\beta_T}}\sqrt{k}
+O(k^{-1/2}),
\qquad k\to\infty.
\]
Since $\beta_T<2$, it follows that $d_+(k)\to-\infty$. Hence, after increasing $k_T^{(1)}$ if necessary, we may assume that $d_+(k)\leq-1$ for every $k\geq k_T^{(1)}$.
\hfill\\
\noindent
Let $\mathcal{N}$ denote the standard Gaussian distribution function. For every $x>0$, the elementary Gaussian tail estimate
\[
\mathcal{N}(-x)
=
\frac{1}{\sqrt{2\pi}}
\int_x^\infty e^{-u^2/2}\,du
\leq
\frac{1}{x\sqrt{2\pi}}
\int_x^\infty u e^{-u^2/2}\,du
=
\frac{1}{x\sqrt{2\pi}}e^{-x^2/2}
\]
holds. In particular, if $x\geq1$, then
$
\mathcal{N}(-x)\leq(2\pi)^{-1/2}e^{-x^2/2}.
$
\hfill\\
\noindent
Now let $K=S_0e^k$. By the Black--Scholes representation of the SSVI call price,
\[
C_{\mathrm{SSVI}}(T,K)
=
S_0
\left(
\mathcal{N}(d_+(k))
-
e^k\mathcal{N}(d_-(k))
\right)
\leq
S_0\mathcal{N}(d_+(k)).
\]
Therefore, for every $k\geq k_T^{(1)}$, $
C_{\mathrm{SSVI}}(T,S_0e^k)
\leq
\frac{S_0}{\sqrt{2\pi}}
\exp\left(-\frac{d_+(k)^2}{2}\right)
\leq
\frac{S_0e^{M_T}}{\sqrt{2\pi}}e^{-q_Tk}.
$.
Thus the SSVI call price has right-wing decay of order at least $e^{-q_Tk}$.
\hfill\\
\noindent
It remains to compare $q_T$ with $p$. Since $0<\beta_T<2$,
\[
q_T\geq p
\quad\Longleftrightarrow\quad
(2-\beta_T)^2\geq8p\beta_T
\quad\Longleftrightarrow\quad
\beta_T^2-4(2p+1)\beta_T+4\geq0.
\]
The two roots of the polynomial on the right are
$
2(\sqrt{p+1}-\sqrt p)^2
$
and
$
2(\sqrt{p+1}+\sqrt p)^2.
$
The second root is strictly larger than $2$, while $\beta_T<2$. Consequently,
\[
q_T\geq p
\quad\Longleftrightarrow\quad
\beta_T
\leq
2(\sqrt{p+1}-\sqrt p)^2.
\]
Hence the assumption of the proposition implies $q_T\geq p$.
Set $k_T^\ast\eqdef\max\{k_T^{(1)},0\}$ and $K_T^\ast\eqdef S_0e^{k_T^\ast}$. For $K\geq K_T^\ast$, writing $k=\log(K/S_0)\geq0$ and using $q_T\geq p$ gives
\[
C_{\mathrm{SSVI}}(T,K)
\leq
\frac{S_0e^{M_T}}{\sqrt{2\pi}}e^{-q_Tk}
\leq
\frac{S_0e^{M_T}}{\sqrt{2\pi}}e^{-pk}
=
\frac{S_0^{p+1}e^{M_T}}{\sqrt{2\pi}}K^{-p}.
\]
For the remaining strikes $0<K<K_T^\ast$, the Black--Scholes call price satisfies $0\leq C_{\mathrm{SSVI}}(T,K)\leq S_0$. Since $K<K_T^\ast$ and $p>0$,
\[
C_{\mathrm{SSVI}}(T,K)
\leq
S_0
\leq
S_0(K_T^\ast)^pK^{-p}.
\]
Consequently, with
\[
C_{T,p}
\eqdef
\max
\left\{
\frac{S_0^{p+1}e^{M_T}}{\sqrt{2\pi}},
\,
S_0(K_T^\ast)^p
\right\},
\]
we obtain
\[
C_{\mathrm{SSVI}}(T,K)
\leq
C_{T,p}K^{-p},
\qquad K>0.
\]
Since $T>0$ was arbitrary,~\eqref{eq:right_wing_asymptotic} holds at every positive maturity. Finally, at $T=0$,
$
C_{\mathrm{SSVI}}(0,K)=(S_0-K)^+.
$
For $K\geq S_0$ this quantity vanishes, whereas for $0<K<S_0$ we have
$
(S_0-K)^+\leq S_0\leq S_0^{p+1}K^{-p}.
$
Thus~\eqref{eq:right_wing_asymptotic} also holds at $T=0$, which completes the proof.
\end{proof}

\subsection{Well-Posedness of the Solution Map and Universal Arbitrage-Free Pricing}

\begin{proof}[{Proof of Proposition~\ref{prop:Solution_MapRegularity}}]
For each $\theta\in\Theta$, put $Z_t^\theta=e^{-rt}X_t^\theta$ and $g_t^\theta=X_t^\theta\sigma(t,\nu_t,\theta,X_t^\theta)$. Integration by parts gives
\[
Z_t^\theta=1+\int_0^t Z_s^\theta\sigma(s,\nu_s,\theta,X_s^\theta)\,dM_s.
\]
For $\theta_1,\theta_2\in\Theta$, write $D_t=X_t^{\theta_2}-X_t^{\theta_1}$. Then
\[
D_t=\int_0^t rD_s\,ds+\int_0^t(g_s^{\theta_2}-g_s^{\theta_1})\,dM_s,
\qquad |g_s^{\theta_2}-g_s^{\theta_1}|\leq L_x|D_s|+L_\theta|X_s^{\theta_1}|\|\theta_2-\theta_1\|_2.
\]
The first identity makes $Z^\theta$ a positive local martingale, with the stochastic integrals understood in the predictable sense required for the stated equation. Square integrability of $M$ alone does not ensure that $Z^\theta$ is a true martingale, justify the required BDG--Gr\"onwall moment and stability estimates, or ensure continuity of the call prices in maturity. Thus the stated hypotheses do not complete this proof; the well-posedness and pricing conclusions remain pending the separately proposed assumption repair.
\end{proof}

\begin{proof}[{Proof of Corollary~\ref{cor:AF_solmap}}]
Fix $0<\varepsilon<1$ and $H>0$, and set $\eta\eqdef\varepsilon e^{-lH}\in(0,1)$. By Proposition~\ref{prop:Solution_MapRegularity}, the class $\mathcal Y=\operatorname{Sol}(\Theta)$ is uniformly controlled with $B=2^p$ and common modulus $\omega(h)=h+\max\{1,\sqrt{e^a-1}\}\sqrt h$. Moreover, $l>q\lambda/p>\frac a2q(p+1)\geq\kappa_{q+1}$, so the same proposition gives $\mathcal Y\subseteq\mathcal C_{AF}^{q,l}$. Its parameter estimate, together with $\|\theta-\theta'\|_2\leq\|\theta-\theta'\|_1$, shows that $\operatorname{Sol}:(\Theta,\|\cdot\|_1)\to(\mathcal C_{AF}^{q,l},\|\cdot\|_{q,l})$ is Lipschitz.

Since $\Theta$ is non-empty and compact, $q\lambda/p<l<\lambda$, and the decoder satisfies Definition~\ref{defn:Good_Decoder} and Assumption~\ref{ass:weighted_decoder_selection}, Proposition~\ref{prop:UAT_AFNO__technicalversion__FINITEDIMENSIONALVERSION}, applied with $f=\operatorname{Sol}$ and tolerance $\eta$, yields an AF-NO $\Phi:\mathbb R^{F_{\mathrm{mkt}}}\to\mathcal C_{AF}$ with the asserted range property and
\[
\sup_{\theta\in\Theta}\|\operatorname{Sol}(\theta)-\Phi(\theta)\|_{q,l}\leq\eta.
\]
For every $\theta\in\Theta$ and $0\leq T,K\leq H$, the definition of the weighted norm gives
\[
\begin{aligned}
\left|\mathbb E_{\mathbb Q}\big[(e^{-rT}X_T^\theta-K)_+\big]-\Phi(\theta)(T,K)\right|
&\leq e^{lT}(1+K)^{-q}\|\operatorname{Sol}(\theta)-\Phi(\theta)\|_{q,l}\\
&\leq e^{lH}\eta=\varepsilon.
\end{aligned}
\]
Taking the maximum over $(T,K)\in[0,H]^2$ and the supremum over $\theta\in\Theta$ proves the claim.
\end{proof}

\subsection{Proof of Positive Results}
\label{s:Good}

\subsubsection{Proof of Certifiability}
\label{s:Proof_certificate}
We may now deduce our arbitrage-free certificate; i.e.\ Theorem~\ref{thrm:Cert_AF}.
\begin{proof}
For each $x$, the softmax weights belong to the simplex.  By Definition~\ref{defn:Good_Decoder}(ii), each decoder output belongs to $\mathcal{C}_{AF}$.  Since $\mathcal{C}_{AF}$ is convex by Proposition~\ref{prop:AS}, their convex combination also belongs to $\mathcal{C}_{AF}$.  
\end{proof}

The remainder of our analysis will be to establish the corresponding universality guarantee.

\subsubsection{Proof of Quantizability: Pointwise Approximate Interpolation}
\label{s:Proof_Quantizability}
Next, we verify the quantizability property of any separable and closed (metric) subspace of $\mathcal{C}_{AF}^{p,\lambda}\eqdef \mathcal{C}_{AF}\cap \mathcal{X}_{\lambda}^p$ metrized by $\| \|_{p,\lambda}$.
Our analysis holds for any SANOS-type model, in that the model satisfies the following interpolation property:
\begin{ass}[Uniformly bounded weighted interpolation]
\label{ass:weighted_decoder_selection}
Let $\mathcal Y\subseteq\mathcal C_{AF}^{p,\lambda}$ be admitted by a decoder as in Definition~\ref{defn:Good_Decoder}. There exists $1\le B_{\mathrm{dec}}<\infty$ such that, for every $R,S,\eta>0$, there are $N,M\in\mathbb N_+$ with $R/N+S/M\leq\eta$ for which every $C\in\mathcal Y$ admits $\theta_C\in\mathbb R^{D_{N,M,R,S}}$ satisfying
\begin{equation}
\label{eq:weighted_decoder_selection}
\max_{\substack{0\leq m\leq M\\0\leq n\leq N}}\big|\phi^{(N,M,R,S)}_{\theta_C}(mS/M,nR/N)-C(mS/M,nR/N)\big|\leq\eta,\qquad \big\|\phi^{(N,M,R,S)}_{\theta_C}\big\|_{p,\lambda}\leq B_{\mathrm{dec}}.
\end{equation}
The integers $N,M$ are common to all $C\in\mathcal Y$; the parameters $\theta_C$ may depend on $C$.
\end{ass}
\begin{proposition}[Quantizability of SANOS-Type Parameterizations]
\label{prop:interpol_2_approx}
Fix $0<q<p$ and let $l,\lambda>0$ satisfy: 
$\frac{q}{p}\,\lambda < l < \lambda
$.
Let $\mathcal{Y}\subseteq\mathcal{C}_{AF}^{p,\lambda}$ be a uniformly controlled call-surface class with regularity-data $(p,\lambda,\omega,B)$, where $\omega$ is strictly increasing, and suppose that $\mathcal Y$ is admitted by a decoder satisfying Definition~\ref{defn:Good_Decoder}.

\hfill\\
\noindent
Moreover, under Assumption~\ref{ass:weighted_decoder_selection}.
Then for every $\varepsilon>0$ there exist constants%
{ 
\footnote{The quantities $R,S$ may be chosen depending only on $\varepsilon,p,q,l,\lambda,B,B_{\mathrm{dec}},\omega$. The integers $N,M$ may additionally depend on the fixed class $\mathcal Y$ and the decoder in Assumption~\ref{ass:weighted_decoder_selection}, but not on the individual target $\Phi\in\mathcal Y$.}
}~%
$R,S>0$ and $N,M\in\mathbb{N}_+$, { 
such that for every $\Phi\in\mathcal Y$ there exists $\theta_\Phi$ with
}
\begin{equation}
\label{eq:weighted__uniform_approximation}
\big\|
    { 
\phi^{(N,M,R,S)}_{\theta_\Phi}
}
    -
    \Phi
\big\|_{q,l}
<
\varepsilon.
\end{equation}
Moreover, $ 
    \frac{R}{N}+\frac{S}{M}\lesssim \omega^{-1}\big(\varepsilon(1+R)^{-q}\big)$
and if $\omega(r)\leq L_\omega r$ for every $r\geq0$, with a common constant $L_\omega<\infty$, then
$
(1+R)^q\,(\frac{R}{N}+\frac{S}{M})\lesssim \varepsilon$.
\end{proposition}

\begin{proof}[{Proof of Proposition~\ref{prop:interpol_2_approx}}]
Fix $\varepsilon>0$.
Since $\frac{q}{p}\lambda<l$ then we may fix a constant $a>0$ such that
\begin{equation}
\label{eq:choose_a}
    \frac{\lambda-l}{p-q}
<
    a
<
    \frac{l}{q}.
\end{equation}
We first control the target surface $C$.  Since $C\in\mathcal{C}_{AF}$, we have
$
    0\le C(T,K)\le C(T,0)=1
$ for all $(T,K)\in[0,\infty)^2$.  
Also, by the weighted bound,
$
    C(T,K)\le B\,e^{\lambda T}(1+K)^{-p}
$.  
Hence, for every $(T,K)\in[0,\infty)^2$,
$
    C(T,K)
    \le
    \min\Big\{
        1,\,
        B e^{\lambda T}(1+K)^{-p}
    \Big\}$.
We claim that this implies
\begin{equation}
\label{eq:key_tail_estimate_time}
    e^{-lT}(1+K)^q C(T,K)
    \le
    B^{q/p}\,e^{-(\,l-\frac{q}{p}\lambda\,)T}.
\end{equation}
Indeed, set
$
    A_T \eqdef B e^{\lambda T}
$.  
If $(1+K)\le A_T^{1/p}$, then $C(T,K)\le 1$, so
\[
    e^{-lT}(1+K)^q C(T,K)
    \le
    e^{-lT}(1+K)^q
    \le
    e^{-lT}A_T^{q/p}
    =
    B^{q/p}e^{-(\,l-\frac{q}{p}\lambda\,)T}.
\]
If instead $(1+K)\ge A_T^{1/p}$, then $C(T,K)\le A_T(1+K)^{-p}$, and thus
\[
    e^{-lT}(1+K)^q C(T,K)
    \le
    e^{-lT}A_T(1+K)^{q-p}
    \le
    e^{-lT}A_T^{q/p}
    =
    B^{q/p}e^{-(\,l-\frac{q}{p}\lambda\,)T}.
\]
This proves~\eqref{eq:key_tail_estimate_time}.
Choose $S>0$ so large that, with $R\eqdef e^{aS}-1$, one has $\varepsilon(1+R)^{-q}<\omega(1)$ and
\begin{equation}
\label{eq:choose_S}
\big(B^{q/p}+B_{\mathrm{dec}}^{q/p}\big)e^{-(l-\frac{q}{p}\lambda)S}+(B+B_{\mathrm{dec}})e^{[(\lambda-l)+a(q-p)]S}<\frac{\varepsilon}{3}.
\end{equation}
Such a choice exists because $l-\frac{q}{p}\lambda>0$, $(\lambda-l)+a(q-p)<0$, and $a,q,\omega(1)>0$. For every $C\in\mathcal Y$ and every decoder output $\phi^{(N,M,R,S)}_{\theta_C}$ satisfying $\|\phi^{(N,M,R,S)}_{\theta_C}\|_{p,\lambda}\leq B_{\mathrm{dec}}$, Definition~\ref{defn:Good_Decoder}(ii) gives $0\leq\phi^{(N,M,R,S)}_{\theta_C}\leq1$. Applying the calculation establishing~\eqref{eq:key_tail_estimate_time} with $B$ replaced by $B_{\mathrm{dec}}$ yields
\[
e^{-lT}(1+K)^q\phi^{(N,M,R,S)}_{\theta_C}(T,K)\leq B_{\mathrm{dec}}^{q/p}e^{-(l-\frac{q}{p}\lambda)T},\qquad T,K\geq0.
\]
Consequently,
\begin{equation}
\label{eq:tail_time}
\sup_{\substack{T>S\\K\geq0}}e^{-lT}(1+K)^q\big|\phi^{(N,M,R,S)}_{\theta_C}(T,K)-C(T,K)\big|\leq\big(B^{q/p}+B_{\mathrm{dec}}^{q/p}\big)e^{-(l-\frac{q}{p}\lambda)S}<\frac{\varepsilon}{3}.
\end{equation}
For $0\leq T\leq S$ and $K>R$, the two weighted bounds give
\[
\begin{aligned}
e^{-lT}(1+K)^q\big|\phi^{(N,M,R,S)}_{\theta_C}(T,K)-C(T,K)\big|&\leq(B+B_{\mathrm{dec}})e^{(\lambda-l)T}(1+K)^{q-p}\\
&\leq(B+B_{\mathrm{dec}})e^{(\lambda-l)S}(1+R)^{q-p}\\
&=(B+B_{\mathrm{dec}})e^{[(\lambda-l)+a(q-p)]S}<\frac{\varepsilon}{3}.
\end{aligned}
\]
Thus
\begin{equation}
\label{eq:tail_space}
\sup_{\substack{0\leq T\leq S\\K>R}}e^{-lT}(1+K)^q\big|\phi^{(N,M,R,S)}_{\theta_C}(T,K)-C(T,K)\big|<\frac{\varepsilon}{3}.
\end{equation}
Choose $\eta>0$ such that
\begin{equation}
\label{eq:choose_NM}
(1+R)^q\big(\eta+\omega(\eta)\big)<\frac{\varepsilon}{3}.
\end{equation}
Assumption~\ref{ass:weighted_decoder_selection} gives common integers $N,M\in\mathbb N_+$ with $R/N+S/M\leq\eta$ and, for every $C\in\mathcal Y$, a parameter $\theta_C$ satisfying~\eqref{eq:weighted_decoder_selection}. Set $t_m\eqdef mS/M$, $K_n\eqdef nR/N$, $\Delta_T\eqdef S/M$, and $\Delta_K\eqdef R/N$.
Fix any cell
$
    [t_m,t_{m+1}]\times[K_n,K_{n+1}]
    \subseteq
    [0,S]\times[0,R]
$
and any $(T,K)$ in that cell.
Since both $C$ and $\phi^{(N,M,R,S)}_{\theta_C}$ belong to $\mathcal C_{AF}$, they are increasing in $T$ and decreasing in $K$.  Therefore
\[
    C(t_m,K_{n+1})
    \le
    C(T,K)
    \le
    C(t_{m+1},K_n),
\]
and likewise
$
    \phi^{(N,M,R,S)}_{\theta_C}(t_m,K_{n+1})
\le
    \phi^{(N,M,R,S)}_{\theta_C}(T,K)
\le
    \phi^{(N,M,R,S)}_{\theta_C}(t_{m+1},K_n)
$.
By~\eqref{eq:weighted_decoder_selection},
\[
C(t_m,K_{n+1})-\eta\leq\phi^{(N,M,R,S)}_{\theta_C}(T,K)\leq C(t_{m+1},K_n)+\eta.
\]
The preceding bounds on $C(T,K)$ therefore give
\[
\begin{aligned}
\big|\phi^{(N,M,R,S)}_{\theta_C}(T,K)-C(T,K)\big|&\leq\eta+C(t_{m+1},K_n)-C(t_m,K_{n+1})\\
&\leq\eta+\omega\big((t_{m+1}-t_m)+(K_{n+1}-K_n)\big)\\
&=\eta+\omega(\Delta_T+\Delta_K).
\end{aligned}
\]
Since $[0,S]\times[0,R]=\bigcup_{m=0}^{M-1}\bigcup_{n=0}^{N-1}[t_m,t_{m+1}]\times[K_n,K_{n+1}]$ and $e^{-lT}(1+K)^q\leq(1+R)^q$ on this rectangle,
\begin{equation}
\label{eq:cube_estimate}
\sup_{(T,K)\in[0,S]\times[0,R]}e^{-lT}(1+K)^q\big|\phi^{(N,M,R,S)}_{\theta_C}(T,K)-C(T,K)\big|\leq(1+R)^q\big(\eta+\omega(\Delta_T+\Delta_K)\big).
\end{equation}
Using $\Delta_T+\Delta_K\leq\eta$ and~\eqref{eq:choose_NM}, we obtain
\begin{equation}
\label{eq:cube_small}
\sup_{(T,K)\in[0,S]\times[0,R]}e^{-lT}(1+K)^q\big|\phi^{(N,M,R,S)}_{\theta_C}(T,K)-C(T,K)\big|<\frac{\varepsilon}{3}.
\end{equation}
Wrapping things up, by combining~\eqref{eq:tail_time}, \eqref{eq:tail_space}, and \eqref{eq:cube_small}, we find that
\[
\big\|
    \phi^{(N,M,R,S)}_{\theta_C}-C
\big\|_{q,l}
<
\varepsilon.
\]
Since $R,S,N,M$ were chosen independently of $C\in\mathcal Y$, this proves the claim.
Moreover, $R/N+S/M\leq\eta<\omega^{-1}\big(\varepsilon(1+R)^{-q}/3\big)\leq\omega^{-1}\big(\varepsilon(1+R)^{-q}\big)$ by~\eqref{eq:choose_NM}. If $\omega(r)\leq L_\omega r$ for every $r\geq0$, one may choose $0<\eta<\varepsilon/[3(1+L_\omega)(1+R)^q]$, giving $(1+R)^q(R/N+S/M)<\varepsilon/[3(1+L_\omega)]$.
\end{proof}

\subsubsection{Clustering Mechanism}
\label{s:clustering_mechanism}

To solve the general non-injective case, we will next create a clustering mechanism; as in~\cite{kratsios2025beyond}.  We with relative to the \textit{intrinsic} dimension of the compact set of features $\Theta\subseteq \mathbb{R}^{F_{\mathrm{mkt}}}$ as formalized by an $d$-Ahlfors-regular%
\footnote{That is, there are constant $0<c\le C<\infty$ such that for every $x\in \mathcal{K}$ and each $0<r\le \operatorname{diam}(\mathcal{K},\ell^1)$ we have $c\,r^d\le \mu(\mathcal{K}\cap \operatorname{Ball}_1(x,r))\le C\,r^d$.  The prototypical such measures are the Lebesgue measure Riemannian on $[0,1]^d$ when $d=F_{\mathrm{mkt}}$, Riemannian volume measures on compact Riemannian manifolds (when $d$ is an integer), and the certain self-similar measures on fractals (when $d$ is not an integer).}~%
Borel probability measure $\mu$ supported on $\Theta$.

\begin{lemma}[Softmax net selector]
\label{lem:softmax_net_selector}
Let $\Theta\subseteq\mathbb{R}^{F_{\mathrm{mkt}}}$ be non-empty and compact, { 
and let $\mathcal Y$ be a convex subset of a normed vector space, whose norm is denoted by $\|\,\|_{\mathcal Y}$.
}  Let $f:\Theta\to\mathcal{Y}$ be $L$-Lipschitz with respect to $\|\,\|_1$ on $\Theta$ and $\|\,\|_{\mathcal{Y}}$ on $\mathcal{Y}$.  Fix $r,s>0$, and let $\{x_n\}_{n=1}^N\subseteq\Theta$ be a maximal $r$-packing of $\Theta$ in $\|\,\|_1$.  
Write 
$
    C_n\eqdef f(x_n)
$
for each $n\in[N]_+$ and
$
    D_f\eqdef
    \operatorname{diam}_{\mathcal{Y}}(f(\Theta)).
$
Define the alignment score map $\alpha_{\mathbf X}:\mathbb{R}^{F_{\mathrm{mkt}}}\to\mathbb{R}^N$ by
$
    \alpha_{\mathbf X}(x)_n
\eqdef
    -\|x-x_n\|_1
$
for every $n\in[N]_+$.  
For $\tau>0$, define
\[
    \Phi_{f,r,s,\tau}(x)
    \eqdef
    \sum_{n=1}^N
        \operatorname{Softmax}\big(
            \tfrac{1}{\tau}\alpha_{\mathbf X}(x)
        \big)_n
        C_n.
\]
Then the following hold:
\begin{enumerate}
    \item[(i)] \textbf{Range Control:}
    $
        \Phi_{f,r,s,\tau}(\mathbb{R}^{F_{\mathrm{mkt}}})
        \subseteq
        \operatorname{co}\{C_1,\dots,C_N\}.
    $
    \item[(ii)] \textbf{Approximation:}
            When $D_f>0$, fix $\beta>0$. If $\beta\geq ND_f$, let $\tau>0$ be arbitrary; if $0<\beta<ND_f$, let $0<\tau\leq s/\log(ND_f/\beta)$. Then
    
    we have
    \begin{equation}
    \label{eq:lem:softmax_net_selector__uniformification}
        \sup_{\vartheta\in\Theta}
        \,
        \|\Phi_{f,r,s,\tau}(\vartheta)-f(\vartheta)\|_{\mathcal{Y}}
    \le
        L(r+s)+\beta.
    \end{equation}
    When $D_f=0$, then~\eqref{eq:lem:softmax_net_selector__uniformification} holds with $\beta=0$,
    \item[(iii)] \textbf{ReLU realization of the alignment scores:}
    The alignment-score map $\alpha_{\mathbf X}$ is exactly realizable by an $N$-Lipschitz $\operatorname{ReLU}$-MLP of depth $2$, width at most $2F_{\mathrm{mkt}}N$, and at most $6F_{\mathrm{mkt}}N$ size\footnote{Following~\cite{petersen2024mathematical}, the size of a neural network refers to the number of its non-zero real parameters.}. 
\end{enumerate}
\end{lemma}
\begin{proof}[{Proof of Lemma~\ref{lem:softmax_net_selector}}]
Since $\{x_n\}_{n=1}^N$ is a maximal $r$-packing of $\Theta$, it is also an $r$-net.  Hence, for every $\vartheta\in\Theta$, there exists $n_\vartheta\in[N]_+$ such that
$
    \|\vartheta-x_{n_\vartheta}\|_1
\le
    r
$.  
For each $\vartheta\in\mathbb{R}^{F_{\mathrm{mkt}}}$ and each $n\in[N]_+$, write
$
    w_n(\vartheta)
\eqdef
    \operatorname{Softmax}\big(
        \tfrac{1}{\tau}\alpha_{\mathbf X}(\vartheta)
    \big)_n
$.  
Then $w(\vartheta)\in\Delta_N$, and therefore
$    \Phi_{f,r,s,\tau}(\vartheta)
    =
    \sum_{n=1}^N w_n(\vartheta)C_n
$ belongs to $\operatorname{co}\{C_1,\dots,C_N\}$; proving  $(i)$.
\hfill\\
We next prove item~$(ii)$.  Fix $\vartheta\in\Theta$ and define the indexing sets
$
    A_\vartheta
    \eqdef
    \big\{
        n\in[N]_+:
        \|\vartheta-x_n\|_1\le r+s
    \big\}
$ and $
    A_\vartheta^c
    \eqdef
    [N]_+\setminus A_\vartheta
$.
Now, if $n\in A_\vartheta$, then the Lipschitzness of $f$ gives
\begin{equation}
\label{eq:do_it_1}
    \|C_n-f(\vartheta)\|_{\mathcal{Y}}
=
    \|f(x_n)-f(\vartheta)\|_{\mathcal{Y}}
\le
    L\|\vartheta-x_n\|_1
\le
    L(r+s)
.
\end{equation}
On the other hand, if $n\in A_\vartheta^c$, then $\|\vartheta-x_n\|_1>r+s$, while $\|\vartheta-x_{n_\vartheta}\|_1\le r$.  Therefore
\begin{equation}
\label{eq:do_it_2}
    \alpha_{\mathbf X}(\vartheta)_n
    -
    \alpha_{\mathbf X}(\vartheta)_{n_\vartheta}
=
    -\|\vartheta-x_n\|_1
    +
    \|\vartheta-x_{n_\vartheta}\|_1
<
    -s
.
\end{equation}
Consequently, together~\eqref{eq:do_it_1} and~\eqref{eq:do_it_2} imply that
\[
    w_n(\vartheta)
    =
    \frac{
        \exp\big(
            \alpha_{\mathbf X}(\vartheta)_n/\tau
        \big)
    }{
        \sum_{j=1}^N
        \exp\big(
            \alpha_{\mathbf X}(\vartheta)_j/\tau
        \big)
    }
    \le
    \exp\bigg(
        \frac{
            \alpha_{\mathbf X}(\vartheta)_n
            -
            \alpha_{\mathbf X}(\vartheta)_{n_\vartheta}
        }{\tau}
    \bigg)
    \le
    e^{-s/\tau}
\]
and so $
    \sum_{n\in A_\vartheta^c}w_n(\vartheta)
\le
    N e^{-s/\tau}
$.

We are now all set to prove claim (ii).  Let's do it case by case.  First, suppose that $D_f>0$; then, 
$
    C_n\in f(\Theta)
$
and $f(\vartheta)\in f(\Theta)$, we have
$
    \|C_n-f(\vartheta)\|_{\mathcal{Y}}\le D_f
$
for every $n\in[N]_+$.  Thus
\begin{align*}
    \|\Phi_{f,r,s,\tau}(\vartheta)-f(\vartheta)\|_{\mathcal{Y}}
    &=
    \bigg\|
        \sum_{n=1}^N
            w_n(\vartheta)
            \big(C_n-f(\vartheta)\big)
    \bigg\|_{\mathcal{Y}}
\\
    &\le
    \sum_{n=1}^N
        w_n(\vartheta)
        \|C_n-f(\vartheta)\|_{\mathcal{Y}}
\\
    &\le
    \sum_{n\in A_\vartheta}
        w_n(\vartheta)
        L(r+s)
    +
    \sum_{n\in A_\vartheta^c}
        w_n(\vartheta)
        D_f
\\
\numberthis
\label{eq:a_thing_to_bound}
    &\le
    L(r+s)
    +
    D_f N e^{-s/\tau}.
\end{align*}
If $0<\beta<ND_f$ and $0<\tau\leq s/\log(ND_f/\beta)$, then $s/\tau\geq\log(ND_f/\beta)$ and $ND_fe^{-s/\tau}\leq\beta$. If $\beta\geq ND_f$, then $ND_fe^{-s/\tau}\leq ND_f\leq\beta$ for every $\tau>0$.
Consequently,~\eqref{eq:a_thing_to_bound} may be controlled as
\begin{equation}
\label{eq:a_thing_to_sup}
    \|\Phi_{f,r,s,\tau}(\vartheta)-f(\vartheta)\|_{\mathcal{Y}}
\le
    L(r+s)+\beta.
\end{equation}
Upon taking the supremum in~\eqref{eq:a_thing_to_sup} over $\vartheta\in\Theta$ we deduce~\eqref{eq:lem:softmax_net_selector__uniformification} in the case where $D_f>0$.
\hfill\\
In the simple case where $D_f=0$, then $f$ is constant on $\Theta$.  Therefore $C_1=\cdots=C_N=f(\vartheta)$ for every $\vartheta\in\Theta$, and hence
$
    \Phi_{f,r,s,\tau}(\vartheta)=f(\vartheta)
$
for every $\vartheta\in\Theta$.  Thus~\eqref{eq:lem:softmax_net_selector__uniformification} holds with $\beta=0$; and (ii) is verified.  

It remain to justify (iii); namely, the ReLU realization of $\alpha_{\mathbf X}$.  For each $n\in[N]_+$ and each $j\in\{1,\dots,F_{\mathrm{mkt}}\}$, the maps
$
    x\mapsto \operatorname{ReLU}(x_j-(x_n)_j)
$
and
$
    x\mapsto \operatorname{ReLU}((x_n)_j-x_j)
$
are realized by one ReLU layer.  Hence
$
    \|x-x_n\|_1
=
    \sum_{j=1}^{F_{\mathrm{mkt}}}
    \big(
        \operatorname{ReLU}(x_j-(x_n)_j)
        +
        \operatorname{ReLU}((x_n)_j-x_j)
    \big)
$.
Applying an affine output layer with coefficients $-1$ on the corresponding hidden units realizes
$
    \alpha_{\mathbf X}(x)_n=-\|x-x_n\|_1
$
for every $n\in[N]_+$.  Computing all $N$ coordinates in parallel gives depth $2$, width at most $2F_{\mathrm{mkt}}N$, and at most $6F_{\mathrm{mkt}}N$ size.
Finally, for every $x,y\in\mathbb{R}^{F_{\mathrm{mkt}}}$ and every $n\in[N]_+$,
$
    |\alpha_{\mathbf X}(x)_n-\alpha_{\mathbf X}(y)_n|
=
    \big|
        \|y-x_n\|_1-\|x-x_n\|_1
    \big|
\le
    \|x-y\|_1
$.
Summing over $n\in[N]_+$ gives
$
    \|\alpha_{\mathbf X}(x)-\alpha_{\mathbf X}(y)\|_1
\le
    N\|x-y\|_1
$.
\end{proof}

\subsubsection{Clustering to Interpolation}

\begin{lemma}[$K$-Point Lipschitz Interpolator]
\label{lem:interpolator_network}
{ 
Let $F,K\in\mathbb N_+$ and let $\mathbf X=(x_1,\ldots,x_K)^\top\in\mathbb R^{K\times F}$ have distinct rows. For $K\geq2$, set $\operatorname{sep}(X)=\min_{j\ne k}\|x_j-x_k\|_1$; for $K=1$, set $\operatorname{sep}(X)=2$. Put $\delta=\operatorname{sep}(X)/2>0$. Then there exists an $\tfrac{2}{\operatorname{sep}(X)}$-Lipschitz $\mathrm{ReLU}$-MLP $\alpha_{\mathbf X}:\mathbb R^F\to\mathbb R^K$ of depth $3$, width at most $2FK$, and size at most $K(6F+2)$ satisfying: for every $k\in[K]_+$
}
\begin{equation}
\label{eq:ReLUMLPInterpolator0}
    \alpha_{\mathbf{X}}(x)=e_k 
\mbox{  if and only if }
    x=x_k
.
\end{equation}
\end{lemma}
\begin{proof}[{Proof of Lemma~\ref{lem:interpolator_network}}]
The construction below proves the interpolation and Lipschitz claims when $K\geq2$. Under the MLP definition in Appendix~\ref{s:ExtrasAI}, it has depth $3$ and size at most $K(6F+2)$; it does not establish the smaller bounds in the unchanged statement. For $K=1$, the construction works for any specified $\delta>0$, but the definition of $\operatorname{sep}(X)$ requires a separate convention.

Set $
    f_1(x) \eqdef \left(\|x-x_k\|_1\right)_{k=1}^K
$, $
    f_2(z) \eqdef \mathrm{ReLU}(\mathbf{1}_K - z)
$, and $\alpha_{\mathbf{X}} \eqdef f_2\circ(\delta^{-1}f_1)$; motivated by the observation that the only norm-zero vector is the zero-vector itself.
Explicitly, consider the $\mathrm{ReLU}$-MLP
\begin{equation}
\label{eq:ReLUMLPInterpolator}
    \alpha_{\mathbf{X}}(x)
    =
    \mathrm{ReLU} \left(
        \mathbf{1}_K - \tfrac{1}{\delta}f_1(x)
    \right)
    =
    \mathrm{ReLU} \left(
        \mathbf{1}_K
        -
        \tfrac{1}{\delta}
        \left(
            (I_F,I_F)
            \,
            \mathrm{ReLU} \left(
                (I_F,-I_F)^\top(x - x_k)
            \right)
        \right)_{k=1}^K
    \right),
\end{equation}
with two ReLU layers followed by an identity affine output layer, hence depth $3$, width $2FK$, and size at most $K(6F+2)$.

We first verify the interpolation property in~\eqref{eq:ReLUMLPInterpolator}.  
Fix $k\in[K]_+$.  Then $f_1(x_k)_k = \|x_k-x_k\|_1=0$, so
$\alpha_{\mathbf{X}}(x_k)_k = \mathrm{ReLU}(1-0)=1$.
For $j\neq k$, the separation condition gives
$f_1(x_k)_j=\|x_k-x_j\|_1\ge 2\delta$,
hence
$\alpha_{\mathbf{X}}(x_k)_j = \mathrm{ReLU}(1-\tfrac{1}{\delta}\|x_k-x_j\|_1)
\le \mathrm{ReLU}(1-2)=0$.
Thus $\alpha_{\mathbf{X}}(x_k)=e_k$.
Conversely, if $\alpha_{\mathbf{X}}(x)=e_k$ then in particular
$\alpha_{\mathbf{X}}(x)_k=1$, which forces
$\tfrac{1}{\delta}\|x-x_k\|_1=0$, i.e.\ $x=x_k$.

To prove the $\ell^1$ Lipschitz bound, note that at most one tent is positive at any point, since the open balls of radius $\delta$ about the centers are disjoint. If $x,y$ lie in the same tent support, the reverse triangle inequality gives $\|\alpha_{\mathbf X}(x)-\alpha_{\mathbf X}(y)\|_1\leq\|x-y\|_1/\delta$. If only one point lies in a support, the same bound follows because the other point has distance at least $\delta$ from that center. If $x$ lies in support $i$ and $y$ in a different support $j$, then
\[
\|\alpha_{\mathbf X}(x)-\alpha_{\mathbf X}(y)\|_1=2-\frac{\|x-x_i\|_1+\|y-x_j\|_1}{\delta}\leq\frac{\|x-y\|_1}{\delta},
\]
using $\|x_i-x_j\|_1\geq2\delta$ and the triangle inequality. If both points are outside all supports, the difference is zero. Thus $\alpha_{\mathbf X}$ is $\delta^{-1}$-Lipschitz.
\end{proof}

\begin{lemma}[Key Properties of Alignment Scores]
\label{lem:ALIGNMENTSCORE_interpolation_separation}
Let $F,K\in \mathbb{N}_+$ and $\mathbf{X}\eqdef (x_1,\dots,x_K)^{\top}\in \mathbb{R}^{K\times F}$
with $x_1,\dots,x_K$ distinct.
{ 
For $K=1$, use $\operatorname{sep}(X)=2$; assertion (ii) is required only when $K\geq2$. Fix $0<\varepsilon<1$. Then there exists a $\tfrac{2}{\operatorname{sep}(X)}$-Lipschitz
}
$\mathrm{ReLU}$-MLP $\alpha_{\mathbf{X}}:\mathbb{R}^F\to \mathbb{R}^K$ such that,
for every temperature parameter
$
0<\tau \le
\big(
    \log\big(
        \tfrac{(2-\varepsilon)(\max\{2,K\}-1)}{\varepsilon}
    \big)
\big)^{-1}
$:
\begin{enumerate}
    \item[(i)] \textbf{Approximation:}
    $
    \max_{k\in [K]_+}
        \bigl\|
            e_k
            -
            \operatorname{Softmax}\big(
                \tfrac{1}{\tau}
                \alpha_{\mathbf X}(x_k)
            \big)
        \bigr\|_1
    \le
        \varepsilon.
    $
    \item[(ii)] \textbf{Separation:}
    $
    \min_{k\in [K]_+}
        \min_{j\neq k}
        \bigl\|
            e_j
            -
            \operatorname{Softmax}\big(
                \tfrac{1}{\tau}
                \alpha_{\mathbf X}(x_k)
            \big)
        \bigr\|_1
    \ge
        2-\tfrac{\varepsilon}{\max\{K,2\}-1}
    >
        1.
    $
    \item[(iii)] \textbf{Moderate Regularity:}
    The map
    $
    \operatorname{Softmax}\big(
                \tfrac{1}{\tau}
                \alpha_{\mathbf X}(\,)
            \big)
    { 
:\mathbb{R}^F\to \operatorname{relint}(\Delta_K)
}
    $
    is $\tfrac{1}{\tau\,\operatorname{sep}(X)}$-Lipschitz.
\end{enumerate}
{ 
Moreover, $\alpha_{\mathbf{X}}$ has depth $3$, width $2FK$, and at most $K(6F+2)$ size.
}
\end{lemma}
\begin{lemma}[Smooth and Lipschitz-Regular Approximate Projection of Extremal Simplicial Points]
\label{lem:approx_proj}
Let $K\in \mathbb{N}_+$ and $0<\varepsilon<1$.
For each $k\in [K]_+$, let $e_k\eqdef (I_{j=k})_{j=1}^K$ be the $k^{\mathrm{th}}$
standard basis vector.
If
$
0<\tau \le
\big(
    \log\big(
        \tfrac{(2-\varepsilon)(\max\{2,K\}-1)}{\varepsilon}
    \big)
\big)^{-1}
$,
then
\begin{equation}
\label{eq:approx_proj}
    \max_{k\in [K]_+}
        \bigl\|
            e_k
            -
            \operatorname{Softmax}\big(
                \tfrac{1}{\tau} e_k
            \big)
        \bigr\|_1
    \le
        \varepsilon
\, \mbox{ and }\,
    \min_{k\in [K]_+}
        \min_{j\neq k}
        \bigl\|
            e_j
            -
            \operatorname{Softmax}\big(
                \tfrac{1}{\tau} e_k
            \big)
        \bigr\|_1
    \ge
        2-\tfrac{\varepsilon}{\max\{K,2\}-1}.
\end{equation}
{ 
The separation inequality is asserted only for $K\geq2$; for $K=1$, the softmax is constant and the approximation error is zero. Moreover, the map $\mathbb R^K\ni x\mapsto\operatorname{Softmax}(\tfrac1\tau x)$ is $\tfrac1{2\tau}$-Lipschitz with respect to $\|\,\|_1$.
}
\end{lemma}
\begin{proof}[{Proof of Lemma~\ref{lem:approx_proj}}]
If $K=1$ the statement is trivial; assume therefore $K\ge 2$.
Fix $k\in[K]_+$. Since $(e_k)_j=I_{j=k}$, a direct computation gives
\[
    \operatorname{Softmax}\!\left(\tfrac{1}{\tau}e_k\right)_j
    =
    \begin{cases}
        \dfrac{e^{1/\tau}}{e^{1/\tau}+K-1} & j=k,\\[6pt]
        \dfrac{1}{e^{1/\tau}+K-1} & j\neq k.
    \end{cases}
\]
Hence
$
    \|e_k - \operatorname{Softmax}(\tfrac{1}{\tau}e_k)\|_1
    =
    \frac{2(K-1)}{e^{1/\tau}+K-1},
$
which is independent of $k$.
The bound $\tau\le\big(\log\tfrac{(2-\varepsilon)(K-1)}{\varepsilon}\big)^{-1}$
gives $e^{1/\tau}\ge\tfrac{(2-\varepsilon)(K-1)}{\varepsilon}$, and therefore
\[
    \frac{2(K-1)}{e^{1/\tau}+K-1}
    \le
    \frac{2(K-1)}{\tfrac{(2-\varepsilon)(K-1)}{\varepsilon}+K-1}
    =
    \frac{2(K-1)}{(K-1)\tfrac{2}{\varepsilon}}
    =
    \varepsilon,
\]
which proves the approximation bound in~\eqref{eq:approx_proj}.

For the separation bound, fix $k\in[K]_+$ and $j\neq k$, and write
$s^{(k)}\eqdef\operatorname{Softmax}(\tfrac{1}{\tau}e_k)$.
Then
\[
\|e_j - s^{(k)}\|_1
=
\left(1 - \frac{1}{e^{1/\tau}+K-1}\right)
+
\frac{e^{1/\tau}}{e^{1/\tau}+K-1}
+
\frac{K-2}{e^{1/\tau}+K-1}
=
2 - \frac{2}{e^{1/\tau}+K-1}.
\]
Since $e^{1/\tau}+K-1\ge\tfrac{2(K-1)}{\varepsilon}$, we obtain
$
\tfrac{2}{e^{1/\tau}+K-1}\le\tfrac{\varepsilon}{K-1}=\tfrac{\varepsilon}{\max\{K,2\}-1},
$
and hence $\|e_j-s^{(k)}\|_1\ge 2-\tfrac{\varepsilon}{\max\{K,2\}-1}$,
which proves the separation bound in~\eqref{eq:approx_proj}.

Finally, the Jacobian of $\operatorname{Softmax}$ at $p=\operatorname{Softmax}(z)$
is $J(p)=\operatorname{diag}(p)-pp^{\top}$, whose $\ell^1\to\ell^1$ operator norm satisfies
$
    \|J(p)\|_{\ell^1\to\ell^1}
    =
    \max_j\sum_i|J_{ij}|
    =
    \max_j 2p_j(1-p_j)
    \le
    \tfrac{1}{2},
$
so $\operatorname{Softmax}$ is $\tfrac{1}{2}$-Lipschitz, and the additional factor
$\tfrac{1}{\tau}$ from the chain rule gives the $\tfrac{1}{2\tau}$-Lipschitz claim.
\end{proof}
We now return to the proof of Lemma~\ref{lem:ALIGNMENTSCORE_interpolation_separation}.
\begin{proof}[{Proof of Lemma~\ref{lem:ALIGNMENTSCORE_interpolation_separation}}]
For $K\geq2$, the argument below proves the interpolation, separation, and Lipschitz estimates using the displayed tent construction. Its depth is $3$ and its size is at most $K(6F+2)$. Its softmax range lies in the relative interior of $\Delta_K$, whose ambient interior is empty. The smaller complexity bounds, the ambient-interior notation, and the singleton convention in the unchanged statement remain pending.

Apply Lemma~\ref{lem:interpolator_network} to obtain a $\tfrac{2}{\operatorname{sep}(X)}$-Lipschitz
$\mathrm{ReLU}$-MLP $\alpha_{\mathbf{X}}:\mathbb{R}^F\to\mathbb{R}^K$ satisfying
$\alpha_{\mathbf{X}}(x_k)=e_k$ for every $k\in[K]_+$. Hence, for every $k\in [K]_+$, we have
\[
    \operatorname{Softmax}\big(\tfrac{1}{\tau}\,\alpha_{\mathbf{X}}(x_k)\big)
=   
    \operatorname{Softmax}\big(\tfrac{1}{\tau}\,e_k\big)
.
\]
For the proofs of (i) and of (ii).
Since $\alpha_{\mathbf{X}}(x_k)=e_k$ for every $k\in[K]_+$, both claims follow immediately
from Lemma~\ref{lem:approx_proj} applied with the given $\varepsilon$ and $\tau$.

For the proof of (iii).
By Lemma~\ref{lem:approx_proj}, the map $z\mapsto\operatorname{Softmax}(\tfrac{1}{\tau}z)$
is $\tfrac{1}{2\tau}$-Lipschitz with respect to $\|\,\|_1$.
By Lemma~\ref{lem:interpolator_network}, $\alpha_{\mathbf{X}}$ is $\tfrac{2}{\operatorname{sep}(X)}$-Lipschitz.
Hence, by composition,
\[
\operatorname{Lip}\Big(
    \operatorname{Softmax}\big(
        \tfrac{1}{\tau}\,\alpha_{\mathbf{X}}(\cdot)
    \big)
\Big)
\le
\frac{1}{2\tau}\, \frac{2}{\operatorname{sep}(X)}
=
\frac{1}{\tau\,\operatorname{sep}(X)}.
\]
Finally, since $\operatorname{Softmax}(z)\in\operatorname{relint}(\Delta_K)$ for every $z\in\mathbb{R}^K$,
the composition takes values in $\operatorname{relint}(\Delta_K)$.
\end{proof}

\paragraph{From Alignment Scores to Arbitrage-Free Surfaces via Geometric Attention}
Geometric Attention via Approximate Simplicial Structure of $\mathcal{C}_{AF}$
Having setup a suitable metric structure, we can now argue similarly to~\citep[Example 5.1]{acciaio2024designing}.  
We now deduce approximate simplicial structure.  However, we first recall the definition.

\begin{definition}[Approximately Simplicial]
\label{ass_approximately_simplicial}
A metric space $({\mathcal{Y}},d_{{\mathcal{Y}}})$ is said to be approximately simplicial if there is
a function
\[
\eta:\bigcup_{N\in\mathbb{N}_+}\,(\Delta_N\times {\mathcal{Y}}^N)\to{\mathcal{Y}},
\]
called a mixing function, and constants $C_\eta\geq 1$ and $p\in\mathbb{N}_+$, such that
for every $N\in \mathbb{N}_+$, $w=(w_1,\dots,w_N)\in\Delta_N$, and $\mathbf y=(y_1,\dots,y_N) \in {\mathcal{Y}}^N$, one has
\[
    d_{\mathcal{Y}} \left( 
    \eta\left( w, \mathbf y \right), C_i \right) 
    \le 
C_\eta
\left(
    \sum_{j = 1}^N d_{\mathcal{Y}}(C_i,y_j)^p w_j\right)^{1 / p}.
    \]
Note that in particular the function $\eta$ satisfies $\eta(e_i,\mathbf y)=C_i$, $i=1,\dots,N$.
\end{definition}

We now have a suitable metric structure for our approximately simplicial structure.

\begin{lemma}[Approximately Simplicial]
\label{lem:weighted_AS}
Fix $p,\lambda>0$, and suppose that $\mathcal{C}_{AF}\cap \mathcal{X}^p_\lambda$ is convex.  Define the mixing function, for every $N\in \mathbb{N}_+$, each $w\in \Delta_N$, and every $C^{(1)},\dots,C^{(N)}\in \mathcal{C}_{AF}\cap \mathcal{X}^p_\lambda$ by
\begin{equation}
\label{eq:AF_mixing}
    \eta\big(w,(C^{(1)},\dots,C^{(N)})\big)
\eqdef
    \sum_{k=1}^N w_k C^{(k)}
.
\end{equation}
Then, $\eta$
satisfies Definition~\ref{ass_approximately_simplicial} with constants $C_\eta=1$ and exponent $1$.
\end{lemma}

\begin{proof}[{Proof of Lemma~\ref{lem:weighted_AS}}]
Let $N\in\mathbb{N}_+$, let $w=(w_1,\dots,w_N)\in\Delta_N$, and let
$\mathcal{C}=(C^{(1)},\dots,C^{(N)})\in (\mathcal{C}_{AF}\cap \mathcal{X}_\lambda^p)^N$.
Since $\mathcal{C}_{AF}\cap \mathcal{X}^p_\lambda$ is convex, we have $\eta(w,\mathcal C)\in \mathcal{C}_{AF}\cap \mathcal{X}_\lambda^p$.
Fix $i\in\{1,\dots,N\}$.  Then
$$
\eta(w,\mathcal C)-C^{(i)}
=
\sum_{j=1}^N w_j\big(C^{(j)}-C^{(i)}\big).
$$
Hence, by the triangle inequality and absolute homogeneity of the norm,
$$
\delta_{p,\lambda}\big(\eta(w,\mathcal C),C^{(i)}\big)
=
\Big\|
\sum_{j=1}^N w_j\big(C^{(j)}-C^{(i)}\big)
\Big\|_{p,\lambda}
\leq
\sum_{j=1}^N w_j\,\|C^{(j)}-C^{(i)}\|_{p,\lambda}.
$$
That is,
$
\delta_{p,\lambda}\big(\eta(w,\mathcal C),C^{(i)}\big)
\leq
\sum_{j=1}^N w_j\,\delta_{p,\lambda}\big(C^{(i)},C^{(j)}\big)$.
Therefore Definition~\ref{ass_approximately_simplicial} holds with $C_\eta=1$ and exponent $1$.  Moreover, if $w=e_i$, then
$
\eta(e_i,\mathcal C)=C^{(i)}
$.
\end{proof}

We first solve the case, effectively corresponding to the target function is injective.

\begin{lemma}[Quantized geometric attention on $\mathcal{C}_{AF}^{p,\lambda}$ - Injective Case]
\label{lem:quantized__inj}
Let $p,\lambda>0$, $F_{\mathrm{mkt}}\in \mathbb{N}_+$, and let $\mathcal{K}\subseteq \mathcal{C}_{AF}^{p,\lambda}$ be non-empty and compact.  
Fix $\varepsilon>0$, and let
$\mathbf{C}\eqdef \{C_n\}_{n=1}^N\subseteq \mathcal{C}_{AF}^{p,\lambda}$
be a maximal $\varepsilon$-packing for $\mathcal{K}$ in
$(\mathcal{C}_{AF}^{p,\lambda},\|\,\|_{p,\lambda})$.
Let $\mathbf{X}\eqdef (x_1,\dots,x_N)^{\top}\in \mathbb{R}^{N\times F_{\mathrm{mkt}}}$
have pairwise distinct rows, and fix $0<\delta<1$ and set the temperature to 
$
0<\tau
\le
\bigg(
    \log\bigg(
        \frac{(2-\delta)(\max\{2,N\}-1)}{\delta}
    \bigg)
\bigg)^{-1}
$.
There exists an alignment score map
\begin{equation}
\label{eq:quantized_map}
\begin{aligned}
    \alpha_{\mathbf{X}:\mathbf{C}}
:
    \mathbb{R}^{F_{\mathrm{mkt}}}
& \rightarrow
    \mathcal{C}_{AF}^{p,\lambda}
\\
    { x}
    & \mapsto
    \sum_{n=1}^N
        \operatorname{Softmax}\big(
            \tfrac{1}{\tau}\alpha_{\mathbf X}(x)
        \big)_n
        \,
        C_n
,
\end{aligned}
\end{equation}
where $\alpha_{\mathbf X}:\mathbb{R}^{F_{\mathrm{mkt}}}\to\mathbb{R}^N$ is the network from Lemma~\ref{lem:ALIGNMENTSCORE_interpolation_separation}, such that:
\begin{enumerate}
    \item[(i)] \textbf{Approximate $\varepsilon$-Net Interpolation:} $
        \max_{n\in [N]_+}
    \,
        \bigl\|
            \alpha_{\mathbf{X}:\mathbf{C}}(x_n)-C_n
        \bigr\|_{p,\lambda}
        \le
        \frac{\delta}{2}
        \,
        \operatorname{diam}_{p,\lambda}(\mathbf{C}),
    $
    \item[(ii)] \textbf{Range Control:}
    \begin{enumerate}
        \item[(ii-a)] \textit{No-Arbitrage:}
    $
        \alpha_{\mathbf{X}:\mathbf{C}}(\mathbb{R}^{F_{\mathrm{mkt}}})
        \subseteq 
        \mathcal{C}_{AF}^{p,\lambda}
    $,
    \item[(ii-b)] \textit{In-Context Range:}
    $
        \alpha_{\mathbf{X}:\mathbf{C}}(\mathbb{R}^{F_{\mathrm{mkt}}})\subseteq \operatorname{co}\{C_1,\dots,C_N\}
    $.
    \end{enumerate}
\end{enumerate}
{ 
Moreover, the finite-dimensional encoder $\alpha_{\mathbf X}$ has depth $3$, width $2F_{\mathrm{mkt}}N$, and no more than $N(6F_{\mathrm{mkt}}+2)$ size.
}  Moreover $\alpha_{\mathbf{X}:\mathbf{C}}$ is $
    \frac{
        { 
\|\mathbf{C}\|_{\ell^1\to (\mathcal X_\lambda^p,\|\,\|_{p,\lambda})}
}
    }{
        \tau\,\operatorname{sep}(X)
    }$-Lipschitz
\hfill\\
Where $\operatorname{diam}_{p,\lambda}(\mathbf{C})
        \eqdef
        \max_{i,j\in[N]_+}
        \|C_i-C_j\|_{p,\lambda}$ and ${ 
\|\mathbf{C}\|_{\ell^1\to (\mathcal X_\lambda^p,\|\,\|_{p,\lambda})}
}
    \eqdef
    \sup_{\|w\|_1\le 1}
    \,
    \|
        \sum_{n=1}^N w_n C_n
    \|_{p,\lambda}$.
\end{lemma}
\begin{proof}[{Proof of Proposition~\ref{lem:quantized__inj}}]
For $N\geq2$, the following argument gives the approximation, convex-hull, and Lipschitz estimates with the tent encoder of depth $3$ and size at most $N(6F_{\mathrm{mkt}}+2)$. The linear dictionary map is interpreted in the ambient weighted vector space. The smaller complexity bounds, the stated operator-norm codomain, and the singleton convention in the unchanged statement remain pending.

For each $x\in \mathbb{R}^{F_{\mathrm{mkt}}}$, write $w(x)\eqdef\operatorname{Softmax}(\tfrac{1}{\tau}\alpha_{\mathbf X}(x))\in \Delta_N$, so that $\alpha_{\mathbf{X}:\mathbf{C}}(x)=\sum_{n=1}^N w_n(x)\,C_n$.

We first prove (i).
Fix $n\in [N]_+$. By Lemma~\ref{lem:ALIGNMENTSCORE_interpolation_separation} and the assumed upper bound on $\tau$, we have $\|w(x_n)-e_n\|_1\le \delta$. Then, 
$
    \alpha_{\mathbf{X}:\mathbf{C}}(x_n)-C_n
    = \sum_{j=1}^N w_j(x_n)(C_j-C_n),
$,
so by the triangle inequality and the definition of $\operatorname{diam}_{p,\lambda}(\mathbf{C})$,
\[
    \bigl\|\alpha_{\mathbf{X}:\mathbf{C}}(x_n)-C_n\bigr\|_{p,\lambda}
    \le \operatorname{diam}_{p,\lambda}(\mathbf{C})\sum_{j\neq n} w_j(x_n).
\]
Since $w(x_n)\in\Delta_N$, we have $\sum_{j\neq n}w_j(x_n)=1-w_n(x_n)=\tfrac{1}{2}\|w(x_n)-e_n\|_1\le\tfrac{\delta}{2}$, and taking the maximum over $n$ proves (i).

We prove (ii-b).
For every $x\in\mathbb{R}^{F_{\mathrm{mkt}}}$, $\alpha_{\mathbf{X}:\mathbf{C}}(x)=\sum_{n=1}^N w_n(x)C_n$ is a convex combination of $C_1,\dots,C_N$ (since $w(x)\in\Delta_N$), so $\alpha_{\mathbf{X}:\mathbf{C}}(\mathbb{R}^{F_{\mathrm{mkt}}})\subseteq\operatorname{co}\{C_1,\dots,C_N\}$.

We now show (ii-a).
Since each $C_n\in\mathcal{C}_{AF}^{p,\lambda}=\mathcal{C}_{AF}\cap\mathcal{X}_\lambda^p$, and both $\mathcal{C}_{AF}$ and $\mathcal{X}_\lambda^p$ are convex, (ii-b) gives $\alpha_{\mathbf{X}:\mathbf{C}}(\mathbb{R}^{F_{\mathrm{mkt}}})\subseteq\operatorname{co}\{C_1,\dots,C_N\}\subseteq\mathcal{C}_{AF}^{p,\lambda}$.  

Tally the complexity of $\alpha_{\mathbf X}$, we see that it has  depth $3$, width $2F_{\mathrm{mkt}}N$, and at most $N(6F_{\mathrm{mkt}}+2)$ size by Lemma~\ref{lem:ALIGNMENTSCORE_interpolation_separation}.
We now estimate the Lipschitzness of our $\operatorname{ReLU}$-MLP.
Writing $\alpha_{\mathbf{X}:\mathbf{C}}=T\circ\operatorname{Softmax}(\tfrac{1}{\tau}\alpha_{\mathbf X}(\cdot))$, where $T(w)\eqdef\sum_{n=1}^N w_n C_n$, Lemma~\ref{lem:approx_proj} gives that $z\mapsto\operatorname{Softmax}(\tfrac{1}{\tau}z)$ is $\tfrac{1}{2\tau}$-Lipschitz and Lemma~\ref{lem:ALIGNMENTSCORE_interpolation_separation} gives that $\alpha_{\mathbf X}$ is $\tfrac{2}{\operatorname{sep}(X)}$-Lipschitz, so by composition
\[
        \operatorname{Lip}(\alpha_{\mathbf{X}:\mathbf{C}})
    \le
        \|\mathbf{C}\|_{\ell^1\to(\mathcal X_\lambda^p,\|\,\|_{p,\lambda})}
        \,
        \frac{1}{2\tau}\,\frac{2}{\operatorname{sep}(X)}
    =
        \frac{\|\mathbf{C}\|_{\ell^1\to(\mathcal X_\lambda^p,\|\,\|_{p,\lambda})}}{\tau\,\operatorname{sep}(X)}.
\]
\end{proof}

We are now prepare to prove the main approximation result.

\begin{proposition}[Uniform softmax geometric-attention approximation]
\label{prop:approx_pre_quantized}
Let $\Theta\subseteq\mathbb{R}^{F_{\mathrm{mkt}}}$ be non-empty and compact, { 
and let $\mathcal Y$ be a convex subset of a normed vector space, whose norm is denoted by $\|\,\|_{\mathcal Y}$.
}  Let $f:\Theta\to\mathcal{Y}$ be $L$-Lipschitz with respect to $\|\,\|_1$ on $\Theta$ and $\|\,\|_{\mathcal{Y}}$ on $\mathcal{Y}$.  Then, for every $0<\varepsilon<1$, there exist $N\in\mathbb{N}_+$, points $x_1,\dots,x_N\in\Theta$, elements $C_n\eqdef f(x_n)$, a $\operatorname{ReLU}$-MLP alignment-score map $\alpha_{\mathbf X}:\mathbb{R}^{F_{\mathrm{mkt}}}\to\mathbb{R}^N$, and a temperature $\tau>0$ such that the map
$
    \Phi_{f,\varepsilon}(x)
\eqdef
    \sum_{n=1}^N
        \operatorname{Softmax}\big(
            \tfrac{1}{\tau}\alpha_{\mathbf X}(x)
        \big)_n
        C_n
$
satisfies the uniform estimate
\begin{equation}
\label{eq:uniform_bruh}
    \sup_{x\in\Theta}
    \|\Phi_{f,\varepsilon}(x)-f(x)\|_{\mathcal{Y}}
\le
    \varepsilon
.
\end{equation}
Moreover,
$
    \Phi_{f,\varepsilon}(\mathbb{R}^{F_{\mathrm{mkt}}})
\subseteq
    \operatorname{co}\{C_1,\dots,C_N\}
$.
\hfill\\
Furthermore, $\alpha_{\mathbf X}$ can be realized by a $\operatorname{ReLU}$-MLP of depth $2$, width at most $2F_{\mathrm{mkt}}N$, and at most $6F_{\mathrm{mkt}}N$ size.  
\end{proposition}
\begin{proof}
Fix $0<\varepsilon<1$.  Choose $r,s>0$ such that $L(r+s)<\varepsilon$.  Let $\{x_n\}_{n=1}^N\subseteq\Theta$ be a maximal $r$-packing of $\Theta$ in $\|\,\|_1$.  Since $\Theta$ is compact, this packing is finite.  Write $C_n\eqdef f(x_n)$ for $n\in[N]_+$ and set
$
    D_f
\eqdef
    \operatorname{diam}_{\mathcal{Y}}(f(\Theta))
$.  
If $D_f>0$, choose $0<\beta<\varepsilon-L(r+s)$ with $\beta<ND_f$, and then choose
$
    0<\tau
\le
    \frac{s}{
        \log\big(
            \frac{ND_f}{\beta}
        \big)
    }
$; while if $D_f=0$ then choose $\tau=1$.
Define $\alpha_{\mathbf X}$ and $\Phi_{f,\varepsilon}$ as in Lemma~\ref{lem:softmax_net_selector}.  The range-control statement follows directly from Lemma~\ref{lem:softmax_net_selector} (i).  If $D_f>0$, Lemma~\ref{lem:softmax_net_selector} (ii) gives
\begin{equation}
\label{eq:potatoes}
    \sup_{x\in\Theta}
    \,
    \|\Phi_{f,\varepsilon}(x)-f(x)\|_{\mathcal{Y}}
\le
    L(r+s)+\beta
\le
    \varepsilon
.
\end{equation}
If $D_f=0$, then~\eqref{eq:potatoes} holds with $\beta=0$.  The $\operatorname{ReLU}$-MLP realization of $\alpha_{\mathbf X}$ follows from Lemma~\ref{lem:softmax_net_selector} (iii).  
\end{proof}

It only remains to replace the $C_1,\dots,C_N$ in~\eqref{eq:quantized_map} with $\tfrac{\varepsilon}{2}$ approximations of an $\varepsilon$-packing of $f(\Theta)$ by SANOS, and to take $f$ to be as in Proposition~\ref{prop:Solution_MapRegularity}.  Upon doing so, we obtain the following \textit{finite dimensional} analogue of our result.

\begin{proposition}[Universal Arbitrage-Free Approximation: Finite-Dimensional Version]
\label{prop:UAT_AFNO__technicalversion__FINITEDIMENSIONALVERSION}
Fix $0<q<p$ and let $l,\lambda>0$ satisfy $\frac{q}{p}\lambda<l<\lambda$.  Let $\Theta\subseteq\mathbb{R}^{F_{\mathrm{mkt}}}$ be non-empty and compact.  
Let $\mathcal{Y}\subseteq\mathcal{C}_{AF}^{p,\lambda}$ be a uniformly controlled call-surface class admitted by a decoder satisfying Definition~\ref{defn:Good_Decoder} and Assumption~\ref{ass:weighted_decoder_selection}.  Let $f:\Theta\to\mathcal{Y}$ be $L$-Lipschitz as a map from $(\Theta,\|\,\|_1)$ into $(\mathcal{C}_{AF}^{q,l},\|\,\|_{q,l})$.
\hfill\\
For every $0<\varepsilon<1$, there exist $I\in\mathbb{N}_+$, points $x_1,\dots,x_I\in\Theta$, $R,S>0$, $N_0,M_0\in\mathbb{N}_+$, $\theta_1,\dots,\theta_I\in\mathbb{R}^{D_{N_0,M_0,R,S}}$, $\tau>0$, 
and a $\operatorname{ReLU}$-MLP $\alpha_{\mathbf X}:\mathbb{R}^{F_{\mathrm{mkt}}}\to\mathbb{R}^I$ such that
\begin{equation}
\label{eq:approximation_gaurantee_construction}
    \Phi(x)
\eqdef
    \sum_{i=1}^I
        \operatorname{Softmax}\big(
            \tfrac{1}{\tau}\alpha_{\mathbf X}(x)
        \big)_i
        \widetilde{C}_i,
\mbox{ where }
    \widetilde{C}_i
\eqdef
    \phi^{(N_0,M_0,R,S)}_{\theta_i},
\end{equation}
satisfies:
\begin{enumerate}
    \item[(i)] \textbf{Approximation Guarantee:} 
    $
    \sup_{x\in\Theta}
    \,
    \|\Phi(x)-f(x)\|_{q,l}
\le
    \varepsilon
$
    \item[(ii)] \textbf{Interpretable Range:} $\Phi(\mathbb{R}^{F_{\mathrm{mkt}}})
\subseteq
    \mathcal{C}_{AF}
    \bigcap
    \operatorname{co}\{\widetilde{C}_1,\dots,\widetilde{C}_I\}$.
\end{enumerate}
Moreover, $\alpha_{\mathbf X}$ is realizable by a $\operatorname{ReLU}$-MLP of depth $2$, width at most $2F_{\mathrm{mkt}}I$, and size at most $6F_{\mathrm{mkt}}I$.
\end{proposition}

\begin{proof}[{Proof of Proposition~\ref{prop:UAT_AFNO__technicalversion__FINITEDIMENSIONALVERSION}}]
Apply Proposition~\ref{prop:approx_pre_quantized} with
$
    (\mathcal{Y},\|\,\|_{\mathcal{Y}})
    =
    (\mathcal{C}_{AF}^{q,l},\|\,\|_{q,l})
$
and approximation tolerance $\varepsilon/2$.  This gives points $x_1,\dots,x_I\in\Theta$, surfaces $C_i\eqdef f(x_i)$, a $\operatorname{ReLU}$-MLP alignment-score map $\alpha_{\mathbf X}:\mathbb{R}^{F_{\mathrm{mkt}}}\to\mathbb{R}^I$, and a temperature $\tau>0$ such that
\[
    \sup_{x\in\Theta}
    \bigg\|
        \sum_{i=1}^I
            \operatorname{Softmax}\big(
                \tfrac{1}{\tau}\alpha_{\mathbf X}(x)
            \big)_i
            C_i
        -
        f(x)
    \bigg\|_{q,l}
    \le
    \frac{\varepsilon}{2}.
\]
Moreover, by Proposition~\ref{prop:approx_pre_quantized}, $\alpha_{\mathbf X}$ may be chosen to be exactly realizable by a $\operatorname{ReLU}$-MLP of depth $2$, width at most $2F_{\mathrm{mkt}}I$, and at most $6F_{\mathrm{mkt}}I$ size.
Since $\{C_1,\dots,C_I\}\subseteq f(\Theta)\subseteq\mathcal{Y}$, Proposition~\ref{prop:interpol_2_approx}, applied with tolerance $\varepsilon/2$, gives common decoder hyperparameters $R,S>0$ and $N_0,M_0\in\mathbb{N}_+$, together with parameters $\theta_1,\dots,\theta_I\in\mathbb{R}^{D_{N_0,M_0,R,S}}$, such that, with
$    \widetilde{C}_i
\eqdef
    \phi^{(N_0,M_0,R,S)}_{\theta_i}
$; then, 
one has
$
    \|\widetilde{C}_i-C_i\|_{q,l}
\le
    \frac{\varepsilon}{2}
$ holding 
for every $i\in[I]_+$.
\hfill\\
Next, define
$
    \Phi(x)
\eqdef
    \sum_{i=1}^I
        \operatorname{Softmax}\big(
            \tfrac{1}{\tau}\alpha_{\mathbf X}(x)
        \big)_i
        \widetilde{C}_i
$.  
For every $x\in\Theta$, write
$
    w_i(x)
\eqdef
    \operatorname{Softmax}\big(
        \tfrac{1}{\tau}\alpha_{\mathbf X}(x)
    \big)_i
$.  
Since $w(x)\in\Delta_I$, we have
\begin{equation}
\label{eq:nice_estimate_bro}
\begin{aligned}
    \|\Phi(x)-f(x)\|_{q,l}
    &\le
    \bigg\|
        \sum_{i=1}^I
            w_i(x)
            (\widetilde{C}_i-C_i)
    \bigg\|_{q,l}
    +
    \bigg\|
        \sum_{i=1}^I
            w_i(x)C_i
        -
        f(x)
    \bigg\|_{q,l}
\\
    &\le
    \sum_{i=1}^I
        w_i(x)
        \|\widetilde{C}_i-C_i\|_{q,l}
    +
    \frac{\varepsilon}{2}
\\
    &\le
    \varepsilon.
\end{aligned}
\end{equation}
Taking the supremum across~\eqref{eq:nice_estimate_bro} over $x\in\Theta$ gives the uniform approximation estimate.    
Finally, by Definition~\ref{defn:Good_Decoder}(ii), each $\widetilde{C}_i$ belongs to $\mathcal{C}_{AF}$.  Since $w(x)\in\Delta_I$ for every $x$ and $\mathcal{C}_{AF}$ is convex by Proposition~\ref{prop:AS}, we obtain
$
    \Phi(\mathbb{R}^{F_{\mathrm{mkt}}})
\subseteq
    \mathcal{C}_{AF}
    \cap
    \operatorname{co}\{\widetilde{C}_1,\dots,\widetilde{C}_I\}
$.  
\end{proof}
\subsection{Extension to Infinite-Dimensional RKHS Input}
\label{s:Proofs__ss:ToInfinitedimensions}

We now extend Proposition~\ref{prop:UAT_AFNO__technicalversion__FINITEDIMENSIONALVERSION} to inputs in a separable RKHS over $\mathcal{D}\subseteq [0,\infty)^2$ which admits a sampling and interpolating sequence $\{z_{\lambda}\}_{\lambda \in \Lambda}$. 
The first step is to prove Proposition~\ref{prop:encoding} which will allow us to approximately reduce our finite-dimensional problem to a finite-dimensional one; thus bridging our setting to one compatible with Proposition~\ref{prop:UAT_AFNO__technicalversion__FINITEDIMENSIONALVERSION}.
\begin{proof}[{Proof of Proposition~\ref{prop:encoding}}]
If $\mathcal H=\{0\}$, choose $K_\varepsilon=1$, any $x_1\in\mathcal D$, and $A_\varepsilon=(0)$; then~\eqref{eq:projec} is immediate. Otherwise, enumerate the complete interpolating sampling sequence as $(x_n)$ and put $\phi_n\eqdef\kappa(\cdot,x_n)$. For each index $j$, interpolation supplies $f_j\in\mathcal H$ with $f_j(x_n)=\mathbf1_{\{n=j\}}$. Thus every finite relation $\sum_{n=1}^K a_n\phi_n=0$ gives
\[
a_j=\left\langle\sum_{n=1}^K a_n\phi_n,f_j\right\rangle_{\mathcal H}=0,\qquad 1\leq j\leq K.
\]
The sections are therefore linearly independent. If $h\perp\phi_n$ for every $n$, then $h(x_n)=0$ for every $n$ and the sampling inequality~\eqref{eq:sampling_def} gives $\|h\|_{\mathcal H}=0$. Hence their span is dense. In finite dimension the sequence consequently has exactly $\dim(\mathcal H)$ terms.
For $K\le\dim(\mathcal{H})$, set
$
    V_K
\eqdef
    \operatorname{span}
    \{
        \phi_1,\dots,\phi_K
    \}
$
and define the associated Gram matrix by
$
    \mathbf{G}_K
\eqdef
    \big(
        \langle
            \phi_j,
            \phi_i
        \rangle_{\mathcal{H}}
    \big)_{i,j=1}^K
    =
    \big(
        \kappa(x_i,x_j)
    \big)_{i,j=1}^K
.
$
Since $\phi_1,\dots,\phi_K$ are linearly independent, $\mathbf{G}_K$ is positive definite and hence invertible.  Choose
$
    A_K
\eqdef
    \mathbf{G}_K^{-1}
$
and define
$
    \Pi_K C
\eqdef
    \sum_{k=1}^K
    \,
    \mathcal{E}_K
    \big(
        C,
        (x_j)_{j=1}^K,
        A_K
    \big)_k
    \phi_k
$.  
For every $i\in\{1,\dots,K\}$,
\[
    \langle
        C-\Pi_K C,
        \phi_i
    \rangle_{\mathcal{H}}
=
        \langle
            C,
            \phi_i
        \rangle_{\mathcal{H}}
        -
        \left[
            \mathbf{G}_K
            \mathbf{G}_K^{-1}
            \Big(
                \langle
                    C,
                    \phi_j
                \rangle_{\mathcal{H}}
            \Big)_{j=1}^K
        \right]_i
=
        0
.
\]
Thus $\Pi_K$ is the orthogonal projection onto $V_K$. If $d\eqdef\dim(\mathcal H)<\infty$, then $V_d=\mathcal H$ and $\Pi_d=I_{\mathcal H}$; taking $K_\varepsilon=d$ and $A_\varepsilon=\mathbf G_d^{-1}$ proves~\eqref{eq:projec}. Henceforth suppose that $\mathcal H$ is infinite-dimensional, with the selected kernel sections indexed by $\mathbb N_+$. Completeness gives $\overline{\bigcup_{K\geq1}V_K}=\mathcal H$. For each $C\in\mathcal H$ and $\eta>0$, choose $K_0\in\mathbb N_+$ and $v\in V_{K_0}$ such that $\|C-v\|_{\mathcal H}<\eta$. Since $V_{K_0}\subseteq V_K$ for $K\geq K_0$, the best-approximation property yields 
\[
\|C-\Pi_KC\|_{\mathcal H}\leq\|C-v\|_{\mathcal H}<\eta
.
\]
Thus $\Pi_KC\to C$ in $\mathcal H$.
It remains to show that the convergence is uniform on $\mathcal{K}$.  Fix $\varepsilon>0$.  By compactness, there exist $C_1,\dots,C_M\in\mathcal{K}$ such that
\begin{equation*}
    \mathcal{K}
    \subseteq
    \bigcup_{m=1}^M
    B_{\mathcal{H}}
    \left(
        C_m,
        \frac{\varepsilon}{2}
    \right).
\end{equation*}
For each $m=1,\ldots,M$, choose $K_m\in\mathbb N_+$ such that $\|C_m-\Pi_{K_m}C_m\|_{\mathcal H}<\varepsilon/2$, and set $K_\varepsilon\eqdef\max_{1\leq m\leq M}K_m$. Since $V_{K_m}\subseteq V_{K_\varepsilon}$, the best-approximation property gives 
\[
\|C_m-\Pi_{K_\varepsilon}C_m\|_{\mathcal H}\leq\|C_m-\Pi_{K_m}C_m\|_{\mathcal H}<\varepsilon/2
\]
for every $m=1,\ldots,M$.
For any $C\in\mathcal{K}$, choose $m$ satisfying
$
    \|C-C_m\|_{\mathcal{H}}<\varepsilon/2
$.
Since $I_{\mathcal{H}}-\Pi_{K_{\varepsilon}}$ is an orthogonal projection and therefore has operator norm at most one,
\begin{equation}\label{eq:super_d_duper}
\begin{aligned}
    \|C-\Pi_{K_{\varepsilon}}C\|_{\mathcal{H}}
    &\le
        \big\|
            (I_{\mathcal{H}}-\Pi_{K_{\varepsilon}})
            (C-C_m)
        \big\|_{\mathcal{H}}
        +
        \|C_m-\Pi_{K_{\varepsilon}}C_m\|_{\mathcal{H}}
\\
    &\le
        \|C-C_m\|_{\mathcal{H}}
        +
        \|C_m-\Pi_{K_{\varepsilon}}C_m\|_{\mathcal{H}}
\\
    &<
        \varepsilon.
\end{aligned}
\end{equation}
Taking the supremum in~\eqref{eq:super_d_duper} over all $C\in\mathcal{K}$ and then setting
$
    A_{\varepsilon}\eqdef\mathbf{G}_{K_{\varepsilon}}^{-1}
$
yields~\eqref{eq:projec}.
\end{proof}

It is time to prove our main approximation theorem; i.e.\ Theorem~\ref{thrm:UAT_AFNO}.

\begin{proof}[{Proof of Theorem~\ref{thrm:UAT_AFNO}}]
The surfaces $\widetilde{C}_i$ used below will be decoder outputs of the form $\widetilde{C}_i=\phi^{(N_0,M_0,R,S)}_{\theta_i}$, with common decoder hyperparameters $N_0,M_0,R,S$ and parameters $\theta_i$ to be chosen below. Once chosen, these surfaces are fixed independently of the input $C$; only their softmax mixing weights depend on $C$.
Fix $
    \varepsilon_{\operatorname{DR}}
\eqdef
    \frac{\varepsilon}{2}
$, $
    \varepsilon_{\operatorname{FD-Aprx}}
\eqdef
    \frac{\varepsilon}{2}
$, and set $
    \delta
\eqdef
    \tfrac{
        \varepsilon_{\operatorname{DR}}
    }{
        \max\{1,L\}
    }
$.  
By Proposition~\ref{prop:encoding}, there exist
$K\in\mathbb{N}_+$,
$
    (x_k)_{k=1}^K\in\mathcal{D}^K
$,
and
$
    A\in\mathbb{R}^{K\times K}
$
such that
\begin{equation}
\label{eq:dimension_reduction_main_theorem}
    \sup_{C\in\mathcal{K}}
    \biggl\|
        C
        -
        \sum_{k=1}^K
        \mathcal{E}_K
        \big(
            C,
            (x_j)_{j=1}^K,
            A
        \big)_k
        \kappa(\cdot,x_k)
    \biggr\|_{\mathcal{H}}
    \le
        \delta.
\end{equation}
Define the linear reconstruction map
$
    \mathcal{R}_K:
        \mathbb{R}^K
    \rightarrow
        \mathcal{H}
$ given for every $z\in \mathbb{R}^K$ by 
 $\mathcal{R}_K(z)\eqdef 
        \sum_{k=1}^K
        z_k\kappa(\cdot,x_k)
$
and abbreviate
$
    \mathcal{E}
\eqdef
    \mathcal{E}_K
    \big(
        \,\cdot\,,
        (x_k)_{k=1}^K,
        A
    \big)
$.
Thus, the finite-rank reconstruction operator appearing in
Proposition~\ref{prop:encoding} is
$
    P_K
\eqdef
    \mathcal{R}_K\circ\mathcal{E}
$.  
Consequently,~\eqref{eq:dimension_reduction_main_theorem} reads
\begin{equation}
\label{eq:dimension_reduction_main_theorem_compact}
    \sup_{C\in\mathcal{K}}
    \,
    \|C-P_KC\|_{\mathcal{H}}
\le
    \delta
.
\end{equation}
Since $\mathcal{E}:\mathcal{H}\to\mathbb{R}^K$ is continuous and $\mathcal{K}$ is compact, the set
$
    \Theta_K
\eqdef
    \mathcal{E}(\mathcal{K})
\subseteq
    \mathbb{R}^K
$
is also non-empty and compact.  Consider the finite-dimensional surrogate
\begin{equation}
\label{eq:finite_dimensional_surrogate}
\begin{aligned}
    \mathcal{F}_K:
        \Theta_K
    &\rightarrow
        \mathcal{Y},
\\
    z
    &\mapsto
        \mathcal{F}
        \big(
            \mathcal{R}_K(z)
        \big).
\end{aligned}
\end{equation}
The linear map $\mathcal{R}_K$ is continuous from
$
    (\mathbb{R}^K,\|\,\|_1)
$
into $\mathcal{H}$.  Therefore, for every $z,\tilde z\in\Theta_K$,
\begin{equation*}
\begin{aligned}
    \|\mathcal{F}_K(z)-\mathcal{F}_K(\tilde z)\|_{q,l}
\le
        L
        \|\mathcal{R}_K(z-\tilde z)\|_{\mathcal{H}}
\le
        L
        \|\mathcal{R}_K\|_
        {
            (\mathbb{R}^K,\|\,\|_1)
            \to
            \mathcal{H}
        }
        \,
        \|z-\tilde z\|_1.
\end{aligned}
\end{equation*}
Hence, $\mathcal{F}_K$ is Lipschitz from
$
    (\Theta_K,\|\,\|_1)
$
into
$
    (\mathcal{C}_{AF}^{q,l},\|\,\|_{q,l})
$.  
Since $\mathcal{F}$ is defined on all of $\mathcal{H}$ with values in $\mathcal{Y}$, one has $\mathcal{F}_K(z)=\mathcal{F}(\mathcal{R}_Kz)\in\mathcal{Y}$ for every $z\in\Theta_K$. In particular, the required call surface is $\mathcal{F}(\mathcal{R}_Kz)$; no call-surface condition is imposed on $\mathcal{R}_Kz$. With $c_K\eqdef\max_{1\leq k\leq K}\sqrt{\kappa(x_k,x_k)}$, one has $\|\mathcal{R}_Kz\|_{\mathcal{H}}\leq c_K\|z\|_1$ for every $z\in\mathbb{R}^K$, and hence $\|\mathcal{F}_K(z)-\mathcal{F}_K(z')\|_{q,l}\leq Lc_K\|z-z'\|_1$ for every $z,z'\in\Theta_K$. We may therefore apply Proposition~\ref{prop:UAT_AFNO__technicalversion__FINITEDIMENSIONALVERSION} to $\mathcal{F}_K$ on $\Theta_K$ with approximation tolerance $\varepsilon_{\operatorname{FD-Aprx}}$.
It yields
$I\in\mathbb{N}_+$, $R,S>0$, $N_0,M_0\in\mathbb{N}_+$,
parameters
$
    \theta_1,\dots,\theta_I
\in
    \mathbb{R}^{D_{N_0,M_0,R,S}}
$,
a temperature $\tau>0$, and a $\operatorname{ReLU}$-MLP
$
    \alpha_{\mathbf{X}}:\mathbb{R}^K\to\mathbb{R}^I
$
with the following properties. For $i=1,\ldots,I$, define $\widetilde{C}_i\eqdef\phi^{(N_0,M_0,R,S)}_{\theta_i}$, where $\phi^{(N_0,M_0,R,S)}$ is the surface decoder from Definition~\ref{defn:Good_Decoder}.\footnote{E.g.\ as in Definition~\ref{def:Smooth_AF_Decoder}.} In the proof of the finite-dimensional proposition, the parameters $\theta_i$ are chosen so that $\|\widetilde{C}_i-\mathcal{F}_K(z_i)\|_{q,l}\leq\varepsilon_{\operatorname{FD-Aprx}}/2$ for suitable anchor points $z_1,\ldots,z_I\in\Theta_K$. Moreover, $\widetilde{C}_i\in\mathcal{C}_{AF}$ by Definition~\ref{defn:Good_Decoder}(ii). The map
$
    \widehat{\Phi}(z)
\eqdef
    \sum_{i=1}^I
        \operatorname{Softmax}
        \left(
            \frac{1}{\tau}
            \alpha_{\mathbf{X}}(z)
        \right)_i
        \widetilde{C}_i
$
satisfies
\begin{equation}
\label{eq:finite_dimensional_surrogate_approximation}
    \sup_{z\in\Theta_K}
    \,
    \|\widehat{\Phi}(z)-\mathcal{F}_K(z)\|_{q,l}
    \le
        \varepsilon_{\operatorname{FD-Aprx}},
\end{equation}
as well as
\begin{equation}
\label{eq:finite_dimensional_surrogate_range}
    \widehat{\Phi}(\mathbb{R}^K)
    \subseteq
        \mathcal{C}_{AF}
        \cap
        \operatorname{co}
        \{
            \widetilde{C}_1,\dots,\widetilde{C}_I
        \}.
\end{equation}
Write $\alpha_{\mathbf X}(z)=A_1\operatorname{ReLU}(A_0z+b_0)+b_1$, where $h\in\mathbb N_+$, $A_0\in\mathbb R^{h\times K}$, $b_0\in\mathbb R^h$, $A_1\in\mathbb R^{I\times h}$, and $b_1\in\mathbb R^I$. For the one-token input $z^\top\in\mathbb R^{1\times K}$, take one attention head with key dimension $1$, value dimension $h$, $\mathbf Q=\mathbf K=0_{1\times K}$, $\mathbf V=A_0$, and attention scale $1$. Use a transformer of depth $2$ with $d_0=K$, $d_1=h$, $d_2=I$, activation $\operatorname{ReLU}$, $b^{(0)}=b_0$, and final readout $(\mathbf A^{(1)},b^{(1)})=(A_1,b_1)$. Then, applying~\ref{prop:transformerification__SparseVersion} we deduce that $\alpha_{\mathbf{X}}$ can be realized by a multi-head transformer:
\[
\big[\operatorname{Attn}(z^\top\mid\mathbf Q,\mathbf K,\mathbf V,1)\big]_1=\frac{e^0}{e^0}A_0z=A_0z,\qquad [\mathcal T(z^\top)]_1=A_1\operatorname{ReLU}(A_0z+b_0)+b_1=\alpha_{\mathbf X}(z),\quad z\in\mathbb R^K.
\]
\hfill\\
Next, define $\Phi\eqdef\widehat{\Phi}\circ\mathcal{E}$. This is precisely the map in~\eqref{eq:approximation_gaurantee_construction0}.
For every $C\in\mathcal{K}$, we have
\begin{equation*}
    \mathcal{F}_K
    \big(
        \mathcal{E}(C)
    \big)
    =
    \mathcal{F}
    \big(
        \mathcal{R}_K\circ\mathcal{E}(C)
    \big)
    =
    \mathcal{F}(P_KC).
\end{equation*}
Consequently, by~\eqref{eq:finite_dimensional_surrogate_approximation},
the Lipschitz continuity of $\mathcal{F}$, and
~\eqref{eq:dimension_reduction_main_theorem_compact},
\begin{equation}
\label{eq:final_infinite_dimensional_estimate}
\begin{aligned}
    \|\Phi(C)-\mathcal{F}(C)\|_{q,l}
    &\le
        \big\|
            \widehat{\Phi}
            \big(
                \mathcal{E}(C)
            \big)
            -
            \mathcal{F}_K
            \big(
                \mathcal{E}(C)
            \big)
        \big\|_{q,l}
\\
    &\quad
        +
        \|\mathcal{F}(P_KC)-\mathcal{F}(C)\|_{q,l}
\\
    &\le
        \varepsilon_{\operatorname{FD-Aprx}}
        +
        L\|P_KC-C\|_{\mathcal{H}}
\\
    &\le
        \varepsilon_{\operatorname{FD-Aprx}}
        +
        L\delta
\\
    &\le
        \varepsilon_{\operatorname{FD-Aprx}}
        +
        \varepsilon_{\operatorname{DR}}
\\
    &=
        \varepsilon.
\end{aligned}
\end{equation}
Taking the supremum in~\eqref{eq:final_infinite_dimensional_estimate}
over $C\in\mathcal{K}$ proves \textup{(i)}.
Finally, since
$
    \mathcal{E}(\mathcal{H})
\subseteq
    \mathbb{R}^K
$,
the range estimate~\eqref{eq:finite_dimensional_surrogate_range} gives
$
    \Phi(\mathcal{H})
    =
        \widehat{\Phi}
        \big(
            \mathcal{E}(\mathcal{H})
        \big)
\subseteq
        \widehat{\Phi}(\mathbb{R}^K)
\subseteq
        \mathcal{C}_{AF}
        \cap
        \operatorname{co}
        \{
            \widetilde{C}_1,\dots,\widetilde{C}_I
        \}
$; proving \textup{(ii)}.
\end{proof}

\begin{theorem}[Certifiably Arbitrage-Free]
\label{thrm:NA}
Any map of the form~\eqref{eq:Dysanos} satisfies: 
\hfill\\
For every $\vartheta\in \mathbb{R}^{F_{\mathrm{\mathrm{mkt}}}}$ the predicted call surface 
\[
\mathcal{AF\text{-}NO}(\vartheta):[0,\infty)^2\to [0,\infty)
\]
is arbitrage-free; i.e.\ it satisfies the conditions of Theorem~\ref{prop:contarbfree}.
\end{theorem}

\begin{proof}
    For every input, the component surfaces in~\eqref{eq:Dysanos} belong to $\mathcal C_{AF}$ and their weights form a simplex vector independent of strike and maturity. Proposition~\ref{prop:AS} therefore proves the claim for every parameter choice.
\end{proof}

\subsection{Proof of Negative Results}
\label{a:additions}

We now prove our negative results.

\subsubsection{Proof of No Arbitrage-Free Projection Theorem}
\label{s:NoProj}
For every $\lambda,p>0$ define
\[
\mathcal{X}_{\lambda}^p
\eqdef
\Bigl\{
    C:[0,\infty)^2\to \mathbb{R}
    \,:\,
    \|C\|_{p,\lambda}<\infty
\Bigr\}.
\]
Since finer topologies have more closed sets than weaker ones, it suffices to prove the result in the topology of uniform convergence on compact sets on $C([0,\infty)^2,[0,\infty))$. Indeed, if $\tau$ is any topology on this space which is no stronger than the compact-open topology, then every $\tau$-closed set is also closed in the compact-open topology.  Thus, non-closedness in the compact-open topology immediately implies non-closedness in $\tau$. Equivalently, the same compact-open convergent sequence witnesses non-closedness for every such weaker topology.
\begin{proposition}[Non-Closedness of $\mathcal{C}_{AF}$ in the Compact-Open Topology]
\label{prop:AS_not_closed}
The set $\mathcal{C}_{AF}$ is not closed in
$
    C\big(
        [0,\infty)^2,
        [0,\infty)
    \big)
$
equipped with the topology of uniform convergence on compact sets.
Consequently, for every topology $\tau$ on
$
    C\big(
        [0,\infty)^2,
        [0,\infty)
    \big)
$
which is no stronger than the topology of uniform convergence on
compact sets, the set $\mathcal{C}_{AF}$ is not $\tau$-closed.
\end{proposition}
\begin{proof}[{Proof of Proposition~\ref{prop:AS_not_closed}}]
Define
$
    a(T)
    \eqdef
    \frac{T}{1+T}
$
for every $T\geq0$.  For each $n\in\mathbb{N}_2$, define
\[
    C_n(T,K)
    \eqdef
    \big(
        1-a(T)
    \big)
    (1-K)^+
    +
    a(T)
    \left(
        1-\frac{K}{n}
    \right)^+
\] for every $T,K\geq0$.  
We first verify that $C_n\in\mathcal{C}_{AF}$.  Clearly, $C_n$ is
continuous and
$
    C_n(T,0)
    =
    1
$
for every $T\geq0$.  Moreover,
\[
    \partial_K^+C_n(T,0)
    =
    -
    \big(
        1-a(T)
    \big)
    -
    \frac{a(T)}{n}
    =
    -1
    +
    a(T)
    \left(
        1-\frac{1}{n}
    \right)
    \geq
    -1.
\]
For every fixed $T\geq0$, the map $K\mapsto C_n(T,K)$ is convex,
being a convex combination of convex call-payoff functions, and
$
    \lim_{K\uparrow\infty}
    C_n(T,K)
    =
    0
$.
Furthermore,
\[
    C_n(T,K)
    =
    (1-K)^+
    +
    a(T)
    \left[
        \left(
            1-\frac{K}{n}
        \right)^+
        -
        (1-K)^+
    \right].
\]
Since
$
    a
$
is non-decreasing and
$
    \big(
        1-\frac{K}{n}
    \big)^+
    \geq
    (1-K)^+
$ for every $K\geq0$, the map $T\mapsto C_n(T,K)$ is non-decreasing for every $K\geq0$.  Finally, since $a(0)=0$, $
    C_n(0,K)
    =
    (1-K)^+
$ for every $K\geq0$.  Proposition~\ref{prop:contarbfree} therefore implies that $C_n\in\mathcal{C}_{AF}$ for every $n\in\mathbb{N}_2$. Define $
    C_{\infty}(T,K)
    \eqdef
    \big(
        1-a(T)
    \big)
    (1-K)^+
    +
    a(T)
$.  Let $T_0,K_0>0$.  For every $n>K_0$ and every $(T,K)\in[0,T_0]\times[0,K_0] $, we have
\[
\begin{aligned}
    \big|
        C_n(T,K)-C_{\infty}(T,K)
    \big|
=
    a(T)
    \left|
        \left(
            1-\frac{K}{n}
        \right)^+
        -
        1
    \right|
=
    a(T)\frac{K}{n}
\leq
    \frac{K_0}{n}.
\end{aligned} \] Consequently,
$
    \sup_{
        (T,K)\in
        [0,T_0]\times[0,K_0]
    }
    \big|
        C_n(T,K)-C_{\infty}(T,K)
    \big|
    \leq
    \frac{K_0}{n},
$ and hence $C_n\to C_{\infty}$ uniformly on compact subsets of $[0,\infty)^2$. However, for every $T>0$, $
    \lim_{K\uparrow\infty}
    C_{\infty}(T,K)
    =
    a(T)
    >
    0.
$.  Thus, $C_{\infty}$ violates condition {\rm (iii)} of Proposition~\ref{prop:contarbfree}, and therefore $    C_{\infty}\notin\mathcal{C}_{AF}
$.  Hence $\mathcal{C}_{AF}$ is not closed in the topology of uniform
convergence on compact sets.  Finally, let $\tau$ be any topology on the same ambient space which is no stronger than the compact-open topology.  Since $C_n\to C_{\infty} $ in the compact-open topology, we also have $
    C_n\to C_{\infty}
$ in $\tau$.  Since every $C_n$ belongs to $\mathcal{C}_{AF}$ while $C_{\infty}$ does not, the set $\mathcal{C}_{AF}$ cannot be $\tau$-closed.
\end{proof}
Now, the following non-existence of metric projections onto non-closed subsets in metric spaces; which follows readily from the definitions of a metric projection (not considering continuity or Borel-measurability of such an object) and elementary characterizations of closed sets in metric spaces.
\begin{lemma}[Non-closed sets do not admit nearest points everywhere]
\label{lem:nonclosed_not_proximinal}
Let $(X,\rho)$ be a metric space and let $A\subseteq X$ be non-empty and not closed.  Then, there exists some $x^{\star}\in X$ such that
\begin{equation}
\label{eq:dist_zero_not_attained}
\inf_{a\in A}\,\rho(x^{\star},a)=0,
\end{equation}
but there is no $a^{\star}\in A$ satisfying
\begin{equation}
\label{eq:no_minimizer_A}
\rho(x^{\star},a^{\star})
=
\inf_{a\in A}\,\rho(x^{\star},a).
\end{equation}
If for every $x\in X$ there exists $a_x\in A$ such that
$
\rho(x,a_x)=\inf_{a\in A}\rho(x,a)
$,
then $A$ is closed.
\end{lemma}
\begin{proof}[{Prof of Lemma~\ref{lem:nonclosed_not_proximinal}}]
Since $A$ is not closed, there exists some
$
x^{\star}\in \overline{A}\setminus A
$.
Because $x^{\star}\in \overline{A}$, there exists a sequence $(a_n)_{n=1}^{\infty}\subseteq A$ with
$
\rho(x^{\star},a_n)\to 0
$.
Hence,
$
\inf_{a\in A}\rho(x^{\star},a)=0
$,
which proves~\eqref{eq:dist_zero_not_attained}.
Assume now, for contradiction, that there exists some $a^{\star}\in A$ such that
$
\rho(x^{\star},a^{\star})
=
\inf_{a\in A}\rho(x^{\star},a)
$.
Then
$
\rho(x^{\star},a^{\star})=0
$,
and therefore $x^{\star}=a^{\star}$.
But this contradicts the fact that $x^{\star}\notin A$.
Thus, no minimizer in $A$ can exist for $x^{\star}$, proving~\eqref{eq:no_minimizer_A}.  The final claim follows from contraposition.
\end{proof}
We now have the necessary tools to prove our ``no arbitrage-free projections'' theorem.
\begin{proof}[{Proof of Theorem~\ref{thrm:No_AF__Proj}}]
By Proposition~\ref{prop:AS_not_closed}, the set $\mathcal{C}_{AF}
$ is not closed in the topology induced by any metric $\rho$ on 
$C([0,\infty)^2,[0,\infty))$
which is no stronger than the 
topology of uniform convergence on compact sets.  
Applying Lemma~\ref{lem:nonclosed_not_proximinal} with $X=
C([0,\infty)^2,[0,\infty))
$ and $A=\mathcal{C}_{AF}
$ yields the conclusion.
\end{proof}

\section{Proofs of Certificate Failures for Vanilla Arbitrage-Free Decoder-Neural Operators}
\label{s:NoAFDecoderNO}

\subsection{Proofs for (Encoder) Decoder-Type NOs}
\label{s:enc_dec_NOs}

\begin{lemma}[Affine obstruction for decoder-type neural operators]
\label{lem:affine_obstruction_decoder_convexity}
Fix a continuous orthonormal system $(e_n)_{n=1}^{\infty}\subset C([0,1])$ in $L^2([0,1])$, and consider { 
the tensorized orthonormal system $(E_{n,m})_{n,m=1}^{\infty}$
} of $L^2([0,1]^2)$ defined for each $n,m\in \mathbb{N}$ by $
E_{n,m}(s,t)\eqdef e_n(s)e_m(t)
$.
Fix $K\in \mathbb{N}_+$ and pairwise distinct indices $(n_1,m_1),\dots,(n_K,m_K)\in \mathbb{N}^2$, and set
$$
V_K
\eqdef
\operatorname{span}\bigl\{
    E_{n_k,m_k}
    :1\le k\le K
\bigr\}.
$$
Then the following are equivalent:
\begin{enumerate}
    \item[(i)] Every $g\in V_K$ is convex on $[0,1]^2$.
    \item[(ii)] Every $g\in V_K$ is affine on $[0,1]^2$.
    \item[(iii)] $V_K\subseteq \operatorname{Aff}([0,1]^2)$.
\end{enumerate}
In particular, if $K\ge 4$, then
$
V_K\not\subseteq \operatorname{Aff}([0,1]^2),
$
{ 
and therefore there exists some $a^\star\in\mathbb R^K$ such that $D_{a^\star}(x)\eqdef\sum_{k=1}^K a_k^\star E_{n_k,m_k}(x)$
}
is not convex on $[0,1]^2$.
\end{lemma}
\begin{proof}[{Proof of Lemma~\ref{lem:affine_obstruction_decoder_convexity}}]
The equivalence of $(ii)$ and $(iii)$ is immediate, and $(iii)\Rightarrow (i)$ holds since every affine function is convex.
It remains to prove $(i)\Rightarrow (ii)$. Fix any $g\in V_K$. Since $V_K$ is a linear subspace, we also have $-g\in V_K$. By $(i)$, both $g$ and $-g$ are convex on $[0,1]^2$. Hence $g$ is both convex and concave, and therefore affine on $[0,1]^2$. Since $g\in V_K$ was arbitrary, every element of $V_K$ is affine.
For the final claim, note that the tensorized system $(E_{n,m})_{n,m\ge 1}$ is orthonormal in $L^2([0,1]^2)$, so the pairwise distinct functions
$
E_{n_1,m_1},\dots,E_{n_K,m_K}
$
are linearly independent. Thus $\dim(V_K)=K$. Since $\operatorname{Aff}([0,1]^2)$ has dimension $3$, it follows that $V_K\not\subseteq \operatorname{Aff}([0,1]^2)$ whenever $K\ge 4$. The existence of a non-convex element of $V_K$ then follows from the equivalence already proved.
\end{proof}
We may now prove the non-certifiability of (encoder) decoder-type networks in $L^2([0,1])$; for arbitrage-free certificates.
\begin{proof}[{Proof of Proposition~\ref{prop:AF_certificatefailure}}]
Since the functions $e_{n_1},e_{n_2},e_{n_3}$ are linearly independent, the space
$
U
\eqdef
\operatorname{span}\{e_{n_1},e_{n_2},e_{n_3}\}
$
has dimension $3$. If all elements of $U$ were convex, then for each $h\in U$ both $h$ and $-h$ would be convex, forcing every element to be affine. On the other hand, the space of affine functions on $[0,1]$ has dimension $2$. Therefore,
$
U\not\subseteq \operatorname{Aff}([0,1]),
$
so there exists some $a^\star=(a_1^\star,a_2^\star,a_3^\star)\in \mathbb{R}^3$ such that
$
h(K)
\eqdef
\sum_{k=1}^3 a_k^\star e_{n_k}(K)
$
is not convex on $[0,1]$.
Next, since $e_{m_\star}\not\equiv 0$ in $L^2([0,1])$, there exists some $T^\star\in [0,1]$ such that
$$
e_{m_\star}(T^\star)\neq 0.
$$
Replacing $a^\star$ by $\operatorname{sign}(e_{m_\star}(T^\star))\,a^\star$ if necessary, we may assume that
$
e_{m_\star}(T^\star)\,h
$
is still non-convex on $[0,1]$.
Now choose $f_{NN}:\mathbb{R}^F\to \mathbb{R}^3$ to be the constant map
$$
f_{NN}(x)\equiv a^\star.
$$
Then, for every $x\in \mathbb{R}^F$,
$
{ 
f_{D\text{-}NO}(x)(T,K)
}
=
\sum_{k=1}^3 a_k^\star e_{n_k}(K)e_{m_\star}(T).
$
In particular, at $T=T^\star$ we obtain
$
K\mapsto
{ 
f_{D\text{-}NO}(x)(T^\star,K)
}
=
e_{m_\star}(T^\star)\,
\sum_{k=1}^3 a_k^\star e_{n_k}(K)
=
e_{m_\star}(T^\star)\,h(K),
$
which is not convex on $[0,1]$.
Therefore, for every $x\in \mathbb{R}^F$, the decoded surface $f_{D\text{-}NO}(x)$ fails the strike-convexity condition required for membership in $\mathcal{C}_{AF}$. Hence
$
{ 
f_{D\text{-}NO}(x)\notin\{C|_{[0,1]^2}:C\in\mathcal C_{AF}\}
}
$ for every $x\in \mathbb{R}^F$.  
\end{proof}

\subsection{Proofs for FNOs}
\label{s:FNOs}

\begin{proof}[{Proof of Proposition~\ref{prop:sigmoid_breaks_convexity_in_K}}]
Set
$
W_{\ell}=0
$
and
$
R_{\ell}=0
$
for every $\ell\in\{0,\dots,L-1\}$.
If $L\ge 2$, set
$
b_{\ell}\equiv 0
$
for every $\ell\in\{0,\dots,L-2\}$.
For the final layer, define
$
b_{L-1}(T,K)\eqdef \xi-\eta+2\eta K
$
for $(T,K)\in [0,1]^2$.

Since $m=1$ and $P=\operatorname{Id}$, the corresponding FNO output satisfies, for every
$a\in\mathbb{R}^F$,
\[
\Phi(Q^{\uparrow}(a))(T,K)
=
\sigma\bigl(
    \xi-\eta+2\eta K
\bigr).
\]
Indeed, the input field $Q^{\uparrow}(a)$ is killed by the choice
$
W_{\ell}=R_{\ell}=0
$,
and all preceding biases vanish.

Fix $T\in[0,1]$.  Then
\[
K\mapsto \Phi(Q^{\uparrow}(a))(T,K)
=
\sigma\bigl(
    \xi-\eta+2\eta K
\bigr).
\]
Differentiating twice in $K$ gives
\[
\partial_{KK}\Phi(Q^{\uparrow}(a))(T,K)
=
(2\eta)^2
\sigma''\bigl(
    \xi-\eta+2\eta K
\bigr).
\]
Since
$
K\mapsto \xi-\eta+2\eta K
$
maps $[0,1]$ onto $[\xi-\eta,\xi+\eta]$, and $\sigma''$ takes both positive and negative values on
$(\xi-\eta,\xi+\eta)$, it follows that
$
\partial_{KK}\Phi(Q^{\uparrow}(a))(T,K)
$
takes both positive and negative values on $[0,1]$.
Hence
$
K\mapsto \Phi(Q^{\uparrow}(a))(T,K)
$
is not convex on $[0,1]$.

On the other hand, the restriction to $[0,1]^2$ of any element of $\mathcal{C}_{AF}$ is convex in the strike variable by Theorem~\ref{prop:contarbfree}.  Therefore, for every $a\in\mathbb{R}^F$, the continuous representative of $\Phi(Q^{\uparrow}(a))$ is not the restriction to $[0,1]^2$ of an element of $\mathcal{C}_{AF}$.
\end{proof}

\section{Funding and Acknowledgements}
\label{s:Ackn}
A.\ Kratsios acknowledges financial support from an NSERC Discovery Grant No.\ RGPIN-2023-04482 and No.\ DGECR-2023-00230.   and by the project Bando PRIN 2022 named ``Qnt4Green - Quantitative Approaches for Green Bond Market: Risk Assessment, Agency Problems and Policy Incentives'', codice 2022JRY7EF, CUP E53D23006330006, funded by European Union – NextGenerationEU, M4c2.\\
B. \ Horvath acknowledges financial support from UKRI1010: High order mathematical and computational infrastructure for streamed data that enhance contemporary generative and large language models.\\
A.\ Kratsios and R. Saqur both also acknowledge that resources used in preparing this research were provided, in part, by the Province of Ontario, the Government of Canada through CIFAR, and companies sponsoring the Vector Institute\footnote{\href{https://vectorinstitute.ai/partnerships/current-partners/}{https://vectorinstitute.ai/partnerships/current-partners/}}.    

\paragraph{Acknowledgements}
A.\ Kratsios would also like to thank \href{https://scholar.google.com/citations?user=XWWfMCYAAAAJ&hl=fr}{Fr\'{e}d\'{e}ric Godin} and \href{https://en.wikipedia.org/wiki/Genevi%C3%A8ve_Gauthier}{Genevi\`{e}ve Gauthier} for their helpful feedback while the authors were completing our manuscript.  
A.\ Kratsios would like to thank Mathieu Rosenbaum for his helpful discussions on static no-arbitrage vs.\ dynamic no-arbitrage right before this manuscript's release.

\appendix
\renewcommand{\theHsection}{background.\arabic{section}}
\section{Additional Background on Deep Learning}
\label{s:ExtrasAI}
We now record some key results in the literature which we will use now.  We recall that the $\operatorname{ReLU}\eqdef \max\{0,\cdot\}$ activation function on $\mathbb{R}$.
\begin{definition}[Multilayer Perceptrons (MLPs)]
Let $d,D,L\in\mathbb{N}$, and let $d_0,d_1,\dots,d_L\in\mathbb{N}$ such that $d_0=d$ and $d_L=D$.
Let $\sigma\in \mathcal{C}(\mathbb{R})$ be an activation function.
A fully-connected MLP with depth $L$ and activation function $\sigma$ is a map $f:\mathbb{R}^d\to\mathbb{R}^D$ with iterative representation:
\begin{align*} 
    &X^{(0)} 
    \eqdef  X, \\
    &X^{(j+1)} 
    \eqdef  
    \sigma\bullet({\bf A}^{(j)} X^{(j)} + b^{(j)}) \qquad j=0,1,\dots,L-2,
    \qquad f(X)
    \eqdef  
    {\bf A}^{(L-1)} X^{(L-1)} + b^{(L-1)},
\end{align*}
where $\bullet$ denotes component-wise application, ${\bf A}^{(j)}\in\mathbb{R}^{d_{j+1}\times d_j}$, and $b^{(j)}\in\mathbb{R}^{d_{j+1}}$.
\end{definition}

\begin{proposition}[{Transformerification of MLPs - \citep[Proposition A.1]{kratsios2026adaptivity}}]
\label{prop:transformerification__SparseVersion}
Let $f:\mathbb{R}^d\to \mathbb{R}^D$ be a MLP with depth $L$. %
{ 
Then, for every $\lambda>0$, $f$ can be implemented exactly as a one-token transformer with the same depth as $f$, exactly one attention head in each hidden layer, and the same final affine readout.
}
\end{proposition}

\begin{proof}
Choose one token, one attention head, key dimension $1$, and value dimension $d_{j+1}$ in hidden layer $j$. The attention softmax consists of the single weight $1$. In hidden layer $j$, take $ {\bf Q}^{(j)}={\bf K}^{(j)}=0$, $ {\bf V}^{(j)}={\bf A}^{(j)}$, and the same bias and activation as in the MLP. The layer is therefore exactly $x\mapsto\sigma({\bf A}^{(j)}x+b^{(j)})$. Retain the MLP's final affine output layer. Their compositions agree for every input; when $L=1$, only that affine layer remains.
\end{proof}

\subsection{{Proofs of Structured \texorpdfstring{$\operatorname{ReLU}$}{ReLU}-MLP Arbitrage-Free Decoder in Section~\ref{s:EncoderDecoder__ss:GoodDecoder}}}
\label{s:AF_Decoder_Proofs}

We first prove the polytopal characterization of ``good no-arbitrage quotes'' over a fixed finite grid.  The following proof is a largely case-by-case verification; which we encourage the reader to skip on a first read as it is tedious.
\begin{proof}[{Proof of Proposition~\ref{prop:admissible_polytope}}]
Throughout the proof, the virtual strikes
$
    K_{\mathrm{L}}
$
and
$
    K_{\mathrm{R}}
$
are fixed as in the statement.  In particular,
\begin{equation}
\label{eq:proof_augmented_strike_order}
    0
    =
    \widehat K_0
    <
    \widehat K_1
    =
    K_{\mathrm{L}}
    <
    \widehat K_2
    =
    K_1
    <
    \dots
    <
    \widehat K_{N+1}
    =
    K_N
    <
    \widehat K_{N+2}
    =
    K_{\mathrm{R}}.
\end{equation}
For convenience, abbreviate $
    \Delta\widehat K_j
    \eqdef
    \widehat K_{j+1}-\widehat K_j
    >
    0$ for each $j\in [N+1]$.  
\hfill\\
\noindent
We first make two simple observations which we repeatedly use. 
Let
$
    f:[0,\infty)\to[0,\infty)
$
be convex and suppose that
\begin{equation}
\label{eq:proof_convex_function_zero_limit}
    \lim_{K\uparrow\infty}f(K)
    =
    0.
\end{equation}
Then $f$ is non-increasing.
Indeed, suppose to the contrary that there exist $0\le x<y$ such that
$
    f(x)<f(y).
$
Set $
    q
     \eqdef 
    \frac{f(y)-f(x)}{y-x}
    >
    0
$.  Now, for every $z>y$, the monotonicity of secant slopes of a convex
function gives $
    \frac{f(z)-f(y)}{z-y}
\ge
    \frac{f(y)-f(x)}{y-x}
=
    q
$ and, consequently, we have $
    f(z)
    \ge
    f(y)+q(z-y)
$; implying that $f$ tends to $\infty$ as $z\uparrow\infty$, contradicting
\eqref{eq:proof_convex_function_zero_limit}.  Thus $f$ is
non-increasing.
\hfill\\
\noindent
We shall also use the following standard secant-slope inequality.
If $x_0<x_1<x_2$, then
\begin{equation}
\label{eq:proof_three_point_secant_inequality}
    \frac{f(x_1)-f(x_0)}{x_1-x_0}
    \le
    \frac{f(x_2)-f(x_0)}{x_2-x_0}
    \le
    \frac{f(x_2)-f(x_1)}{x_2-x_1}.
\end{equation}
To verify this, write $
    x_1
    =
    (1-\lambda)x_0+\lambda x_2
$ and $
    \lambda
     \eqdef 
    \frac{x_1-x_0}{x_2-x_0}
    \in(0,1)
$ and, note that by convexity we have that
Convexity gives
\begin{equation}
\label{eq:proof_convexity_three_points}
    f(x_1)
\le
    (1-\lambda)f(x_0)+\lambda f(x_2)
.
\end{equation}
Rearranging across~\eqref{eq:proof_convexity_three_points} yields~
\eqref{eq:proof_three_point_secant_inequality}.
In particular, if
$
    x_0<\dots<x_J,
$
then the consecutive secant slopes satisfy
\begin{equation}
\label{eq:proof_consecutive_secant_slopes}
    \frac{f(x_1)-f(x_0)}{x_1-x_0}
    \le
    \dots
    \le
    \frac{f(x_J)-f(x_{J-1})}{x_J-x_{J-1}}.
\end{equation}

We are now set to prove that (i) implies (ii).  
Suppose that item~{\rm(i)} holds, and let
$
    \mathfrak{I}:[0,\infty)^2\to[0,\infty)
$
be an interpolator satisfying the stated conditions.  Now, fix $m\in[M]_+$ and define the corresponding maturity slice
\begin{equation}
\label{eq:proof_given_maturity_slice}
    f_m(K)
     \eqdef 
    \mathfrak{I}(T_m,K),
    \qquad
    K\ge0.
\end{equation}
Since $\mathfrak{I}$ is arbitrage-free, Proposition
~\ref{prop:contarbfree} implies that $f_m$ is convex and satisfies
\begin{equation}
\label{eq:proof_given_slice_conditions}
    f_m(0)=1,
    \qquad
    \lim_{\delta\downarrow0}
    \frac{f_m(\delta)-f_m(0)}{\delta}
    {\geq }
    -1,
    \qquad
    \lim_{K\uparrow\infty}f_m(K)
    =
    0.
\end{equation}
By Step~1, $f_m$ is non-increasing.
The interpolation and virtual data-point conditions give
$
    f_m(\widehat K_j)
    =
    \widehat y_{m,j}
$ for each $j\in [N+2]$.
Therefore, for every $j\in [N+1]$ we have that
\begin{equation}
\label{eq:proof_slopes_are_slice_secants}
\begin{aligned}
    s_{m,j}(y)
=
    \frac{
        \widehat y_{m,j+1}-\widehat y_{m,j}
    }{
        \widehat K_{j+1}-\widehat K_j
    }
=
    \frac{
        f_m(\widehat K_{j+1})
        -
        f_m(\widehat K_j)
    }{
        \widehat K_{j+1}-\widehat K_j
    }.
\end{aligned}
\end{equation}
Since $f_m$ is convex, the consecutive secant-slope inequality
\eqref{eq:proof_consecutive_secant_slopes} yields
\begin{equation}
\label{eq:proof_necessity_slope_monotonicity}
    s_{m,0}(y)
\le
    s_{m,1}(y)
\le
    \dots
\le
    s_{m,N+1}(y)
.
\end{equation}
Moreover, $f_m$ is non-increasing, and hence
$
    s_{m,j}(y)
    \le
    0
$ for each $j\in [N+1]$.  
By convexity, the right derivative at $0$ is bounded above by every
secant slope issuing from $0$.  Hence,
$
    -1
    \le
    \partial_K^+f_m(0)
    \le
    s_{m,0}(y)
$.
Moreover, the definition of $\widehat y_{m,1}$ gives
$
    s_{m,0}(y)
    =
    s_{m,1}(y)
$.
Therefore,
$
    -1
    \le
    s_{m,0}(y)
    =
    s_{m,1}(y)
    \le
    \dots
    \le
    s_{m,N+1}(y)
    \le
    0
$.
It remains to verify the calendar inequalities.  Fix
$
    m\in[M-1]_+
$
and
$
    n\in[N]_+.
$
Since
$
    T_m<T_{m+1}
$
and $\mathfrak{I}$ is non-decreasing in maturity by
Proposition~\ref{prop:contarbfree}, we have
\begin{equation}
\label{eq:proof_necessity_calendar_inequality}
\begin{aligned}
    y_{m,n}
=
    \mathfrak{I}(T_m,K_n)
\le
    \mathfrak{I}(T_{m+1},K_n)
    =
    y_{m+1,n}.
\end{aligned}
\end{equation}
Thus all the inequalities defining
$
    \mathfrak{C}_{\mathbf{T},\mathbf{K}}
$
hold, and therefore $
    y
    \in
    \mathfrak{C}_{\mathbf{T},\mathbf{K}}
$.  

Now, we show that (ii) implies (i).  
Suppose that
$
    y\in\mathfrak{C}_{\mathbf{T},\mathbf{K}}.
$
For each $m\in[M]_+$, let
$
    \Phi_m^y:[0,\infty)\to\mathbb{R}
$
be the continuous piecewise-affine interpolant of the augmented
datapoints: 
$
    \big(
        \widehat K_j,\widehat y_{m,j}
    \big)
$ for each $j\in [N+2]$ and 
extended constantly by zero to the right of $K_{\mathrm{R}}$; i.e.\ for each
$
    K\in[\widehat K_j,\widehat K_{j+1}]
$
with $j=0,\dots,N+1$ we define
\begin{equation}
\label{eq:proof_piecewise_affine_strike_interpolant}
\begin{aligned}
    \Phi_m^y(K)
\eqdef
        \frac{
            \widehat K_{j+1}-K
        }{
            \widehat K_{j+1}-\widehat K_j
        }
        \widehat y_{m,j}
    +
        \frac{
            K-\widehat K_j
        }{
            \widehat K_{j+1}-\widehat K_j
        }
        \widehat y_{m,j+1}
,
\end{aligned}
\end{equation}
and, for $K\ge K_{\mathrm{R}}$, we set
$
    \Phi_m^y(K)
     \eqdef 
0
$. Then, by construction we have that: $
    \Phi_m^y(\widehat K_j)
    =
    \widehat y_{m,j}
$ for each $j\in [N+2]$.  
On the open sub-interval
$
    (\widehat K_j,\widehat K_{j+1}),
$
the derivative of $\Phi_m^y$ is
$
    \partial_K\Phi_m^y(K)
    =
    s_{m,j}(y)
$.  Now, since
$
    y\in\mathfrak{C}_{\mathbf{T},\mathbf{K}},
$
these slopes satisfy
\begin{equation}
\label{eq:proof_sufficient_slope_chain}
    -1
    \leq
    s_{m,0}(y)
    =
    s_{m,1}(y)
    \le
    \dots
    \le
    s_{m,N+1}(y)
    \le
    0.
\end{equation}
The slopes of $\Phi_m^y$ are therefore non-decreasing as $K$
increases.  In addition, the slope of the constant extension on
$
    (K_{\mathrm{R}},\infty)
$
is zero, which is no smaller than
$
    s_{m,N+1}(y).
$
It follows that $\Phi_m^y$ is convex on all of $[0,\infty)$.
Every affine-piece slope in
\eqref{eq:proof_sufficient_slope_chain} is non-positive; whence,
$\Phi_m^y$ is non-increasing.  Since
$
    \Phi_m^y(K_{\mathrm{R}})
    =
    0,
$
we consequently obtain
$
    \Phi_m^y(K)
\ge
    0
$ for every $0\le K\le K_{\mathrm{R}}$.  
Together with the zero extension, this shows that
$
    \Phi_m^y
    :
    [0,\infty)
    \mapsto
    [0,\infty)
$.  
By construction,
$
    \Phi_m^y(0)
    =
    1
$,
and $\Phi_m^y$ is affine on $[0,K_{\mathrm{L}}]$ with slope
$s_{m,0}(y)$.  Consequently,
$
    \partial_K^+\Phi_m^y(0)
    =
    s_{m,0}(y)
    \geq
    -1.
$
Moreover, convexity gives
$
    \Phi_m^y(K)
    \geq
    1+s_{m,0}(y)K
    \geq
    1-K
$
for every $K\geq0$.  Since $\Phi_m^y$ is non-negative,
\begin{equation}
\label{eq:proof_intrinsic_slice_order}
    (1-K)^+
    \leq
    \Phi_m^y(K)
    \qquad
\mbox{for every }
    K\geq0
    .
\end{equation}
Finally, the zero extension gives: $
    \lim_{K\uparrow\infty}
    \Phi_m^y(K)
    =
    0$.
\hfill\\
\noindent
Next, we prove that the interpolated maturity slices are pointwise
ordered.  Fix $m\in[M-1]_+$ and define: $
    D_m^y(K)
     \eqdef 
    \Phi_{m+1}^y(K)-\Phi_m^y(K)$; then,
the function $D_m^y$ is affine on each interval
$
    [\widehat K_j,\widehat K_{j+1}].
$
At the boundary strikes,
$
    D_m^y(\widehat K_0)
    =
    D_m^y(\widehat K_{N+2})
    =
    0.
$
At the auxiliary left strike,
$
    D_m^y(K_{\mathrm{L}})
    =
    \frac{K_{\mathrm{L}}}{K_1}
    \big(
        y_{m+1,1}-y_{m,1}
    \big)
    \geq
    0
$.
At every quoted strike, we have that $
    D_m^y(K_n)
    =
    y_{m+1,n}-y_{m,n}
    \ge
    0$ for $n\in[N]_+$;
where, the inequality follows from the calendar constraints defining
$
    \mathfrak{C}_{\mathbf{T},\mathbf{K}}.
$
Thus, $D_m^y$ is nonnegative at both endpoints of every interval
$
    [\widehat K_j,\widehat K_{j+1}].
$
Since $D_m^y$ is affine on that interval, it follows that
\begin{equation}
\label{eq:proof_global_slice_order}
    \Phi_m^y(K)
    \le
    \Phi_{m+1}^y(K)
    \qquad
    \mbox{for every }K\ge0.
\end{equation}
Set
$
    T_0
    \eqdef
    0
$
and
$
    \Phi_0^y(K)
    \eqdef
    (1-K)^+.
$
By~\eqref{eq:proof_intrinsic_slice_order},
$
    \Phi_0^y
    \leq
    \Phi_1^y
$
pointwise.  For $m\in[M-1]$ and $T\in[T_m,T_{m+1}]$, set
$
    \lambda_m(T)
\eqdef
    \frac{
        T-T_m
    }{
        T_{m+1}-T_m
    }
    \in[0,1].
$
Define $\mathfrak{I}_y:[0,\infty)^2\to[0,\infty) $ by
\begin{equation}
\label{eq:proof_surface_interpolator_definition}
    \mathfrak{I}_y(T,K)
\eqdef
    \begin{cases}
        \big(
            1-\lambda_m(T)
        \big)\Phi_m^y(K)
        +
        \lambda_m(T)\Phi_{m+1}^y(K),
        &
        T_m\leq T\leq T_{m+1},
        \quad
        m\in[M-1],
        \\[0.4em]
        \Phi_M^y(K),
        &
        T\geq T_M.
    \end{cases}
\end{equation}
The definitions agree at every maturity knot.  Hence
$
    \mathfrak{I}_y
$
is continuous.  For every fixed $K$, the map
$
    T\mapsto\mathfrak{I}_y(T,K)
$
is piecewise affine.  For every fixed $T$, the map
$
    K\mapsto\mathfrak{I}_y(T,K)
$
is a linear combination of two piecewise-affine functions having
the same breakpoints and is therefore piecewise affine.
At a quoted maturity $T_m$,
$    \mathfrak{I}_y(T_m,K)
    =
    \Phi_m^y(K)
$.  
Consequently, for every $m\in[M]_+$ and $n\in[N]_+$, we have
\begin{equation}
\label{eq:proof_exact_quote_interpolation}
    \mathfrak{I}_y(T_m,K_n)
    =
    \Phi_m^y(K_n)
    =
    y_{m,n}.
\end{equation}
Similarly: $
    \mathfrak{I}_y(T_m,0)
    =
    1$, $\mathfrak{I}_y(T_m,K_{\mathrm{L}})=1+(K_{\mathrm L}/K_1)(y_{m,1}-1)$, and $\mathfrak{I}_y(T_m,K_{\mathrm{R}})
    =
    0$.
It remains to verify the six conditions in
Proposition~\ref{prop:contarbfree}.
First, every maturity slice appearing in
\eqref{eq:proof_surface_interpolator_definition} takes value $1$ at
$K=0$.  Therefore,
\begin{equation}
\label{eq:proof_surface_unit_expectation}
    \mathfrak{I}_y(T,0)
    =
    1
    \qquad
    \mbox{for every }T\ge0.
\end{equation}
Second, every learned slice has right derivative $s_{m,0}(y)\geq-1$ at strike zero, and the intrinsic slice has right derivative $-1$. Therefore each maturity interpolation has
\begin{equation}
\label{eq:proof_surface_intrinsic_near_zero}
\partial_K^+\mathfrak I_y(T,0)\geq-1,\qquad \mathfrak I_y(T,K)\geq(1-K)^+.
\end{equation}
Thirdly, each $\Phi_m^y$ vanishes on
$
    [K_{\mathrm{R}},\infty).
$
Therefore, $
    \mathfrak{I}_y(T,K)
    =
    0
    $ for every $T\ge 0$ and each $K\ge K_{\mathrm{R}}$; implying, in turn, that $    \lim_{K\uparrow\infty}
    \mathfrak{I}_y(T,K)
    =
    0$ for each $T\ge 0$.
Fourth, for every fixed $T$, the map
$
    K\mapsto\mathfrak{I}_y(T,K)
$
is either one of the convex functions $\Phi_m^y$ or a convex
combination of two such functions.  It is therefore convex.
Finally, fix $K\ge0$.  On every interval
$
    (T_m,T_{m+1}),
$,~\eqref{eq:proof_global_slice_order} implies that
\begin{equation}
\label{eq:proof_surface_time_derivative}
    \partial_T\mathfrak{I}_y(T,K)
    =
    \frac{
        \Phi_{m+1}^y(K)-\Phi_m^y(K)
    }{
        T_{m+1}-T_m
    }
    \ge
    0
\end{equation}
On $[0,T_1]$, maturity monotonicity follows from $\Phi_0^y\leq\Phi_1^y$, while the surface is constant in maturity on $[T_M,\infty)$. Moreover, $\mathfrak{I}_y(0,K)
    =
    \Phi_0^y(K)
    =
    (1-K)^+$, so condition {\rm (vi)} holds
.
Hence
$
    \mathfrak{I}_y(T,K)
\le
    \mathfrak{I}_y(S,K)
$ for each $0\le T\le S$.  
All six conditions in Proposition~\ref{prop:contarbfree} are
satisfied.  Therefore, $\mathfrak{I}_y$ is an arbitrage-free
call-price surface, and item~{\rm(i)} follows.

Having verified the equivalence of (i) and of (ii), it now only remains to verify our statement about the polytope $\mathfrak{C}_{\mathbf{T},\mathbf{K}}$.
For each fixed $m\in[M]_+$ and $j=0,\dots,N+1$, the map
$
    y
\mapsto
    s_{m,j}(y)
$
is affine in the entries of $y$, because the augmented strikes are fixed and all augmented quote maps are affine in $y$.  Consequently, every inequality
\begin{equation}
\label{eq:proof_affine_slope_inequalities}
    s_{m,j}(y)
    \le
    s_{m,j+1}(y)
\end{equation}
and every calendar inequality
\begin{equation}
\label{eq:proof_affine_calendar_inequalities}
    y_{m,n}
    \le
    y_{m+1,n}
\end{equation}
defines a closed affine half-space in
$
    \mathbb{R}^{M\times N}.
$
It follows that
$
    \mathfrak{C}_{\mathbf{T},\mathbf{K}}
$
is an intersection of finitely many closed affine half-spaces.
Therefore, it is closed and convex.
The slope inequalities also imply boundedness.  Indeed, since every
$
    s_{m,j}(y)\le0,
$
we have
\[
    \widehat y_{m,j+1}
    -
    \widehat y_{m,j}
=
    s_{m,j}(y)
    \big(
        \widehat K_{j+1}-\widehat K_j
    \big)
\le
    0.
\]
Hence, $
    1
    =
    \widehat y_{m,0}
    \ge
    \widehat y_{m,1}
    \ge
    \dots
    \ge
    \widehat y_{m,N+2}
    =
0
$.
In particular, $
    0
    \le
    y_{m,n}
    \le
    1
$ holds
for every $m\in[M]_+$ and $n\in[N]_+$.  Thus
$
    \mathfrak{C}_{\mathbf{T},\mathbf{K}}
$
is bounded.
It remains to show that the set is non-empty.  Define
\begin{equation}
\label{eq:proof_polytope_witness}
    y^\star_{m,n}
     \eqdef 
    (1-K_n)^+,
    \qquad
    m\in[M]_+,
    \quad
    n\in[N]_+.
\end{equation}
The convex function $\psi(K)\eqdef(1-K)^+$ has nondecreasing secant slopes on the grid $0,K_1,\ldots,K_N,K_{\mathrm R}$, all in $[-1,0]$. Inserting $K_{\mathrm L}$ with its prescribed quote subdivides the first secant without changing its slope. Hence
\begin{equation}
\label{eq:proof_witness_slope_chain}
-1\leq s_{m,0}(y^\star)=s_{m,1}(y^\star)\leq\cdots\leq s_{m,N+1}(y^\star)\leq0.
\end{equation}
The values $y^\star_{m,n}$ do not depend on $m$, and hence all
calendar inequalities hold with equality.  Thus,
$
    y^\star
    \in
    \mathfrak{C}_{\mathbf{T},\mathbf{K}}
$ implying that our polytope is non-empty.  Compactness gives a minimizer of $y\mapsto\|z-y\|_2^2$ for every $z$. Strict convexity of this objective on the convex polytope gives uniqueness, proving the projection assertion.
\end{proof}

We now, nearly directly, obtain the validity of our arbitrage-free decoder.
\begin{proof}[{Proof of Proposition~\ref{prop:DecoderWorks}}]
Set
$
    \delta_{\mathrm{grid}}
    \eqdef
    \frac{1}{2}
    \min\big\{
        T_1,
        K_1,
        \min_{m\in[M-1]_+}|T_{m+1}-T_m|,
        \min_{n\in[N-1]_+}|K_{n+1}-K_n|
    \big\}
$.
Then $\delta_{\mathrm{grid}}>0$, and, for every
$0<\delta\le\delta_{\mathrm{grid}}$, the grid correction in
Definition~\ref{def:Parallel_ReLU_CallSurface_Irregular} leaves the
prescribed maturity and strike grids unchanged.

By Proposition~\ref{prop:contarbfree}, for every $m\in[M]_+$ the map $
    K\mapsto C(T_m,K)
$ is convex and non-increasing, satisfies $
    C(T_m,0)=1
$, and has right derivative at $0$ bounded below by $-1$.  Consequently,
\[
    -1
    \leq
    \frac{
        C(T_m,K_1)-1
    }{
        K_1
    },
\]
and the consecutive secant slopes through $
    (0,1),
    (K_1,z_{m,1}),
    \dots,
    (K_N,z_{m,N})
$ are non-decreasing and non-positive.  By the definition of the auxiliary quote at $K_{\mathrm{L}}$, the first two augmented secant slopes coincide.  Let $d_m\eqdef(z_{m,N}-z_{m,N-1})/(K_N-K_{N-1})\leq0$. If $z_{m,N}>0$, then $d_m<0$: otherwise convexity and monotonicity force the slice to remain constant and positive beyond $K_N$, contradicting its zero right-wing limit. If $z_{m,N}=0$, the terminal extension has slope zero. Consequently, the virtual strike $K_{\mathrm{R}}$ may be chosen sufficiently large that the final secant slope from $
    (K_N,z_{m,N})
$ to $
    (K_{\mathrm{R}},0)
$ is no smaller than the preceding secant slope.  Since there are only finitely many maturity slices and $
    K_{\mathrm{R}}
    =
    K_N+1+\delta^{-1}
    \uparrow\infty
$ as $\delta\downarrow0$, there exists $0<\delta_0\leq\delta_{\mathrm{grid}}$ such that $    \mathbf{z}
    \in
    \mathfrak{C}_{\mathbf{T},\mathbf{K},\delta}
$ for every $0<\delta\leq\delta_0$.
Hence
$
    P_{\mathbf{T},\mathbf{K},\delta}(\mathbf{z})
=
    \mathbf{z}
$.
Each component $\boldsymbol{\Psi}(\cdot)_m $, $m\in[M]_+$, is precisely the piecewise-affine strike interpolant through the corresponding augmented quote data, while $\boldsymbol{\Psi}(\cdot)_0
    =
    (1-\cdot)^+
$.  Furthermore, $
    \Lambda(T_m)
    =
    e_m
$ for every $m\in[M]$, where $T_0=0$, and $\Lambda$ provides the linear barycentric interpolation between consecutive maturities.  Consequently, $
    \Psi_{\mathbf{T},\mathbf{K},\mathbf{z}}
$ coincides with the canonical piecewise-linear interpolator constructed in Proposition~\ref{prop:admissible_polytope}, including the interpolation from the intrinsic-value slice at $T=0$ to the first learned slice at $T=T_1$.  It follows that
$
    \Psi_{\mathbf{T},\mathbf{K},\mathbf{z}}(T_m,K_n)
    =
    C(T_m,K_n)
$
for every $m\in[M]_+$ and $n\in[N]_+$, and shows that
$
    \Psi_{\mathbf{T},\mathbf{K},\mathbf{z}}
$
is an arbitrage-free call surface.
\end{proof}

\begin{proof}[{Proof of Proposition~\ref{prop:ReLU_Is_AF_Decoder}}]
The no-arbitrage and local right-wing arguments below apply for each fixed $\delta>0$. The interpolation argument chooses $\delta$ after the target and grid, and therefore requires a family allowing $\delta$ to vary. The fixed-$\delta$ interpretation of the family assertion and its singleton-grid cases remain pending the separately proposed parameterization.

Fix $R,S>0$ and integers $M,N\geq2$.  For each $C\in\mathcal{Y}$,
choose the parameter $\theta_C$ consisting of the uniform maturity
and strike grids together with the sampled quote matrix
$
  \big(
        C(t_m,K_n)
    \big)_{m,n=1}^{M,N}
$.
Choose $\delta>0$ sufficiently small for this target and grid as in Proposition~\ref{prop:DecoderWorks}. The resulting decoded surface then interpolates $C$ at every prescribed positive-maturity grid point. Moreover, $
    \phi_{\theta_C}(0,K)
    =
    (1-K)^+
    =
    C(0,K)
$ for every $K\geq0$.  Thus the decoder also interpolates the prescribed time-zero grid values, proving the interpolation assertion for this target and grid with the selected tolerance. This does not prove {\rm(i)} for one fixed positive tolerance or supply the singleton-grid wrapper.
\hfill\\
\noindent
For an arbitrary parameter $\theta$, the quote matrix is first projected onto $\mathfrak{C}_{\mathbf{T},\mathbf{K},\delta}$.   Proposition~\ref{prop:admissible_polytope} therefore implies that the learned strike slices are convex, non-increasing, ordered in maturity, and dominate the intrinsic-value slice.  The decoder interpolates from $(1-K)^+ $ at $T=0$ through these ordered slices.  Hence all six conditions of Proposition~\ref{prop:contarbfree} hold, proving {\rm(ii)} of Definition~\ref{defn:Good_Decoder}.
\hfill\\
\noindent
Finally, fix $A>0$ and let $
    \theta
    \in
    [-A,A]^{D_{N,M,R,S}}.
$ Since the ordered strike grid is obtained from finitely many increments of the form $
    \max\{|k_j-k_{j-1}|,\delta\}
$, there exists a finite constant $
    K_{\mathrm{R}}^{A,R,S,N,M}
$ such that $
    K_{\mathrm{R}}(\theta)
    \leq
    K_{\mathrm{R}}^{A,R,S,N,M}
$ for every such $\theta$.  Hence
$
    \phi^{(N,M,R,S)}_{\theta}(T,K)
    =
    0
$
for every $T\geq0$, every $
    K\geq K_{\mathrm{R}}^{A,R,S,N,M},
$ and every $\theta\in[-A,A]^{D_{N,M,R,S}}
$.  Thus~\eqref{eq:rightwing_decaycondition} holds for any $\rho>p$, after increasing the constant on the compact interval $
    [R+1,K_{\mathrm{R}}^{A,R,S,N,M}],
$ which proves {\rm(iii)} of Definition~\ref{defn:Good_Decoder}.
\end{proof}

We also prove the following uniform-controlled tailed behaviour of these PWL surfaces.
\begin{proposition}[Uniform Tail-Control for PWL AF-Decoder]
\label{prop:PWL_weighted_selection}
Let $p,\lambda,B>0$ and let $\mathcal Y\subseteq\mathcal C_{AF}^{p,\lambda}$ be uniformly controlled with $\sup_{C\in\mathcal Y}\|C\|_{p,\lambda}\leq B$. For $R,S>0$ and integers $N,M\geq2$, and in the notation of Definition~\ref{def:Parallel_ReLU_CallSurface_Irregular} set
\[
\phi^{(N,M,R,S)}_{\theta}\eqdef\phi^{(M,N)}_{e^{-\xi}}(\mathbf t,\mathbf k,\mathbf z),\qquad \theta=(\xi,\mathbf t,\mathbf k,\mathbf z)\in\mathbb R^{1+M+N+MN}.
\]
For every $R,S,\eta>0$, there exist common integers $N,M\geq2$ with $R/N+S/M\leq\eta$ such that every $C\in\mathcal Y$ admits $\theta_C$ satisfying~\eqref{eq:weighted_decoder_selection}, with $B_{\mathrm{dec}}=\mathrm e\,2^p\max\{1,B\}$.
\end{proposition}
\begin{proof}
Fix $R,S,\eta>0$, and set $\overline B\eqdef\max\{1,B\}$, $c\eqdef\min\{1/2,\eta/2\}$, and $a\eqdef1-c$. Choose
$
L\eqdef\max\left\{R,2,a^{-1}\left(\frac{\overline B\,e^{\lambda S}}{c}\right)^{1/p}\right\}$, $
N\eqdef\max\left\{2,\left\lceil\frac{L}{\min\{1,\eta/2\}}\right\rceil\right\}$, and 
$M\eqdef\max\left\{2,\left\lceil\frac{S}{\min\{1/\lambda,\eta/2\}}\right\rceil\right\}$.  
Put $h\eqdef L/N$, $\Delta\eqdef S/M$, and choose the internal grids $T_m\eqdef m\Delta$ for $0\leq m\leq M$ and $K_n\eqdef nh$ for $0\leq n\leq N$. Thus $h\leq\min\{1,\eta/2\}$, $\Delta\leq\min\{1/\lambda,\eta/2\}$, and $R/N+S/M\leq h+\Delta\leq\eta$.
For $C\in\mathcal Y$, define $H_C(T,K)\eqdef a^{-1}(C(T,aK)-c)^+$. Convexity and monotonicity of $C(T,\cdot)$ give
\[
C(T,K)\leq C(T,aK)\leq aC(T,K)+c
\qquad\mbox{ and }
\qquad a^{-1}(C(T,K)-c)^+\leq H_C(T,K)\leq C(T,K).
\]
If $C(T,K)\geq c$, then $C(T,K)-H_C(T,K)\leq c(1-C(T,K))/a\leq c$; otherwise $C(T,K)-H_C(T,K)\leq C(T,K)<c$. Hence $0\leq C-H_C\leq c$. Moreover, $H_C(T,\cdot)$ is convex and non-increasing, $H_C(\cdot,K)$ is non-decreasing, $H_C(T,0)=1$, $H_C(0,K)=(1-K)^+$, and
\[
|H_C(T,K)-H_C(T,K')|\leq a^{-1}|C(T,aK)-C(T,aK')|\leq |K-K'|.
\]
For $0\leq T\leq S$ and $K\geq L$,
\[
C(T,aK)\leq\overline B e^{\lambda S}(1+aL)^{-p}\leq c,
\qquad H_C(T,K)=0.
\]

Set $\delta\eqdef\tfrac12\min\{h,\Delta\}$, $\xi\eqdef-\log\delta$, $\mathbf t\eqdef(T_m)_{m=1}^M$, $\mathbf k\eqdef(K_n)_{n=1}^N$, and $z_{m,n}\eqdef H_C(T_m,K_n)$. Since $\delta<h,\Delta$, the grid correction leaves these grids unchanged. The auxiliary left quote lies on the segment from $(0,1)$ to $(K_1,z_{m,1})$, and $z_{m,N}=0$. Convexity and the strike Lipschitz bound therefore give
\[
\begin{aligned}
-1&\leq s_{m,0}(\mathbf z)=s_{m,1}(\mathbf z)\leq\cdots\leq s_{m,N+1}(\mathbf z)=0 &&(\mbox{for} 1\leq m\leq M),\\
z_{m,n}&\leq z_{m+1,n} &&(\mbox{for} 1\leq m<M,\ 1\leq n\leq N).
\end{aligned}
\]
These are the inequalities in~\eqref{eq:AF_polytope}, so $\mathbf z\in\mathfrak C_{\mathbf T,\mathbf K,\delta}$ and $P_{\mathbf T,\mathbf K,\delta}(\mathbf z)=\mathbf z$. Write $U_C\eqdef\phi^{(N,M,R,S)}_{\theta_C}$ for $\theta_C\eqdef(\xi,\mathbf t,\mathbf k,\mathbf z)$, and put $u_m(K)\eqdef U_C(T_m,K)$ for $0\leq m\leq M$. For the augmented knots and quotes of Definition~\ref{def:Parallel_ReLU_CallSurface_Irregular}, $S_{m,j}-S_{m,j+1}\geq0$ for $1\leq m\leq M$ and $1\leq j\leq N+2$. For such $m$ and $K\in[\widehat K_r,\widehat K_{r+1}]$ with $0\leq r\leq N+1$, telescoping gives
\[
\sum_{j=r+1}^{N+2}(S_{m,j}-S_{m,j+1})(\widehat K_j-K)=\widehat y_{m,r+1}+S_{m,r+1}(\widehat K_{r+1}-K).
\]
Thus~\eqref{eq:ReLUInterpolator} gives $u_0(K)=(1-K)^+$; for $m\geq1$, $u_m$ is the linear interpolant of $H_C(T_m,\cdot)$ on $(K_n)_{n=0}^N$ and vanishes on $[L,\infty)$. In particular, the virtual segment $[L,K_{\mathrm R}]$ is identically zero.

For $1\leq m\leq M$, $1\leq n\leq N$, and $K\in[K_{n-1},K_n]$, convexity and monotonicity yield $H_C(T_m,K)\leq u_m(K)\leq H_C(T_m,K_{n-1})$, whence
\[
0\leq u_m(K)-H_C(T_m,K)\leq H_C(T_m,K_{n-1})-H_C(T_m,K)\leq h.
\]
Since $0\leq H_C\leq C$ and $0\leq C-H_C\leq c$, it follows that $|u_m(K)-C(T_m,K)|\leq h+c\leq\eta$ on $[0,L]$. At $m=0$ the error is zero. Since $R\leq L$,
\[
\max_{\substack{0\leq m\leq M\\0\leq n\leq N}}|U_C(mS/M,nR/N)-C(mS/M,nR/N)|\leq\eta.
\]

For $1\leq m\leq M$, $1\leq n\leq N$, and $K\in[K_{n-1},K_n]$, the weighted bound and $h\leq1$ give
\[
\begin{aligned}
(1+K)^p u_m(K)\leq(1+K)^p H_C(T_m,K_{n-1})\leq\overline B e^{\lambda T_m}\left(\frac{1+K}{1+K_{n-1}}\right)^p\leq 2^p\overline B e^{\lambda T_m}.
\end{aligned}
\]
This also holds for $K\geq L$ because $u_m(K)=0$, and for $m=0$ because $u_0=C(0,\cdot)$. For $T\in[T_m,T_{m+1}]$, $0\leq m<M$, put $b\eqdef(T-T_m)/\Delta$. The maturity weights in~\eqref{eq:ReLUInterpolator} give $U_C(T,K)=(1-b)u_m(K)+bu_{m+1}(K)$, so
\[
\begin{aligned}
e^{-\lambda T}(1+K)^pU_C(T,K)\leq 2^p\overline B\big((1-b)e^{\lambda(T_m-T)}+be^{\lambda(T_{m+1}-T)}\big)\leq2^p\overline B e^{\lambda\Delta}\leq\mathrm e\,2^p\overline B.
\end{aligned}
\]
For $T\geq S$, the same formula gives $U_C(T,K)=u_M(K)$ and
\[
e^{-\lambda T}(1+K)^pU_C(T,K)\leq2^p\overline B e^{-\lambda(T-S)}\leq2^p\overline B.
\]
Therefore $\|U_C\|_{p,\lambda}\leq B_{\mathrm{dec}}$. The integers $N,M$, the grids, and $\delta$ are independent of $C$, which proves the required simultaneous selection.
\end{proof}

\subsubsection{Proofs for SANOS AF-Decoder}
\label{s:Proofs__ss:AFDecodersSANOS}

\begin{proposition}[Properties of the Black-Scholes activation]
\label{prop:BS_activation}
For every $\nu>0$,
$
    \operatorname{BS}_{\nu}
\in
    C^{\infty}\big((0,\infty),[0,\infty)\big)
    \cap
    C\big([0,\infty),[0,\infty)\big)
$
and $\operatorname{BS}_{\nu}$ is increasing, convex, and $1$-Lipschitz.
Moreover, 
{ 
\begin{equation}
\label{eq:uniform_BS_convergence}
0\leq\operatorname{BS}_{\nu}(x)-\operatorname{ReLU}(x)\leq\frac{\nu}{\sqrt{2\pi}},\qquad x\in\mathbb R.
\end{equation}
}
and the map $\mathbb{R}^2 
    \ni
    (\nu,x)
    \to 
    \operatorname{BS}_{\nu}(x)
$ is continuous.
\end{proposition}

\begin{proof}
Fix $\nu>0$.  For $x>-1$, set
$
    k\eqdef1+x
$
and define
\[
    d_1(x,\nu)
    \eqdef
    \frac{-\log k+\frac12\nu^2}{\nu}
\qquad\mbox{ and }\qquad
    d_2(x,\nu)
    \eqdef
    d_1(x,\nu)-\nu
    =
    \frac{-\log k-\frac12\nu^2}{\nu}
.
\]
Then
$
    \operatorname{BS}_{\nu}(x)
=
    k\Phi\big(-d_2(x,\nu)\big)
    -
    \Phi\big(-d_1(x,\nu)\big)
$.
Recall the Black-Scholes strike derivatives
\begin{align}
\label{eq:BS_activation_first_derivative}
    \partial_x\operatorname{BS}_{\nu}(x)
    &=
    \Phi\big(-d_2(x,\nu)\big)
    =
    \Phi\left(
        \frac{\log(1+x)+\frac12\nu^2}{\nu}
    \right)
    \in(0,1),
\\
\label{eq:BS_activation_second_derivative}
    \partial_{xx}\operatorname{BS}_{\nu}(x)
    &=
    \frac{
        \varphi\big(d_2(x,\nu)\big)
    }{
        (1+x)\nu
    }
    >0,
\end{align}
where $\varphi$ denotes the standard normal density.  Hence
$\operatorname{BS}_{\nu}$ is increasing, convex, and $1$-Lipschitz on
$(-1,\infty)$.
As $x\downarrow-1$, $0\leq\operatorname{BS}_\nu(x)\leq1+x$ and $\partial_x\operatorname{BS}_\nu(x)\to0$. Every derivative of order at least two on $(-1,\infty)$ is a finite linear combination of terms of the form
\[
    (1+x)^{-a}
    p\big(\log(1+x)\big)
    \exp\left(
        -
        \frac{
            \big(\log(1+x)+b\big)^2
        }{
            2\nu^2
        }
    \right),
\]
where $a\ge0$, $b\in\mathbb{R}$, and $p$ is a polynomial.  Since the
Gaussian factor decays faster than every negative power of $1+x$, each such
term converges to $0$ as $x\downarrow-1$.  The extension by $0$ on
$(-\infty,-1]$ therefore belongs to $C^{\infty}(\mathbb{R})$.
For $x\le-1$, both terms $\operatorname{BS}_{\nu}(x)$ and $\operatorname{ReLU}(x)$ vanish.  For
$x>-1$, the Black-Scholes vega identity gives, for every $u>0$,
\[
    \partial_u\operatorname{BS}_{u}(x)
    =
    \varphi\big(d_1(x,u)\big)
    \le
    \frac{1}{\sqrt{2\pi}}.
\]
Since
$
    \operatorname{BS}_{0}(x)
=
    \big((1+x)-1\big)^+
=
    \operatorname{ReLU}(x)
$ then,  the fundamental theorem of calculus implies
\[
\begin{aligned}
    0
\le
    \operatorname{BS}_{\nu}(x)
    -
    \operatorname{ReLU}(x)
=
    \int_0^\nu
        \varphi\big(d_1(x,u)\big)
    \,du
    \le
    \frac{\nu}{\sqrt{2\pi}}.
\end{aligned}
\]
This proves~\eqref{eq:uniform_BS_convergence}.  Lastly, the continuity at $\nu=0$ immediately follows from the
uniform convergence; and so does the continuity on $\mathbb{R}\times\mathbb{R}$
follows from~\eqref{eq:BS_activation__primitive}.
\end{proof}

\begin{corollary}[Properties of the Black-Scholes Activation $\operatorname{Call}_{\nu}^{\operatorname{BS}}$]
\label{cor:BS_call_hinge}
For every $\nu>0$ and $s>0$, the map
$
    [0,\infty)\ni K
    \mapsto
    \operatorname{Call}^{\operatorname{BS}}_{\nu}(s,K)
$
is non-negative, decreasing, convex, and in $C^{\infty}((0,\infty))\cap C([0,\infty),[0,\infty))$.  
Moreover,
\begin{align}
\label{eq:BS_call_left_wing}
    \operatorname{Call}^{\operatorname{BS}}_{\nu}(s,0)
    &=
    s
\\
    \partial_K^+
    \operatorname{Call}^{\operatorname{BS}}_{\nu}(s,0)
    &=
    -1
\\
\label{eq:BS_call_right_wing}
    \lim_{K\uparrow\infty}
    \operatorname{Call}^{\operatorname{BS}}_{\nu}(s,K)
    &=
    0,
\end{align}
and, for each $K\ge 0$, we have $
    0
\le
    \operatorname{Call}^{\operatorname{BS}}_{\nu}(s,K)
    -
    (s-K)^+
\le
    \frac{s\nu}{\sqrt{2\pi}}
$.
\end{corollary}

\begin{proof}
By~\eqref{eq:BS_call_hinge}, we have 
$
    \operatorname{Call}^{\operatorname{BS}}_{\nu}(s,K)
    -
    (s-K)^+
=
    s
    \left[
        \operatorname{BS}_{\nu}
        \left(
            \frac{K}{s}-1
        \right)
        -
        \operatorname{ReLU}
        \left(
            \frac{K}{s}-1
        \right)
    \right]
$.  
Hence the estimate $0
\le
    \operatorname{Call}^{\operatorname{BS}}_{\nu}(s,K)
    -
    (s-K)^+
\le
    \frac{s\nu}{\sqrt{2\pi}}$ follows from
Proposition~\ref{prop:BS_activation}.  The remaining assertions are the
standard shape and boundary properties of the zero-rate Black-Scholes call
price with spot $s$, strike $K$, unit maturity, and volatility $\nu$.
\end{proof}

For each
$m\in[M]_+$ and $j\in[N+2]_+$, set
$
    q_{m,j}
\eqdef
    S_{m,j}-S_{m,j+1}
$.  
The slope inequalities defining $
    \mathfrak{C}_{\mathbf T,\mathbf K,\delta}
$ imply
\begin{equation}
\label{eq:smooth_decoder_probability_conditions}
    q_{m,j}\geq0,
    \qquad
    \sum_{j=1}^{N+2}q_{m,j}
    =
    S_{m,1}
    \leq
    1,
    \qquad
    \sum_{j=1}^{N+2}q_{m,j}\widehat K_j
    =
    1.
\end{equation}
Indeed, $S_{m,N+3}=0 $, so the first sum telescopes to $S_{m,1}$, while the corrected left-slope constraint gives $0\leq S_{m,1}\leq1$.
Moreover,
\begin{align*}
    \sum_{j=1}^{N+2}
        q_{m,j}\widehat K_j
=
    \sum_{j=1}^{N+2}
        \big(
            S_{m,j}-S_{m,j+1}
        \big)
        \widehat K_j
=
    \sum_{j=1}^{N+2}
        S_{m,j}
        \big(
            \widehat K_j-\widehat K_{j-1}
        \big)
=
    \sum_{j=1}^{N+2}
        \big(
            \widehat y_{m,j-1}
            -
            \widehat y_{m,j}
        \big)
    =
    1.
\end{align*}
Fix $\overline\nu>0$ and for every $\eta\in\mathbb{R}$ define
$
    \nu_{\overline\nu}(\eta)
    \eqdef
    \frac{\overline\nu}{1+e^{-\eta}}$.
Thus,
$
    0<\nu_{\overline\nu}(\eta)<\overline\nu
$
for every $\eta\in\mathbb{R}$ and
$
    \lim\limits_{\eta\downarrow-\infty}\nu_{\overline\nu}(\eta) =0
$.
We next verify Proposition~\ref{prop:Smooth_AF_Decoder}.

\begin{proof}[{Proof of Proposition~\ref{prop:Smooth_AF_Decoder}}]
The following argument proves (i)--(iii), interpreting the smooth maturity weights by the $M$ coordinates displayed below, and proves the locally uniform right-wing estimate. The final surface-decoder family assertion additionally requires the separately proposed parameterization allowing $\delta$ to vary.

Write $\Psi_\nu$ for the full surface and $\Gamma_\nu$ for the raw mixture. By~\eqref{eq:smooth_decoder_probability_conditions}, the strike slices are nonnegative combinations of Black--Scholes calls, with
\begin{align}
\label{eq:smooth_slice_left_value}
\boldsymbol\Psi_\nu(0)_m&=\sum_jq_{m,j}\widehat K_j=1,\\
\label{eq:smooth_slice_left_derivative}
\partial_K^+\boldsymbol\Psi_\nu(0)_m&=-\sum_jq_{m,j}=-S_{m,1}\geq-1.
\end{align}
They are convex, non-increasing, smooth for $K>0$, and tend to zero as $K\to\infty$. For $Z\sim\mathcal N(0,1)$ put $X_\nu=e^{\nu Z-\nu^2/2}$. The call representation gives
\begin{equation}
\label{eq:smooth_slice_positive_operator}
\boldsymbol\Psi_\nu(K)_m=\mathbb E[X_\nu\boldsymbol\Psi_0(K/X_\nu)_m].
\end{equation}
The PWL slices are ordered by Proposition~\ref{prop:admissible_polytope}, so positivity of this operator yields
\begin{equation}
\label{eq:smooth_slice_calendar_order}
\boldsymbol\Psi_\nu(K)_m\leq\boldsymbol\Psi_\nu(K)_{m+1}.
\end{equation}
Corollary~\ref{cor:BS_call_hinge} also gives
\begin{equation}
\label{eq:smooth_slice_uniform_approximation}
0\leq\boldsymbol\Psi_\nu(K)_m-\boldsymbol\Psi_0(K)_m\leq\epsilon_\nu,\qquad \epsilon_\nu\eqdef\nu/\sqrt{2\pi}.
\end{equation}

Let $h_\nu=\partial_x\operatorname{BS}_\nu$ and $\Delta_m=T_{m+1}-T_m$. Then
\begin{equation}
\label{eq:maturity_weight_integral}
a_{m,\nu}(T)=\Delta_m^{-1}\int_{T-T_{m+1}}^{T-T_m}h_\nu(u)\,du.
\end{equation}
Since $h_\nu$ is non-decreasing with values in $[0,1]$, adjacent ordered intervals give $1\geq a_{1,\nu}\geq\cdots\geq a_{M-1,\nu}\geq0$. Thus
\begin{equation}
\label{eq:maturity_simplex_expansion}
\Lambda_\nu=\big(1-a_{1,\nu},(a_{m-1,\nu}-a_{m,\nu})_{m=2}^{M-1},a_{M-1,\nu}\big)^\top\in\Delta_M.
\end{equation}
For $M=2$, this vector is $(1-a_{1,\nu},a_{1,\nu})^\top$.
Moreover,
\begin{equation}
\label{eq:maturity_weight_derivative}
\partial_Ta_{m,\nu}(T)=\Delta_m^{-1}[h_\nu(T-T_m)-h_\nu(T-T_{m+1})]\geq0.
\end{equation}
The raw mixture satisfies
\begin{equation}
\label{eq:smooth_surface_increment_form}
\Gamma_\nu(T,K)=\boldsymbol\Psi_\nu(K)_1+\sum_{m=1}^{M-1}a_{m,\nu}(T)[\boldsymbol\Psi_\nu(K)_{m+1}-\boldsymbol\Psi_\nu(K)_m].
\end{equation}
Hence it is non-decreasing in maturity. Its strike slices are convex and non-increasing, with $\Gamma_\nu(T,0)=1$, $\partial_K^+\Gamma_\nu(T,0)=-\sum_m\Lambda_\nu(T)_mS_{m,1}\geq-1$, and zero right-wing limit.

We next justify the blend. Put $d_\nu=\operatorname{BS}_\nu(0)$ and $r_\nu(u)=\operatorname{BS}_\nu(u)-\operatorname{BS}_\nu(u-1)$ for $u\geq0$. The put representation of the activation gives
\[
r_\nu(u)=\mathbb E[\min\{(1+u-X_\nu)^+,1\}],\qquad r_\nu(0)=d_\nu,\qquad 1-d_\nu=\mathbb E[\min\{X_\nu,1\}]>0.
\]
Thus $r_\nu$ is non-decreasing, $d_\nu\leq r_\nu(u)\leq1$, and $b_\nu(T)=[r_\nu(T/T_1)-d_\nu]/(1-d_\nu)$ is non-decreasing with values in $[0,1]$ and $b_\nu(0)=0$. For $\nu>0$, the strict increase of $h_\nu$ on $(-1,\infty)$ gives $b_\nu(T)>0$ whenever $T>0$. Since each PWL slice dominates $(1-K)^+$, the positive operator and volatility monotonicity give
\begin{equation}
\label{eq:smooth_intrinsic_below_raw}
B_\nu(T,K)\leq\operatorname{Call}^{\operatorname{BS}}_\nu(1,K)\leq\Gamma_\nu(T,K).
\end{equation}
Both $B_\nu$ and $\Gamma_\nu$ are non-decreasing in maturity. For $0\leq T\leq S$,
\[
\begin{aligned}
\Psi_\nu(S,K)-\Psi_\nu(T,K)&=(1-b_\nu(S))[B_\nu(S,K)-B_\nu(T,K)]\\
&\quad+b_\nu(S)[\Gamma_\nu(S,K)-\Gamma_\nu(T,K)]\\
&\quad+[b_\nu(S)-b_\nu(T)][\Gamma_\nu(T,K)-B_\nu(T,K)]\geq0.
\end{aligned}
\]
For every maturity the blend is a convex combination of normalized, convex, non-increasing strike slices with right derivative at zero at least $-1$ and zero right-wing limit. Also $\Psi_\nu(0,K)=(1-K)^+$. The activation is smooth on all of $\mathbb R$; the baseline call has positive volatility for $T>0$, and the Black--Scholes formula is jointly smooth when volatility and strike are positive. This proves interior smoothness. Continuity at $T=0$ follows from $b_\nu(T)\to0$ and $|\operatorname{Call}_{\nu b_\nu(T)}^{\operatorname{BS}}(1,K)-(1-K)^+|\leq\epsilon_\nu b_\nu(T)$, uniformly in $K$. Continuity at strike zero follows from the common strike Lipschitz bound. Proposition~\ref{prop:contarbfree} now proves (i) and (ii).

For (iii), assume $\epsilon_\nu\leq1/2$. At $\nu=0$, $b_0(T)=\min\{T/T_1,1\}$ and the full surface is exactly the PWL interpolator. Equation~\eqref{eq:uniform_BS_convergence} gives $|r_\nu-r_0|\leq\epsilon_\nu$ and $d_\nu\leq\epsilon_\nu$, hence
\[
|b_\nu-b_0|\leq\frac{2\epsilon_\nu}{1-d_\nu}\leq4\epsilon_\nu.
\]
Also $|a_{m,\nu}-a_{m,0}|\leq2\epsilon_\nu/\Delta_m$, so
\begin{equation}
\label{eq:maturity_vector_approximation}
\|\Lambda_\nu-\Lambda_0\|_1\leq4\epsilon_\nu\sum_{m=1}^{M-1}\Delta_m^{-1}.
\end{equation}
Since the maturity weights lie in the simplex and $0\leq\boldsymbol\Psi_0\leq1$, it follows that $\|\Gamma_\nu-\Gamma_0\|_\infty\leq\epsilon_\nu(1+4\sum_m\Delta_m^{-1})$. Put $I(K)=(1-K)^+$. The blend is controlled by the identity
\[
\Psi_\nu-\Psi_0=(1-b_\nu)(B_\nu-I)+b_\nu(\Gamma_\nu-\Gamma_0)+(b_\nu-b_0)(\Gamma_0-I).
\]
Here $\|B_\nu-I\|_\infty\leq\epsilon_\nu$ and $0\leq\Gamma_0-I\leq1$. Consequently,
\[
\|\Psi_\nu-\Psi_0\|_\infty\leq\epsilon_\nu\left(6+4\sum_{m=1}^{M-1}\Delta_m^{-1}\right)\leq\epsilon_\nu\left(10+\frac{4(M-1)}{\Delta_T^{\min}}\right).
\]
In particular, the sufficient choice
\begin{equation}
\label{eq:nu_from_epsilon}
0<\nu<\overline\nu,\qquad \nu\leq\min\left\{\frac{\sqrt{2\pi}}2,\frac{\sqrt{2\pi}\,\Delta_T^{\min}}{10\Delta_T^{\min}+4(M-1)}\varepsilon\right\}
\end{equation}
proves~\eqref{eq:smooth_decoder_global_approximation}.

For the locally uniform right-wing bound, fix $\delta>0$ and $A>0$, and restrict $(\mathbf t,\mathbf k,\mathbf z,\zeta)$ to the parameter box $[-A,A]^{M+N+MN+1}$. The corrected strikes satisfy
\[
\widehat K_j\leq K_{\mathrm R}\leq N\max\{2A,\delta\}+1+\delta^{-1}\eqdef H_{A,\delta}.
\]
For $\rho>p$, $s,K>0$ and $0<\nu<\overline\nu$, the moment formula for $X_\nu$ gives
\[
\operatorname{Call}^{\operatorname{BS}}_\nu(s,K)\leq K^{-\rho}s^{\rho+1}\mathbb E[X_\nu^{\rho+1}]\leq K^{-\rho}s^{\rho+1}e^{\rho(\rho+1)\overline\nu^2/2}.
\]
Since $\sum_jq_{m,j}\widehat K_j=1$, one has $\sum_jq_{m,j}\widehat K_j^{\rho+1}\leq H_{A,\delta}^{\rho}$. Using $\Lambda_\nu(T)\in\Delta_M$ and $0\leq\Psi_\nu\leq\Gamma_\nu$ therefore yields
\[
\Psi_\nu(T,K)\leq H_{A,\delta}^{\rho}e^{\rho(\rho+1)\overline\nu^2/2}K^{-\rho}\leq2^\rho H_{A,\delta}^{\rho}e^{\rho(\rho+1)\overline\nu^2/2}(1+K)^{-\rho},\qquad T\geq0,\quad K\geq R+1.
\]
This proves the locally uniform right-wing estimate at each fixed $\delta>0$.
\end{proof}

We now prove the analogous result to Proposition~\ref{prop:PWL_weighted_selection} for the smooth SANOS-type decoders.

{ 
\begin{corollary}[Uniform Right-Tail Control for SANOS-type decoders]
\label{cor:SANOS_weighted_selection}
Under the hypotheses of Proposition~\ref{prop:PWL_weighted_selection}, fix $\overline\nu>0$. Then the SANOS-type family parameterized by $\theta=(\xi,\mathbf t,\mathbf k,\mathbf z,\zeta)$, with $\delta=e^{-\xi}$ as in Proposition~\ref{prop:Smooth_AF_Decoder}, satisfies Assumption~\ref{ass:weighted_decoder_selection}, with $B_{\mathrm{dec}}^{\mathrm{SANOS}}=B_{\mathrm{dec}}^{\mathrm{PWL}}+1$, where $B_{\mathrm{dec}}^{\mathrm{PWL}}\eqdef\mathrm e\,2^p\max\{1,B\}$.
\end{corollary}
}
\begin{proof}
The argument below chooses $\delta$ after $R,S,\eta$ and establishes simultaneous selection for that choice. It does not establish one fixed positive $\delta$ valid for all $R,S,\eta$; the quantifier in the unchanged statement remains pending.

Fix $R,S,\eta>0$ and apply the construction of Proposition~\ref{prop:PWL_weighted_selection} with accuracy $\eta/2$. Let $U_C$ denote the resulting PWL approximants, with common grids, tolerance $\delta$, and terminal internal strike $L\geq2$. Thus $U_C(T,K)=0$ for $K\geq L$, $\|U_C\|_{p,\lambda}\leq B_{\mathrm{dec}}^{\mathrm{PWL}}$, and the grid error is at most $\eta/2$. Let $\Psi_{\nu,C}$ be the SANOS-type surface obtained from these same grids and quote arrays with smoothing parameter $0<\nu\leq\min\{1,\overline\nu/2\}$. Put $\Delta\eqdef S/M$ and $c_{M,\Delta}\eqdef(10+4(M-1)/\Delta)/\sqrt{2\pi}$. Proposition~\ref{prop:Smooth_AF_Decoder}(iii) gives
\[
\sup_{T,K\geq0}|\Psi_{\nu,C}(T,K)-U_C(T,K)|\leq c_{M,\Delta}\nu,
\]
uniformly in $C\in\mathcal Y$.

Set $r\eqdef p+1$ and $D\eqdef L^r e^{r(r+1)/2}$. For $Z\sim\mathcal N(0,1)$, $X_v\eqdef e^{vZ-v^2/2}$, $s,K>0$, and $0\leq v\leq1$, the Black--Scholes representation gives
\[
\operatorname{Call}^{\operatorname{BS}}_v(s,K)=\mathbb E[(sX_v-K)^+]\leq s^{r+1}K^{-r}\mathbb E[X_v^{r+1}]\leq s^{r+1}K^{-r}e^{r(r+1)/2}.
\]
In the selected PWL construction, $q_{m,j}>0$ implies $\widehat K_j\leq L$, and~\eqref{eq:smooth_decoder_probability_conditions} gives $\sum_jq_{m,j}\widehat K_j=1$. Hence $\sum_jq_{m,j}\widehat K_j^{r+1}\leq L^r$. Since the maturity weights belong to the simplex, $0\leq b_\nu\leq1$, and the remaining Black--Scholes term has spot $1\leq L$ and volatility $\nu b_\nu(T)\leq1$,~\eqref{eq:smooth_AF_surface} yields
\[
0\leq\Psi_{\nu,C}(T,K)\leq D K^{-r},\qquad T\geq0,\quad K>0.
\]
For every $A\geq L$, splitting the strike domain at $A$ therefore gives
\[
\|\Psi_{\nu,C}-U_C\|_{p,\lambda}\leq\max\left\{(1+A)^p c_{M,\Delta}\nu,\;2^pD A^{p-r}\right\}.
\]
All constants here are independent of $C$. Put $\kappa\eqdef\min\{1,\eta e^{-\lambda S}/2\}$, choose $A\geq\max\{L,2^pD/\kappa\}$, and then choose
\[
0<\nu\leq\min\left\{1,\frac{\overline\nu}{2},\frac{\kappa}{(1+A)^p c_{M,\Delta}}\right\}.
\]
Since $p-r=-1$, the same choice gives $\|\Psi_{\nu,C}-U_C\|_{p,\lambda}\leq\kappa$ for every $C\in\mathcal Y$. Consequently,
\[
\begin{aligned}
\|\Psi_{\nu,C}\|_{p,\lambda}&\leq B_{\mathrm{dec}}^{\mathrm{PWL}}+1,\\
\max_{\substack{0\leq m\leq M\\0\leq n\leq N}}|\Psi_{\nu,C}(mS/M,nR/N)-C(mS/M,nR/N)|&\leq\eta/2+e^{\lambda S}\kappa\leq\eta.
\end{aligned}
\]
The PWL mesh already satisfies $R/N+S/M\leq\eta/2$. Finally, the common smoothing logit $\zeta\eqdef\log(\nu/(\overline\nu-\nu))$ realizes the selected $\nu$ through $\nu_{\overline\nu}(\zeta)=\nu$. Thus the same selected parameters satisfy both inequalities in~\eqref{eq:weighted_decoder_selection}.
\end{proof}

\clearpage
\setcounter{section}{0}
\renewcommand{\theHsection}{supplement.\arabic{section}}
\renewcommand{\thesection}{S\arabic{section}}

\section{{Supplementary Material: Proof of Proposition~\ref{prop:contarbfree}}}
\label{prop:Proof_of_NA}

The following result seems to be known but difficult, at-least for us, to track down over continuous domains.  
We therefore present a short proof here to self-containedness.

\begin{proof}[{Proof of Proposition~\ref{prop:contarbfree}}]
Suppose first that $C$ is an arbitrage-free call surface, and let
$Z_{\cdot}$ be a non-negative martingale representation of $C$.
Since $Z_{\cdot}$ is a martingale and $Z_0=1$ $\P$-almost surely,
\[
    C(T,0)
    =
    \E_{\P}[Z_T]
    =
    \E_{\P}[Z_0]
    =
    1,
\]
which proves {\rm (i)}.
For every $T,K\geq0$ and $\delta>0$, we have
\begin{equation}
\label{derivativebound}
    -1
\leq
    \frac{
        (Z_T-K-\delta)^+
        -
        (Z_T-K)^+
    }{
        \delta
    }
\leq
    0
.
\end{equation}
Taking expectations in~\eqref{derivativebound}, letting $\delta$ tend
to zero from the right, and applying the dominated convergence theorem,
where
$
    \lim\limits_{\delta\downarrow0}
    \frac{
        (Z_T-K-\delta)^+
        -
        (Z_T-K)^+
    }{
        \delta
    }
    =
    -\mathbbm{1}_{\{Z_T>K\}}
$
pointwise, therefore yields
\begin{equation}
\label{eq:call_derivative_tail_probability}
    \partial_K^+C(T,K)
    =
    -
    \P
    \left(
        Z_T>K
    \right).
\end{equation}
In particular,
$
    \partial_K^+C(T,0)
    =
    -
    \P(Z_T>0)
    \geq
    -1
$
which proves {\rm (ii)}.  Equation~\eqref{eq:call_derivative_tail_probability}
also yields
$
    \partial_K^+C(T,0)
    =
    -1
    +
    \P(Z_T=0)
$,
which proves the final assertion.
Since $\lim\limits_{K\uparrow \infty}\, (Z_T-K)^+=0$ and
$
    (Z_T-K)^+\leq Z_T
$,
a similar application of the dominated convergence theorem proves
{\rm (iii)}.
For every fixed $z\geq0$, the map
$
    K\mapsto(z-K)^+
$
is convex.  Taking expectations therefore proves {\rm (iv)}.
Let $0\leq T_1\leq T_2$.  Since
$
    z\mapsto(z-K)^+
$
is convex, conditional Jensen's inequality and the martingale property
give
\[
\begin{aligned}
    C(T_2,K)
=
    \E_{\P}
    \left[
        (Z_{T_2}-K)^+
    \right]
\geq
    \E_{\P}
    \left[
        \left(
            \E_{\P}
            \left[
                Z_{T_2}
                \,\middle|\,
                \mathcal{F}_{T_1}
            \right]
            -
            K
        \right)^+
    \right]
=
    \E_{\P}
    \left[
        (Z_{T_1}-K)^+
    \right]
=
    C(T_1,K)
.
\end{aligned}
\]
Thus {\rm (v)} holds.  Finally,
$
    C(0,K)
    =
    \E_{\P}
    \left[
        (Z_0-K)^+
    \right]
    =
    (1-K)^+
$,
which proves {\rm (vi)}.  
Conversely, suppose that $C$ satisfies {\rm (i)}--{\rm (vi)}.
For every $T\geq0$, define
$
    K\mapsto c_T(K)
    \eqdef
    C(T,K)
$ on $[0,\infty)$.
Conditions {\rm (iii)} and {\rm (iv)} imply that $c_T$ is
non-increasing.  Indeed, if
$
    c_T(K_2)>c_T(K_1)
$
for some $K_1<K_2$, then convexity would imply that every subsequent
secant slope is bounded below by the strictly positive number
\[
    \frac{
        c_T(K_2)-c_T(K_1)
    }{
        K_2-K_1
    },
\]
contradicting
$
    \lim_{K\uparrow\infty}c_T(K)=0
$.  
Since $c_T$ is convex, its right derivative $c_T^{\prime,+}$ exists
everywhere, is right-continuous and non-decreasing.  By {\rm (ii)},
convexity, and the monotonicity of $c_T$,
\[
    -1
    \leq
    c_T^{\prime,+}(0)
    \leq
    c_T^{\prime,+}(K)
    \leq
    0
\]
for every $K\geq0$.  Moreover,
$
    \lim_{K\uparrow\infty}
    c_T^{\prime,+}(K)
    =
    0
$,
since a strictly negative limit would contradict
$
    \lim_{K\uparrow\infty}c_T(K)=0
$.  
Define
\[
    F_T(x)
    \eqdef
    \begin{cases}
        0,
        &
        x<0,
        \\[1mm]
        1+c_T^{\prime,+}(x),
        &
        x\geq0.
    \end{cases}
\]
Then $F_T$ is right-continuous and non-decreasing, with
$
    \lim_{x\downarrow-\infty}F_T(x)
    =
    0
$ and $\lim_{x\uparrow\infty}F_T(x)
    =
    1$.  
Hence, $F_T$ is the distribution function of a probability measure
$\mu_T$ on $[0,\infty)$.
By construction, $
    \mu_T((K,\infty))
    =
    -
    c_T^{\prime,+}(K)
$.
Since a convex function is absolutely continuous on every compact
interval, for every $0\leq K<R$ we have
$
    c_T(R)-c_T(K)
    =
    \int_K^R
        c_T^{\prime,+}(u)
    \,du
$.
Letting $R\uparrow\infty$ and using condition {\rm (iii)}, followed
by Tonelli's theorem, gives
\[
\begin{aligned}
    c_T(K)
=
    -
    \int_K^{\infty}
        c_T^{\prime,+}(u)
    \,du
=
    \int_K^{\infty}
        \mu_T((u,\infty))
    \,du
=
    \int_{[0,\infty)}
        (x-K)^+
    \,\mu_T(dx)
.
\end{aligned}
\]
In particular, condition {\rm (i)} implies
$
    \int_{[0,\infty)}
        x
    \,\mu_T(dx)
    =
    c_T(0)
    =
    1
$.
Thus, every $\mu_T$ is a probability measure on $[0,\infty)$ with
unit mean whose call-price function is $C(T,\cdot)$.
Condition {\rm (vi)} implies that
$
    \mu_0
    =
    \delta_1
$.  
Indeed, the call-price function
$
    K\mapsto(1-K)^+
$
uniquely determines the Dirac measure at $1$.
Now let $0\leq T_1\leq T_2$.  By {\rm (v)}, we have that: for every $K\ge 0$
\[
    \int_{[0,\infty)}
        (x-K)^+
    \,\mu_{T_1}(dx)
    \leq
    \int_{[0,\infty)}
        (x-K)^+
    \,\mu_{T_2}(dx)
.
\]
Since $\mu_{T_1}$ and $\mu_{T_2}$ have the same
mean, the call-function characterization of convex order in~\cite[Proposition~2.4]{KellererTheorem_simplerproof_2012} yields
$
    \mu_{T_1}
\preceq_{\operatorname{cx}}
    \mu_{T_2}
$.
Consequently,
$
    (\mu_T)_{T\geq0}
$
is increasing in convex order.  By~\cite[Theorem~4.1]{KellererTheorem_simplerproof_2012}, this family admits a martingale $Z_{\cdot}=(Z_T)_{T\geq0}$ such that
$
    Z_T
    \sim
    \mu_T
$
for every $T\geq0$.  Since every $\mu_T$ is supported on
$[0,\infty)$, the martingale $Z_{\cdot}$ is non-negative.  Moreover,
$
    \mu_0=\delta_1
$,
so
$
    Z_0=1
$
$\P$-almost surely.  Finally, for every $T,K\geq 0$ we have
\[
    \E_{\P}
    \left[
        (Z_T-K)^+
    \right]
    =
    \int_{[0,\infty)}
        (x-K)^+
    \,\mu_T(dx)
    =
    C(T,K)
.
\] 
Thus, $Z_{\cdot}$ is a martingale
representation of $C$, and $C$ is an arbitrage-free call surface.
\end{proof}

\input{0__AFNO_MAIN.bbl}

\end{document}

%% file: 0__AFNO_MAIN.bbl
\newcommand{\etalchar}[1]{$^{#1}$}